# Multimodal operando microscopy reveals that interfacial chemistry and nanoscale performance disorder dictate perovskite solar cell stability

**Authors**

Kyle Frohna[1,2,†], Cullen Chosy[1,2,†], Amran Al-Ashouri[3], Florian Scheler[3], Yu-Hsien Chiang[2], Milos Dubajic[1], Julia E. Parker[4], Jessica M. Walker[4], Lea Zimmermann[3], Thomas A. Selby[1], Yang Lu[1,2], Bart Roose[1], Steve Albrecht[3], Miguel Anaya[1,2,*], Samuel D. Stranks[1,2,*]

**Affiliations**

[1] Department of Chemical Engineering and Biotechnology, University of Cambridge, Cambridge CB3 0AS, UK

[2] Cavendish Laboratory, University of Cambridge, Cambridge CB3 0HE, UK

[3] Division Solar Energy, Helmholtz-Zentrum Berlin für Materialien und Energie GmbH, Hahn-Meitner-Platz 1, 14109 Berlin, Germany

[4] Diamond Light Source, Harwell Science and Innovation Campus, Didcot, OX11 0DE, United Kingdom

[†] These authors contributed equally to this work

*email: sds65@cam.ac.uk, ma811@cam.ac.uk

**Abstract**

Next-generation low-cost semiconductors such as halide perovskites exhibit optoelectronic properties dominated by nanoscale variations in their structure[1,2], composition[3,4] and photophysics[5,6]. While microscopy provides a proxy for ultimate device function, past works have focused on neat thin-films on insulating substrates, missing crucial information about charge extraction losses and recombination losses introduced by transport layers[7-9]. Here we use a multimodal *operando* microscopy toolkit to measure nanoscale current-voltage curves, recombination losses and chemical composition in an array of state-of-the-art perovskite solar cells before and after extended operational stress. We apply this toolkit to the same scan areas before and after extended operation to reveal that devices with the highest performance have the lowest initial performance spatial heterogeneity – a crucial link that is missed in conventional microscopy. We find that subtle compositional engineering of the perovskite has surprising effects on local disorder and resilience to operational stress. Minimising variations in local efficiency, rather than compositional disorder, is predictive of improved performance and stability. Modulating the interfaces with different contact layers or passivation treatments can increase initial performance but can also lead to dramatic nanoscale, interface-dominated degradation even in the presence of local performance homogeneity, inducing spatially varying transport, recombination, and electrical losses. These *operando* measurements of full devices act as screenable diagnostic tools, uniquely unveiling the microscopic mechanistic origins of device performance losses and degradation in an array of halide perovskite devices and treatments. This information in turn reveals guidelines for future improvements to both performance and stability.

**Main**

Compositional engineering[10-13], contact engineering[8,14-16] and surface passivation[5,17-20] are established strategies to increase performance of halide perovskite solar cells. However, the effects of bulk and interface modulation across different length scales on perovskite solar cell performance and stability are poorly understood. To gain a complete, nanoscale understanding of device performance and degradation of next-generation optoelectronic technologies including halide perovskites, it is crucial to develop microscopy techniques capable of measuring complete devices under operational conditions. The current-voltage (JV) curve is an essential macroscopic measure of both the recombination and transport losses in a solar cell. Measuring full device stacks under operation at different points on the JV curves is essential to reveal information about charge transport and extraction in addition to non-radiative power loss channels. Several techniques have been demonstrated to probe particular points on the JV curve microscopically[21]: the short circuit current ($J_{SC}$)[4,22,23], open circuit voltage[3,7,24] ($V_{OC}$) or, by fitting a pre-determined diode model, the entire JV curve[25]. Here, we rapidly extract local, microscopic JV curves (without a preconceived diode model[26-28]) on operating solar cells by employing voltage-dependent photoluminescence (PL) microscopy. We combine this voltage-dependent PL with absolutely calibrated hyperspectral PL[3,24] and synchrotron X-ray nanoprobe fluorescence[29,30] (nXRF) to map the optoelectronic properties and chemical composition of the halide-perovskite absorber layer on the same scan area. We apply this powerful multimodal microscopy suite to an array of state-of-the-art, alloyed halide perovskite absorber layers fabricated into device stacks relevant for tandem solar cells[14,31,32] before and after accelerated operational stress with industry standard protocols (see Methods)[33] to reveal how the microscale distributions of composition, recombination, and charge transport play critical roles in dictating both device performance and stability.

We have developed a platform to measure local, spectrally resolved PL on devices under bias, allowing extraction of device performance parameters combined with local chemical composition from synchrotron nXRF mapping on the same scan area (Figure 1a). We use this to first study perovskite solar cell devices fabricated on [2-(9H-carbazol-9-yl)ethyl]phosphonic acid (2PACz), a state-of-the-art, self-assembled monolayer (SAM) hole transporting layer (HTL), with the complete device stack consisting of glass/indium tin oxide (ITO)/SAM/perovskite/$C_{60}$/$SnO_2$/Cu (Figure 1a and Methods). We employ a double-cation double-halide (DCDH) $FA_{0.83}Cs_{0.17}Pb(I_{0.83}Br_{0.17})_3$ perovskite composition that has previously been incorporated into high-efficiency single-junction and tandem solar cells[34] and reproducibly demonstrates high performance (Supplementary Figure 31). X-ray diffraction (XRD) measured of the DCDH devices shows the expected pseudocubic perovskite pattern (Supplementary Figure 32). PL centre of mass energy (COM) maps, showing the spectrally weighted average emission energy, are extracted from the locally extracted PL spectra at each point (Figures 1b and e). The COM map shows the presence of a distinct wrinkled morphology that imprints itself onto the emission energy of the perovskite (Supplementary Note 5). Interestingly, wrinkled areas exhibit red-shifted emission that correlates with a reduction in the relative Br content as revealed by nXRF on the same region (Figures 1b and 1f), although we note that the red-shifted emission may be partially explained by photon re-absorption in the thicker wrinkles[35,36]. While these measurements were performed with the device held at open circuit voltage ($V_{OC}$), the quenching of PL as a function of bias gives information about charge carrier extraction. By sweeping the voltage and comparing the broadband PL intensity ($I_{PL}$) at each point to its value at open circuit, the current extraction efficiency – the fraction of carriers extracted by the contacts, $\Phi_{PL}(V)$ – and corresponding optical-JV (J(V)) curves normalised by the generation current ($J_{gen}$) at each point can be extracted[27] (see Methods, Supplementary Note 2 and Supplementary Figures 1-7):

$$\Phi_{PL}(V) = \frac{I_{PL}(V_{OC}) - I_{PL}(V)}{I_{PL}(V_{OC})} \approx \frac{J(V)}{J_{gen}}$$

The spatially averaged optical JV and electrical JV measurements show excellent agreement (Figure 1c; see also Supplementary Note 2 for further analysis). From the hyperspectral PL and optical-JV curves extracted from each point, we extract local device figures of merit. Fitting PL spectra with the generalised Planck's law enables the extraction of the quasi-Fermi level splitting ($\Delta\mu$) and the bandgap ($E_g$)[37]. $\Delta\mu$ is a measure of the internal voltage of a solar cell and in the absence of energetic offsets and recombination at the contacts[8], should approximate the electrically measured external $V_{OC}$. In Figure 1f, we show a map of $\Delta\mu$ for the DCDH solar cell at open circuit. Comparing these values (mean of ~1.15 eV) to the electrically measured $V_{OC}$ (1.15 V) we determine that there is a negligible energetic offset between the perovskite and the contacts[8,38]. Strikingly, the wrinkled areas do not appear to negatively affect the $\Delta\mu$ values (see Supplementary Note 5). PL spectra from this region (Figure 1b) reveal these wrinkled regions have sufficiently increased PL intensity to almost exactly counteract the expected $\Delta\mu$ loss caused by their reduced bandgap compared to surrounding regions and are therefore benign to local voltage losses (Supplementary Notes 4 and 5). Figure 1h shows that the optical current extraction efficiency $\Phi_{PL}(0V)$ is also spatially homogeneous, although a subset of the wrinkles exhibit slightly worse charge extraction. In Figure 1i, we define the optical power conversion efficiency (PCE) as the product of the maximum power point voltage and the current extraction efficiency at this voltage ($\Phi_{PL}(V_{MPP}) \times V_{MPP}$, units of V). We find a tight optical PCE spatial distribution of less than ±5% relative, particularly striking given the morphological and optoelectronic variations of the perovskite itself. Representative JV curves extracted from pristine and wrinkled areas are shown in Figure 1d highlighting the spatial PCE homogeneity and the relatively small impact wrinkles have on PCE; such a conclusion is in contrast to previous works suggesting that these wrinkles may be detrimental to device performance and longevity[39,40]. This overall good spatial homogeneity is a hallmark of highly efficient, stable devices as we will discuss further below. The correlation between Br:Pb and several device figures of merit summarises our findings of the relationship between local chemistry and performance: while the local Br:Pb ratio does modulate the bandgap and the non-radiative voltage loss, it has little effect on either $\Delta\mu$ or PCE (Supplementary Note 4), showing that this device stack is very tolerant to chemical disorder.

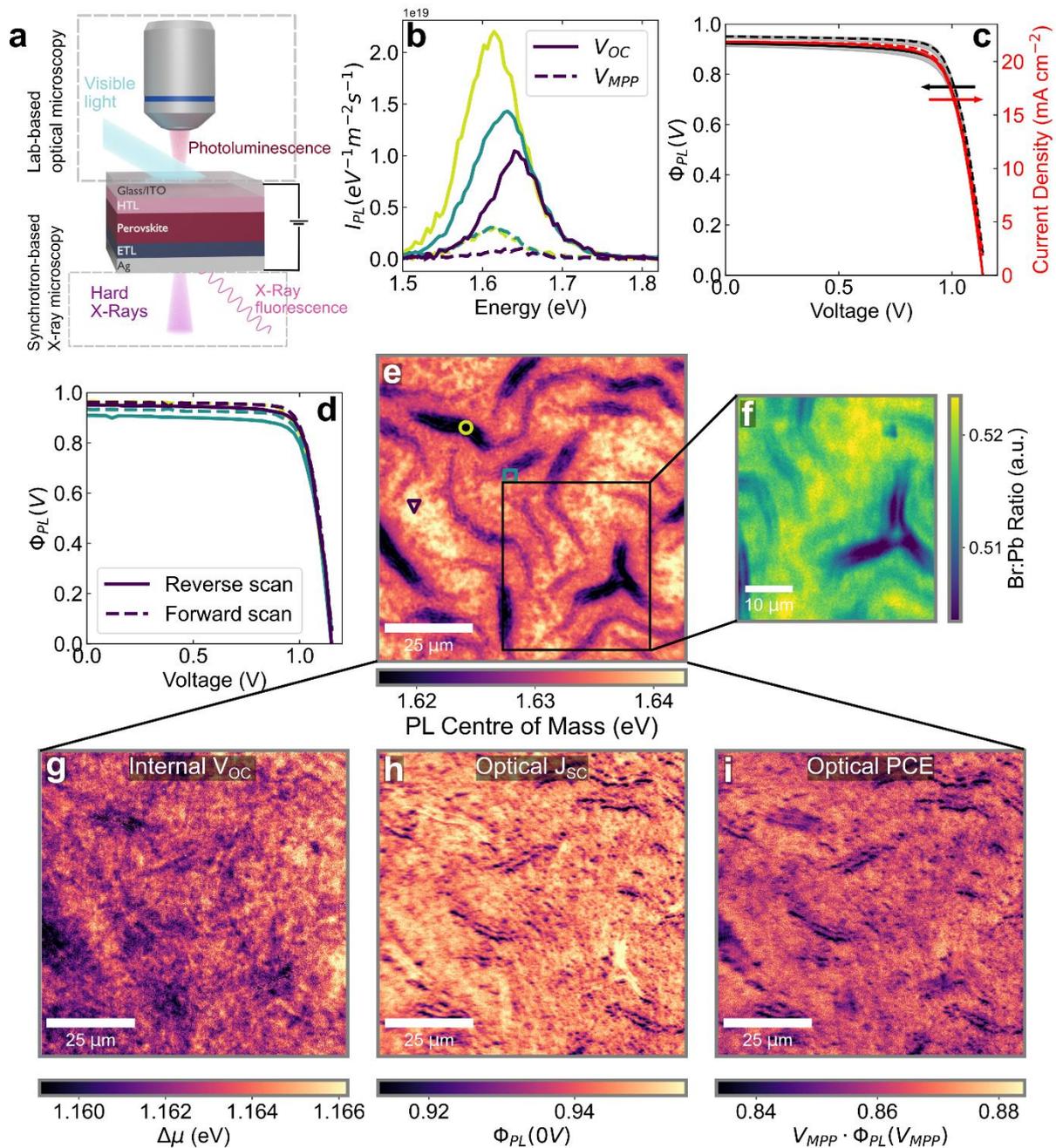

**Figure 1. Device operando microscopy reveals DCDH solar cell performance is tolerant to even dramatic spatial optoelectronic and chemical heterogeneity.** a) Schematic of perovskite solar cell under bias being illuminated either by a white light LED array for the luminescence measurements or monochromatic hard X-rays for the nXRF measurements. b)

hyperspectral PL spectra (at $V_{OC}$ and $V_{MPP}$) of regions marked in panel e. c) Comparison of electrical JV curve (red) and area averaged optical JV curve (black) of DCDH solar cell. Gray shaded areas show distribution of JV curves across the map. d) Optical JV curves of the marked regions in e. e) PL centre of mass (COM) energy plot of a region of a DCDH solar cell at $V_{OC}$. f) Br:Pb map from the marked region in b extracted by nXRF. g) Internal $V_{OC}$ ($\Delta\mu$), h) optical short circuit current extraction efficiency ($\Phi_{PL}(0V)$) and i) Optical PCE ($V_{MPP} * \Phi_{PL}(V_{MPP})$) of the same region as shown in e.

*Microscopic effects of device operation and degradation*

We now subject the devices to extended operational stress to probe the interplay between microscopic changes to solar cell performance, and local composition/optoelectronic properties under operational stress. Specifically, the unencapsulated cells are held at open-circuit voltage under continuous 1 sun illumination 65 °C and continuous nitrogen flow for 100 hours following the standardised stability test ISOS-L-2I (ref. [33]) (see Methods). Holding cells at open circuit under illumination and added external heat is a considerably greater stability challenge compared to a maximum power point track so is ideal for an accelerated operational stress test[41,42]. Figure 2a shows that the initial spatial variation in optical PCE is extremely small (±2% relative). However, upon remeasuring the same area after the operational stress test (Figure 2b), a large global drop in optical PCE is observed along with a striking increase in spatial PCE heterogeneity characterised by a diagonal progression across the sample. The local optical JV curves extracted from the regions marked in the PCE maps reveal that the operation-stressed samples exhibit spatially varying hysteresis behaviour between the forward and reverse voltage scans (Figure 2e), which was absent before operation (Figure 2f). Some regions such as the one marked in the bottom right corner of Figure 2b display a large s-kink in the JV curve where effectively no charge extraction is observed up to 0.3 V below $V_{OC}$. Other regions (square and circle markers) display substantial, spatially varying degrees of apparent series resistance, a key parameter to which our technique is sensitive (Supplementary Note 2). This extraction issue is mainly present in the reverse scan whereas much more spatially homogeneous, higher optical PCE is observed in the forward scan (Supplementary Note 6 and Supplementary Figure 33). However, we see no shifting in the local PL peak position of the perovskite before and after operation and a global increase in PL intensity at $V_{OC}$ is apparent (Figures 2 g and h, Supplementary Figure 34); these results suggest that the bulk of the perovskite absorber layer itself has not been degraded after the extended operational stress and that the issues lies at the interfaces with the contacts, as we will discuss further in a later section.

This result is further supported by compositional maps in Figures 2c and d displaying regions of the scan area that exhibit starkly different optical PCE distributions but very similar chemical distributions, with spatial chemical distributions (Br, Pb, I, Cs and ratios thereof) of the perovskite in fresh and operated devices showing no appreciable change (Supplementary Figures 35-36). Intriguingly, the wrinkles also appear to be enriched with Cs as has been proposed previously[43,44]. These same results are seen in multiple measurements across different scan areas, devices and batches (Supplementary Figures 37-43). Results such as these demonstrate the power of our *operando* microscopy technique over conventional microscopy at open circuit which would be blind to such degradation; further, PL at open-circuit cannot be used alone as a performance metric.

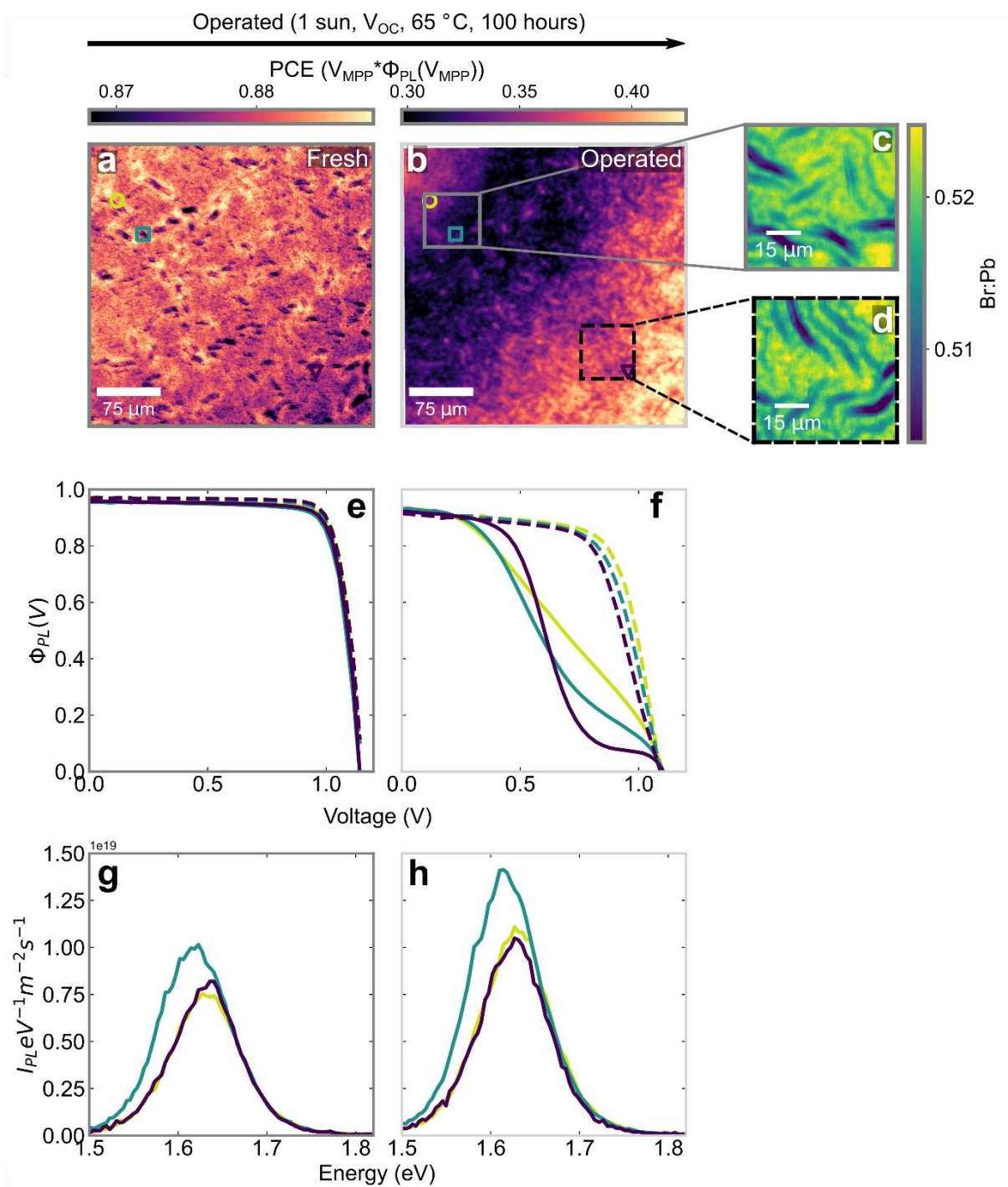

**Figure 2. Local reductions in performance are evident in microscopic JV curves of DCDH perovskite solar cells after extended operation.** Optical PCE maps of the same area of a fresh a) and operated b) DCDH solar cell after 100 hours at $V_{OC}$, 65 °C and 1 sun illumination. c) and d) show Br:Pb ratio maps extracted from nXRF from the two regions marked in panel b) after the 100 hours of operation. e) and f) show optical JV curves before and after ageing from the points marked in panels a and b. Solid lines are reverse scans, dashed lines are forward

scans. g) and h) show PL spectra from the same marked areas before and after ageing respectively.

The stark spatial variation of the voltage-dependent PL shows a degradation front extending from the edge of the active area defined by the overlap of the metal and ITO electrodes. Recent reports have shown that mobile ionic species in the perovskite are particularly problematic at the edges of active device areas and could cause degradation consistent with what we observe in Figure 2b and d[45]. Large area scans of whole device pixels highlight that this degradation front appears from the edges and can even appear in pristine, as-made devices if there are issues with fabrication (Supplementary Figure 44-45). Macroscopic hysteresis in the JV characteristics of perovskite solar cells has been attributed to the presence of mobile ions that can screen electric fields[46,47]. Given that the observed difference between local forward and reverse optical JV scans is large, we attribute this difference to spatially varying concentrations of mobile ions rather than a resistive effect. Looking further away from the edge of the contacts, the spatial variations in optical PCE and hysteresis are reduced (Supplementary Figures 46-48. The macroscopic electrical JV curve in Supplementary Figure 49 shows a hysteresis in between these two extremes, highlighting that the macroscopic hysteresis is an ensemble average of many locally varying mobile ionic concentrations. These observations suggest that cell and module designs preventing edge effects will also play an important role to boost the stability of perovskite solar cells.

*Microscopic degradation in compositionally tuned perovskite compositions for tandem architectures*

To investigate how compositional tuning affects microscopic device performance and stability, we now consider the same device architecture but with the double-cation triple-halide (DCTH) $FA_{0.83}Cs_{0.17}Pb(I_{0.81}Br_{0.16}Cl_{0.03})_3$ composition in which Cl is present in the system introduced through $PbCl_2$ in the precursor solution. The incorporation of Cl increases the bandgap for use as the top cell in high-performance multijunction devices[32,48] (Supplementary Figure 50). Macroscopic electrical JV measurements show an increase in $V_{OC}$ and corresponding decrease in $J_{SC}$, compared to the DCDH analogue, as expected with the increased bandgap. However, there is also a reduction in fill factor, suggesting that the additional compositional alloying has had a small negative effect on carrier transport and extraction (Supplementary Figure 31).

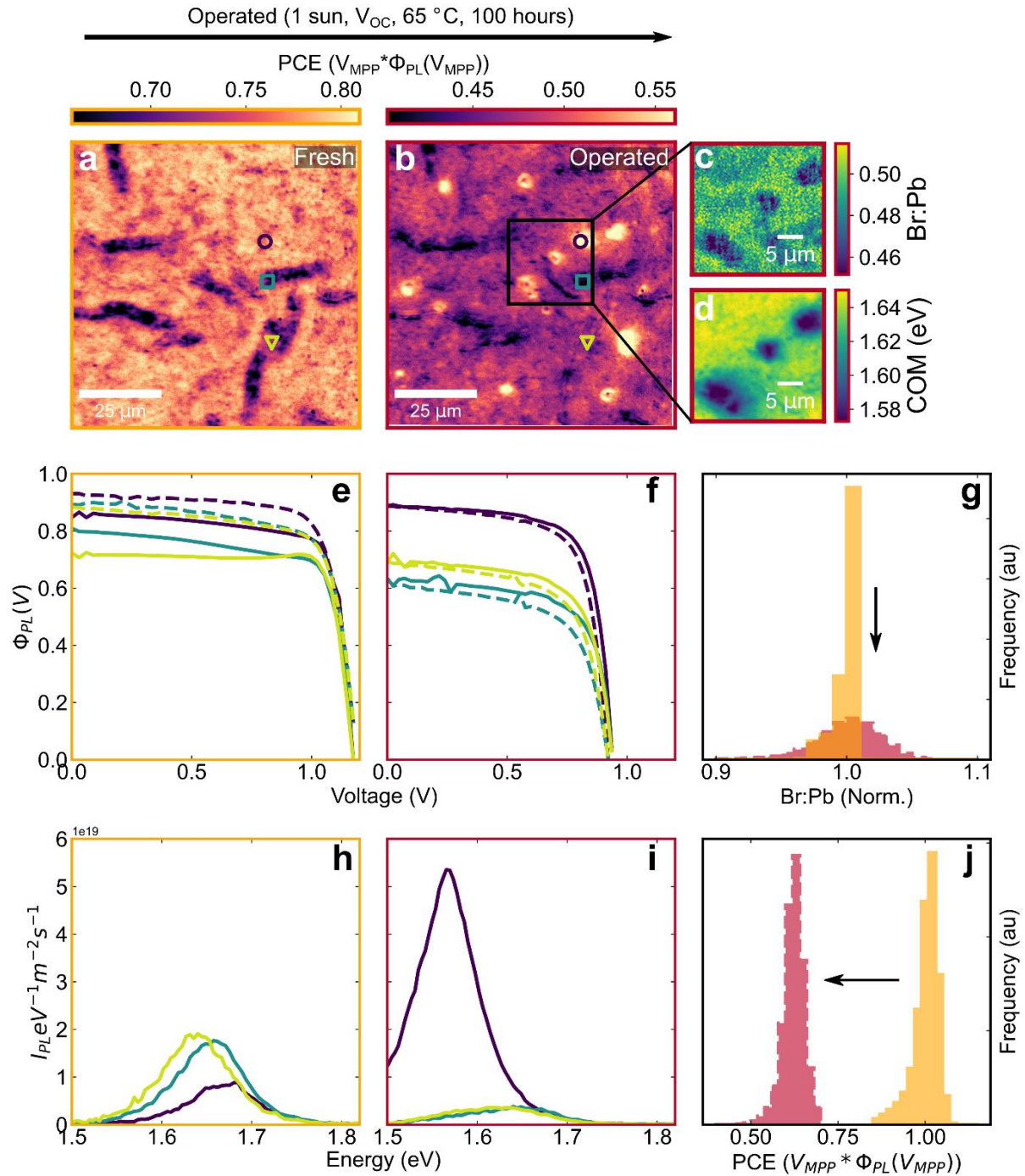

**Figure 3. Multimodal microscopy on DCTH perovskite solar cells reveals reduced device stability and increased microscopic phase segregation compared to DCDH analogues.** Optical PCE maps of the same area of a fresh a) and operated b) DCTH solar cell after 100 hours at $V_{OC}$, 65 °C and 1 sun illumination. c) Br:Pb ratio and d) PL centre of mass energy maps extracted from the region marked in panel b after the 100 hours of operation. Optical JV curves e) before and f) after operational stress from the points marked in panels a and b. Solid lines are reverse scans, dashed lines are forward scans. g) Normalised Br:Pb ratio histograms

for a pristine sample (orange) and the marked region of the operated (red) sample. PL spectra extracted from the same marked areas h) before and i) after ageing respectively. j) Optical PCE histogram for this DCTH solar cell before (orange) and after (red) operation.

A wrinkled morphology with similar surface area coverage and size to the DCDH solar cells is also observed in the DCTH devices (Supplementary Note 5). The optoelectronic behaviour of the wrinkled areas is more varied as some wrinkles now show increased Δμ compared to their surroundings, whereas other regions are reduced (Supplementary Figure 51). However, interestingly, the wrinkles have a more significant impact on the local PCE with relative reductions of up to 20% compared to pristine areas (Figure 3a). This is emphasised by the much greater spatial variance in pseudo-JV curves extracted from across the pristine device (Figure 3e). There is therefore a much greater spatial PCE heterogeneity in the DCTH sample (Figure 3j) compared with the tight PCE distribution of the relatively homogeneous DCDH (Supplementary Figure 52). The fill factor values extracted from the optical JV curves in Figure 3c are also markedly reduced, indicating that the chloride introduces additional nanoscale disorder that hampers charge extraction while increasing non-radiative recombination, suggesting that alloying needs to be carefully managed in halide perovskites on both macro and nanoscales.

After the same extended operational stability test, we see substantially greater degradation in these solar cells than the DCDH analogues. There is a global reduction in relative PCE of >40% (Figure 3j), which is accompanied by the emergence of regions 1-10 µm in size where performance is strikingly higher than their surroundings albeit still reduced from their initial values, and exhibit a five-fold increase in PL intensity and large PL peak redshift of 0.1 eV (Figure 3b, h and i). Combining nXRF and hyperspectral PL, we find that these areas show a large local reduction in Br content and assert that this is due to exaggerated local halide segregation[3,49] driven by the combination of light and heat stressors in the operational test. In most regions, there is a reduction in both optical short circuit current and $V_{OC}$ (Figure 3f), meaning an increase in losses caused by both charge extraction and non-radiative recombination. By contrast, the phase segregated regions display a higher charge extraction efficiency and PCE than their surroundings, suggesting a mitigation of degradation pathways related to charge extraction.

The relative increase in PL intensity in the phase segregated areas is not sufficient to compensate for the voltage loss caused by the dramatic red shifting of the PL peak and so these regions exhibit a lower $\Delta\mu$ compared to their surroundings (Supplementary Figures 51, 53-54). The segregation persists even after extended storage in the dark, as nXRF maps show clear compositional segregation and increased spatial variance in the Br:Pb signal when measured several days after the operational stability test (Figure 3c and 3d). Here, the microscopy suite has revealed the microscopic impact of phase segregation on device stability and performance – an ongoing issue for attaining stable tandems[50]. Unlike the DCDH composition, the DCTH devices also show a transient increase in PL intensity during voltage sweeps, causing $I_{PL}(V_{OC})$ to differ in the reverse and forward scans (a vertical offset close to 0 V) and affecting the observed optical JV curves. We refer to this phenomenon as 'optical hysteresis' and it is clearly observable in optical JV curves extracted from the DCTH sample as shown in Figure 3e (see Supplementary Note 6 for discussion of hysteresis). Taken together, our measurements on the DCTH samples show that the addition of the $PbCl_2$ into the otherwise unchanged precursor solution meaningfully increases the bandgap of the resulting perovskite, but at the cost of increased spatial PCE disorder and hysteresis, ultimately hampering charge-carrier extraction, phase and device stability relative to the control DCDH devices.

As a further compositional tuning step, we added MACl to the DCTH solution to produce triple-cation triple-halide (TCTH, $MA_{0.03}FA_{0.81}Cs_{0.16}Pb(I_{0.81}Br_{0.16}Cl_{0.03})_3$) solar cells – the composition used in many of the highest performing perovskite tandems including our own[18,32,48]. The additional compositional disorder in the precursor solution corresponds to slightly increased spatial disorder in the device (Supplementary Figure 56) and a corresponding slight reduction in initial device performance (Supplementary Figure 31). However, surprisingly, the incorporation of MA results in a large increase in both phase and device stability. After operational stress, there is little to no loss in either open circuit PL intensity or $J_{SC}$, and phase segregation is effectively suppressed (very minor effect on $\Delta\mu$) although not eliminated entirely (Supplementary Figure 57-59) – a marked improvement over the DCTH in all aspects. Overall, we have revealed that even minor compositional variations can have substantial impact on the microscopic and thus macroscopic behaviour of wide bandgap perovskite solar cells under operational stress.

*Interfacial Engineering and Passivation: A Double-Edged Sword for Microscopic Performance and Stability*

To further understand performance issues and instabilities in high efficiency devices, we now explore contact engineering and surface passivation. Here we fix the TCTH perovskite composition due to its stability and utility in tandems and study the impacts of modulating both surfaces independently. We vary the HTL from 2PACz to MeO-2PACz and Me-4PACz, two other SAM HTLs that have been used in high efficiency solar cells[18,31] and shown to affect charge extraction, interfacial recombination, energetic alignment and substrate hydrophobicity[3,8,16,31]. We additionally fix the 2PACz HTL while passivating the perovskite/$C_{60}$ interface with either a thin LiF interlayer, or the ionic liquid piperazinium iodide (PI), both of which have been shown to substantially boost $V_{OC}$, adjust surface charge and modulate charge extraction[18,31,51-53]. We show optical PCE maps of the 2PACz control, Me-4PACz and 2PACz+PI passivation TCTH devices in Figure 4 a-c (Supplementary Figure 60-66 for MeO-2PACz, Me-4PACz and 2PACz+LiF passivation). While the samples display similar morphology, the optical PCE maps of the Me-4PACz and 2PACz+PI/LiF passivated devices show higher mean values than the 2PACz control and exhibit lower spatial variation in performance (see Supplementary Figures 67-69 for comparisons). This is reflected in the electrical JV curves of the devices (Figure 4d) where both passivation strategies produce a $V_{OC}$ of ~1.25 V, a boost of 0.1 V over the control and the Me-4PACz device shows a more modest spatial homogenisation of optical PCE and boost in $V_{OC}$. We note that the optical PCE of the PI-passivated devices is partially overestimated due to shunting (Supplementary Note 2), while this is not the case for the LiF devices.

To quantify the spatial heterogeneity in the optical PCE distributions for each sample, we fit optical PCE histograms with a Gaussian function and define the 'Initial PCE Disorder' as the full width at half maximum (FWHM) of the fitted Gaussian peak. Histograms and corresponding fits are shown in Figure 4e. These data lead to a key general observation that lower optical PCE disorder correlates with higher performing devices. We perform this analysis across our entire device range consisting of 3 perovskite compositions, 3 transport layers and 2 surface passivation treatments and find that this observation holds across this wide device space covering state-of-the-art perovskite developments for tandem solar cells (Figure 4f).

We then expose the interface-modified devices to the same operational stress test as the compositionally engineered samples (Supplementary Figures 60-66 for maps). JV curves after stress are shown in Figure 4g. We find that the control device almost fully retains its original $J_{SC}$ but has lost voltage without any loss in PL intensity – suggesting some losses at the contacts while the bulk perovskite remains relatively pristine. By contrast, the device with the adjusted HTL Me-4PACz has substantially degraded electronically (from JV scan) together with a loss in PL intensity (although on its own not sufficient to explain the drop in $V_{OC}$, Supplementary Figure 70-71) suggesting the interface is to blame for these losses. The PI- and LiF-passivated devices show an equivalent $V_{OC}$ loss (and a corresponding loss in PL intensity), while the LiF device retains substantially more $J_{SC}$. We note that neither the controls nor the passivated devices show any substantial changes in the XRD pattern of the absorber layers[54], consistent with the issues being localised to interfaces rather than any bulk structural changes (Supplementary Figure 72).

To summarise our stability findings, we find that all device types containing Cl exhibited increased transient electronic behaviour after the operational stress test (Supplementary Figure 73) and a larger loss in $V_{OC}$ compared to the loss of internal voltage $\Delta\mu$ (or lack thereof in the 2PACz/TCTH case, Supplementary Figure 74). The transient behaviour and mismatch between internal and external voltages suggests that mobile ions in our Cl-containing devices play a larger role after operational stress compared to double-halide equivalents. In the perovskite compositional tuning series with fixed, relatively stable interfaces, initial optical PCE disorder is a good predictor of overall device stability (Figure 4h), and one could therefore screen and predict for stable devices by looking at the optical PCE distributions in the as-made devices. However, the same is not necessarily true for interfacial modification. The array of interfacial modification methods we tested resulted in improved initial electrical performance and reduced optical PCE disorder, but most, apart from LiF passivation, also caused dramatically poorer stability (Figure 4i). Modifying either interface while keeping both the perovskite composition and the rest of the device stack constant can drastically change the stability of the device. Such a conclusion is consistent with recent observations about a trade-off between efficiency and stability in perovskite LEDs[55].

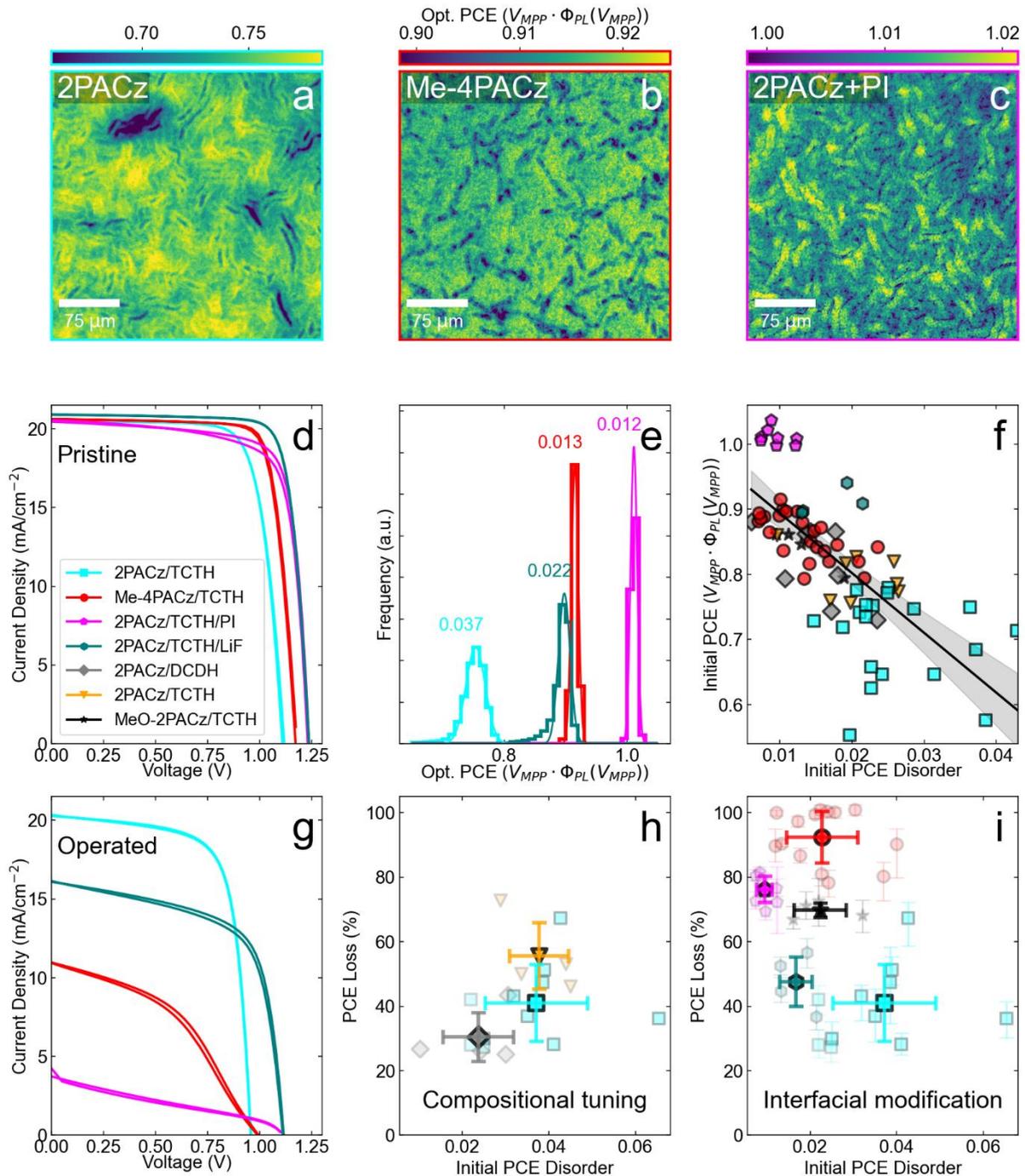

**Figure 4. Interfacial chemistry and spatial PCE disorder predict performance and stability of mixed-cation, mixed-halide perovskite solar cells.** Optical PCE maps of pristine a) control 2PACz/TCTH, b) Me-4PACz/TCTH and c) 2PACz/TCTH + PI passivation devices. d) Representative JV curves of the pristine interface modified devices mapped in panels a-c. e) Optical PCE distributions and corresponding Gaussian fits of the interface modified devices. The numbers over the distributions represent the FWHM of the distribution. f) Initial optical PCE disorder (FWHM of PCE distribution) versus initial PCE (mean of PCE distribution) for a range of perovskite devices, where each point is an individual device. Linear regression

shows a Pearson's r-value of -0.71, Spearman's r-value of -0.77 and a p-value of << 0.01. Shaded regions represent the 95% confidence interval of the linear fit. g) JV curves after operational stress of the representative devices shown in panel d. h) Scatter plot of initial PCE disorder versus PCE loss (%) during operation for the perovskite composition series. i) Scatter plot of initial PCE disorder versus PCE loss (%) during operation for the interfacial modification series. Solid markers are the mean of a given device type, transparent markers are individual devices. Error bars are standard deviations.

**Discussion**

Our microscopic *operando* studies of a wide device space now allow us to propose general design rules towards practical realisation of efficient and stable perovskite solar cells. The combination of local chemistry, emission properties and optoelectronic performance allow us to discern which features are causing degradation and which are most important to address microscopically – providing critical impact on overall performance and stability. The first is that interfacial stability must be the top priority understanding and stabilising the interfaces that impact the long-term stability of the device is crucial. We have shown that initial PCE can be rapidly lost if the interfacial stability is compromised to achieve this end, and this effect dominates over any initial high-performance metrics or PCE homogeneity. This highlights an ongoing challenge in the field to achieve an appropriate passivation for wide bandgap perovskite top cells to achieve both high efficiency and stability simultaneously and reproducibility. If stable interfaces are achieved, the second crucial factor affecting performance and stability is the local optical PCE disorder, which can be tuned via a combination of perovskite composition and interfacial chemistry. In the most severe cases, increasing initial compositional disorder can seed the formation of micron sized, extremely phase segregated domains and our techniques have demonstrated the unexpected charge extraction behaviour of these domains. This multimodal *operando* methodology has revealed that variations in composition, morphology and bandgap can be tolerated in efficient devices with stable interfaces while losses in charge extraction manifesting in optical PCE disorder cannot be, highlighting the utility of this approach over macroscopic measurements and conventional PL microscopy which would miss such a conclusion. *Operando* microscopy of devices can therefore be a powerful predictive screening tool and time-saver for long term operational stability. Our multi-modal toolkit additionally enables detailed 'post-mortem' examinations of devices after extended stress testing – revealing detailed information on the chemical, optical and electronic properties of the entire device stack and what has degraded, ultimately guiding process optimisation.

Taken together, our measurements show that subtle changes to perovskite composition and interfacial chemistry have serious consequences for the degradation mechanisms at play in these devices, and we pinpoint the location and cause of the degradation. Changes in the defect density at an interface may render a kinetically slow process suddenly favourable, or changes in the hydrophobicity of the contact layer may change the crystallisation kinetics of the active layer sufficiently to cause large-scale heterogeneities that can seed degradation[56]. This emphasises the paramount need to be optimising devices simultaneously for both performance and robust, reproducible stability. It also urges caution for the generality of proposed degradation mechanisms because minor variations may alter the 'weak link' of stability in the devices of interest. These measurements show that the nanoscale device performance landscape is remarkably complex in perovskite solar cells. Microscopic platforms such as that demonstrated here that can measure local optoelectronic and charge extraction properties are crucial to understanding the degradation pathways in disordered semiconductor devices as they generate powerful insights[22] and unveil mechanistic pathways which need to be addressed for the success of these devices. Metal-halide perovskites were used as a case study in this work, but the techniques developed for the multi-modal, microscopic toolbox are generalisable to a broader class of light harvesting and emitting devices based on disordered materials, and exciting insights may be gained in areas such as organic and copper-indium-gallium selenide photovoltaics as well as InGaN light emitting diodes.

**Acknowledgements**

K.F. acknowledges an Engineering and Physical Sciences Research Council (EPSRC) Doctoral Prize Postdoctoral Fellowship, George and Lilian Schiff Studentship, Winton Sustainability Fund Studentship, and an EPSRC studentship. C.C. acknowledges the support of a Marshall Scholarship and Winton Sustainability Fund Studentship. The authors acknowledge the Diamond Light Source (Didcot, Oxfordshire, UK) for providing beamtime at the I14 Hard X-ray Nanoprobe facility through proposal MG30427 and MG31964. M.A. acknowledges support by the Royal Academy of Engineering under the Research Fellowship programme. S.D.S. acknowledges the Royal Society and Tata Group (UF150033). Funding for A.A.A.,

F.S., L.Z. and S.A. was provided by the Federal Ministry of Education and Research (BMBF) through the project PEROWIN (grant no. 03SF0631) and by the Helmholtz Association within the projects HySPRINT Innovation lab, the EU Partnering project TAPAS and the project "Zeitenwende – Tandem Solarzellen". T.A.S. Acknowledges funding from EPSRC Cambridge NanoDTC, EP/S022953/1. The work has received funding from the European Research Council under the European Union's Horizon 2020 research and innovation programme (HYPERION - grant agreement number 756962). The authors acknowledge the EPSRC (EP/R023980/1 and EP/T02030X/1) for funding. The authors acknowledge Rebecca A. Belisle, Stuart Macpherson, Tracy Schloemer and Monique Merchant for helpful discussion and comments on the manuscript. The authors acknowledge Eunyoung Choi for assistance with synchrotron measurements. For the purpose of open access, the authors have applied a Creative Commons Attribution (CC BY) licence to any Author Accepted Manuscript version arising from this submission.

**Author Contributions Statement**

K.F., M.A. and S.D.S. conceived the project. K.F., C.C., M.A. and S.D.S. designed the hyperspectral and operando PL mapping experiments, and K.F., C.C., M.A. performed them. K.F., C.C., M.A. and M.D. performed the synchrotron X-ray microscopy with assistance from J.E.P. and J.W.M. C.C. performed the automated image registration. K.F. and C.C. processed, analysed, and interpreted, with input from M.A. and S.D.S., the optical and X-Ray microscopy data. A.A.A, F.S. and L.Z. supervised by S.A., and Y-H.C. fabricated the solar cells and performed macroscopic current-voltage testing. B.R. constructed the stability setup employed by K.F. and C.C. K.F. performed XRD measurements. T.A.S. and Y.L. performed SEM measurements. M.A. and S.D.S. supervised the project. S.D.S. funded the work. K.F. wrote the manuscript with input from C.C., M.A. and S.D.S. All authors commented on the final version of the manuscript.

**Competing Interests**

S.D.S. is a co-founder of Swift Solar.

**Data Availability Statement**

All of the data supporting this work is available at the repository [DOI to be added before publication].

**Code Availability Statement**

The code to register microscopy images is included in the repository [link to be added before publication].

**Materials and Methods**

Materials

Lead salts were purchased from TCI Chemicals (>99.99% purity trace metals basis), FAI and MABr from Dyenamo (>99% purity), CsI from abcr GmbH (>99% purity) and MACl (>99% purity) from Sigma Aldrich. The salts were used as received without further purification. Solvents were purchased from Merck (>99.8% purity, anhydrous) and were used as received without further drying or purification. Piperazinium iodide was synthesized in house as described in ref. [18].

Film and Device Fabrication

Solar cells were fabricated on ITO substrates, which were subsequently cleaned in a Mucasol solution (2% in DI-water), deionized water, acetone, and 2-propanol, each for 15 minutes in an ultrasonic bath. Afterwards, the surface was "activated" for 10-15 minutes in an UV-O3 cleaner (FHR UVOH 150 Lab), which is a crucial step before SAM deposition. The SAM solutions (1 mM/l in ethanol) was spin-coated at 3000 rpm for 10s, after which the substrate was annealed at 100°C for 3-10 min. All spin-coating layer deposition steps were conducted in a nitrogen atmosphere.

The cell configuration is ITO/SAM/Perovskite/C60(20 nm)/SnO2(20 nm)/Cu(100 nm), where the $C_{60}$ and Ag were deposited by thermal evaporation and the SnO2 layer was deposited by atomic layer deposition in an Arradiance GEMStar reactor. There was no air exposure between any of the layer deposition processes. Tetrakis(dimethylamino)tin(IV) (TDMASn) was used as the Sn precursor and was held at 60 °C in a stainless-steel container. Water was used as the oxidant from a stainless-steel container without active heating, whereas the precursor delivery manifold was heated to 115 °C. For the deposition at 80 °C, the TDMASn/purge1/H2O/purge2 times were 1 s/10 s/0.2 s/15 s with corresponding nitrogen flows of 30 sccm/90 sccm/90 sccm/90 sccm. With this, 140 cycles lead to 20 nm of SnO2.

Perovskite layers: First, a 1.4 M "FACs" solution (FA, Cs, PbI2, PbBr2; 22% Cs & 15% Br) in 3:1 DMF:DMSO was shaken at room temperature overnight (solution for the DCDH perovskite). For the DCTH composition, this solution was transferred into another vial that contained PbCl2 powder and shaken for 1 h at 60°C before perovskite layer deposition, with a nominal molar Cl percentage of 5%. For the TCTH composition, this second vial contained both PbCl2 and MACl. Exemplary amounts of chemicals for 1 ml of 1.4 M solution: 500 mg PbI2, 116 mg PbBr2, 188 mg FAI, 80 mg CsI (weighed into one vial) + 4.7 mg MACl, 19.5 mg PbCl2 (in another vial).

The perovskite solution was spin-coated at 3500 rpm for 40 s and 250 µL anisole as the antisolvent was dripped at 28 s after start of the spinning, followed by 20 min annealing on a hotplate at 100°C in $N_2$.

For interface modification, either PI or LiF were desposited on the perovskite prior to C60 evaporation. For the PI treatment, 100 µL of a solution of 0.3mg/mL piperazinium iodide in 2-propanol were dynamically spin-coated at 5000 rpm on the perovskite, followed by 2 min annealing at 100 °C. The sample was then washed with 100 µL 2-propanol and again annealed for 2 min at 100 °C. For LiF, 1 nm was thermally evaporated without breaking the vacuum.

Bulk Solar Cell Characterisation

The ISOS-L-2I accelerated degradation protocol was carried out under an LED array solar simulator (see Supplementary Figure 75 for spectral details). Unencapsulated devices were placed into an O-ring sealed chamber with a quartz window. The chamber was heated above by the LED array and below by a PID temperature probe-controlled hotplate. The hotplate heat output was adjusted until the internal temperature of the chamber was 65 ± 1 °C as measured intermittently by an infrared temperature probe. Devices were kept at open circuit voltage due to the more stressful nature of this ageing protocol to devices versus maximum power point[22].

The solar cells contained each 6 pixels with an active area of 0.16 cm² (overlap of patterned ITO and the Cu stripe, area confirmed with optical microscope), measured with an Oriel LCS-100 ABB solar simulator and Keithley 2400 source-measure unit inside a $N_2$ glovebox. The JV was scanned in 20 mV steps with 20 ms integration time and 20 ms delay time between each voltage step and measurement. JV testing was performed without a mask. The solar simulator was calibrated using a reference KG3 filtered silicon solar cell calibrated by Fraunhofer ISE. The spectral mismatch between the desired spectrum and the solar simulator is ~0.997, within experimental error so no correction is applied.

XRD

XRD patterns were obtained using a Bruker D8 ADVANCE and a Copper X-ray tube operating at 40 kV with Ka emission wavelength of 1.54 Å. The samples were measured in ambient air. The scan range for 2θ was from 7º to 40º with a step size of 0.01º and a dwell time of 0.55 s per angle.

SEM

SEM micrographs were acquired using an FEI Helios FIB/SEM operated with an accelerating voltage of 2kV; current of 0.2nA; and working distance of approximately 4 mm. An Everhart–Thornley detector acquired multiple secondary electron images consecutively (typically 32) at a pixel dwell time of 300 ns which were overlaid to improve the signal to noise ratio. In some micrographs the sample stage was tilted to 40 degrees such that the out-of-plane morphology could be observed.

Hyperspectral Operando Luminescence Microscopy

Hyperspectral microscopy is performed using a Photon Etc. IMA microscopy system. 20x (Nikon TU Plan Fluor, 0.45 NA) and 63x glass collar corrected objectives (Zeiss LD Plan-Neofluar 63x/0.75 Corr M27) with appropriate chromatic aberration corrections were used for all measurements due to their ability to focus through our ~1.1 mm thickness ITO/glass device substrates. The samples are stored in a nitrogen filled glovebox before being transferred to the microscope for measurements. The devices are held in a custom 3D-printed device holder allowing individual pixels to be biased with the use of a Keithley 2450 sourcemeter. The devices are optically excited using a fibre-coupled LED array consisting of variable-power red, green, and blue LEDs from Thorlabs (M455L4, M530L4, M617L3 respectively). The devices are illuminated with this LED array at an acute angle rather than at normal incidence through the objective. This is to ensure that the entire pixel was illuminated uniformly as illuminating a small spot on the device while leaving the rest in the dark caused considerable artefacts in the resulting data such as artificially reduced apparent optical current densities. This acute angle does not cause significant artefacts in our data due to the large refractive index or the wide bandgap perovskite across the visible range. Assuming a refractive index of the perovskite of 2.55(ref. [57]), even light incident at close to 90 degrees from the azimuth is bent to an angle of 22 degrees from the normal, resulting in minimal shading from the wrinkles. The resulting relative path length increase is only 1.086. The power of the LEDs is adjusted so that the measured short circuit current density under the microscope matched the short circuit current of the device measured under our standard LED solar simulator. The powers are adjusted so that one third of the current came from each of the red, green and blue LEDs in order to crudely approximate a white light, solar illumination profile. An appropriate long pass filter (Semrock BLP01-664R) is placed in the luminescence collection path to remove any scattered excitation.

For the hyperspectral measurements, the device is held at a particular voltage (either open circuit or maximum power voltage). The device is illuminated as described above, emitted light is collected in the objective and is incident upon a volume Bragg grating which splits the light spectrally onto a high sensitivity CCD camera (Hamamatsu ORCA Flash 4.0 V3 sCMOS camera) with 2048×2048 6.5×6.5 $\mu m^2$ pixels that is thermoelectrically cooled to -10 °C. A hyperspectral image is created by scanning the angle of the grating with respect to the emitted light. The microscope is calibrated for absolute number of photons to extract quantitative PL spectra using the methodology previously reported[3]. Details of data fitting to extract the quasi-Fermi level splitting and centre of mass are included in Supplementary Note 1.

Voltage-dependent photoluminescence mapping is performed using the Photon Etc. widefield microscope equipped with the large-area LED illumination area as described above. The setup was used in broadband mode where the grating is rotated to the zeroth order diffraction and acts simply as a mirror. A Keithley 2450 sourcemeter is used to step the applied voltage between image acquisitions, performing sweeps from open circuit to short circuit, then back in a total of 80 steps. Scan rates are always kept to 0.01 V s$^{-1}$ or below in order to avoid additional scan-rate dependent hysteresis and to allow transient changes in luminescence and extracted current to stabilise at a given voltage[27]. While the voltage dependent PL measurement is occurring, a simultaneous macroscopic electrical JV measurement is also performed. Full details of data treatment to extract solar cell performance metrics are included in Supplementary Note 2.

Nanoprobe Synchrotron X-ray Fluorescence

Synchrotron measurements were performed on the I14 hard X-ray nanoprobe beamline at Diamond Light Source Ltd., Didcot, UK. Samples were stored in an Ar filled glovebox prior to measurements. The full experimental setup has been described elsewhere[58] and the experimental is very similar to that we have reported previously[3]. X-rays from an undulator source are monochromated to produce a 15 keV X-ray beam which is focused by a pair of Kirkpatrick-Baez mirrors to produce a beam with a FWHM of approximately $50 \times 50$ nm at its focus. For mounting of the large solar cell substrates, a custom designed, 3D printed sample holder was used to enable reproducible mounting on the nanoprobe endstation stages. The sample is placed at the focus and is laterally scanned across this focal point to produce the final maps. The energy-resolved nXRF signal is collected with a 4-element silicon drift detector in a back-scattering geometry. The data was analysed in part with the open-source Python package Hyperspy[59]. Integrated XRF peak intensities were extracted the spectra at each point. For the Br:Pb maps, the Br K$_\alpha$ and Pb L$_\alpha$ peak intensities were used. The ratios shown are a direct ratio of these peak intensities, not a quantitative measure of Br and Pb concentrations. Although the maps were not corrected for self-absorption, we performed checks to ensure that the self-absorption of the perovskite and of the other layers in the device stacks were not substantially skewing the data. We performed complete self-absorption correction using the python package PyMca[60]. This complete correction accounts for the thickness and self-absorption of each layer in the stack, the incident and outcoupled angles of the X-ray beams, the path length of the X-rays through air after fluorescence which can absorb low energy X-rays, as well as the material and thickness of the detector. We see negligible differences in the spatial variation of the chemical composition after having applied the self-absorption correction (Supplementary Figure 76).

Supplementary Information for

# Multimodal operando microscopy reveals that interfacial chemistry and nanoscale performance disorder dictate perovskite solar cell stability

Kyle Frohna[†], Cullen Chosy[†], Amran Al-Ashouri, Florian Scheler, Yu-Hsien Chiang, Milos Dubajic, Julia E. Parker, Jessica M. Walker, Lea Zimmermann, Thomas A. Selby, Yang Lu, Bart Roose, Steve Albrecht, Miguel Anaya[*], Samuel D. Stranks[*]

[†] These authors contributed equally to this work

*email: sds65@cam.ac.uk, ma811@cam.ac.uk

**This pdf file contains:**

Supplementary Notes 1-6
Supplementary Figures 1-76

Supplementary Note 1: Hyperspectral Data Analysis

Quasi-Fermi level splittings were determined by fitting absolute-intensity PL spectra with the generalised Planck's law[1,2] using full peak fitting and models of below-bandgap absorption developed by Katahara and Hillhouse[3,4]. We have previously used this method to extract QFLS from hyperspectral maps of perovskites successfully[5]. The absolute intensity of photoluminescence ($I_{PL}(E)$) can be modelled as the product of the absorptance spectrum of the material, the photon density of states ($\rho(E)$) and a Bose-Einstein ($f_{BE}(E)$) occupation function with a finite quasi-Fermi-level splitting ($\Delta\mu$) as:

$$I_{PL}(E) = \rho(E) \times f_{BE}(E) \times a(E)$$

In the absence of large doping and solar relevant carrier densities, one can ignore occupation corrections to the absorptance spectra and approximate $f_{BE}$ as a Boltzmann type occupation function giving the following expression for $I_{PL}$:

$$I_{PL}(E) = \frac{2\pi E^2}{h^3 c^2} \times a(E) \times exp\left(-\frac{E - \Delta\mu}{kT}\right)$$

The approach by Katahara and Hillhouse involves a model for the absorption coefficient to fit the entire peak to extract $\Delta\mu$. This involves a convolution of an above bandgap square root density of states type absorption coefficient dependence:

$$\alpha = \alpha_0 \sqrt{E - E_g}$$

$\alpha_0$ is a parameter that depends on the oscillator strength of the material and $E_g$ is the bandgap, and a below bandgap exponential tail absorption coefficient:

$$\alpha \propto exp\left(\frac{E_g - E}{\gamma}\right)^\theta$$

Where $\theta$ is the power of the exponential tail and $\gamma$ is its characteristic energy broadening (the Urbach energy when $\theta$ is 1). The convolution integral looks as follows:

$$\alpha(\varepsilon) = \frac{\alpha_0 \sqrt{\gamma}}{2\Gamma\left(1 + \frac{1}{\theta}\right)} \int_{-\infty}^{\Delta\varepsilon} exp(-|\Delta\varepsilon'|^\theta) \sqrt{\Delta\varepsilon - \Delta\varepsilon'}\, d\Delta\varepsilon'$$

where Γ is the gamma function Rather than explicitly evaluating this convolution at each point, an approximation can be made with the assistance of a lookup table provided by Braly et al.[4] where they have explicitly evaluated the convolution integral $G(\frac{E-E_g}{\gamma}, \theta)$ for a range of parameters. The resulting approximation for the absorption coefficient is:

$$\alpha = \alpha_0 \sqrt{\gamma} G(\frac{E-E_g}{\gamma}, \theta)$$

And substituting this expression into the above expression for $I_{PL}$ assuming a Beer-Lambert type exponential relationship between a(E) and α(E) and film thickness d:

$$I_{PL}(E) = \frac{2\pi E^2}{h^3 c^2} \times (1 - \exp(-\alpha_0 d \sqrt{\gamma} G(\frac{E-E_g}{\gamma}, \theta))) \times exp(-\frac{E-\Delta\mu}{kT})$$

α₀d was fixed at 10 as is commonplace for perovskites and has little impact on the resulted fits[4], and T was set to 300K. The remaining parameters were fit using a Levenberg-Marquardt, non-linear least squares fitting protocol implemented in Python. As is common for multi-variable, non-linear least squares fitting, suitable initial guesses are critical and so the average of each scan area was fit and manually inspected to ensure the quality of the fit before using these output parameters as the initial conditions for the automatic fitting procedure.

The spectral centre of mass for each point was calculated by interpolating the data along a uniform energy axis and then calculating as:

$$COM = \frac{\sum E_n \cdot I_{PL}(E)}{\sum I_{PL}(E)}$$

Supplementary Note 2: Voltage-Dependent Photoluminescence Data Analysis

The measurements produced a voltage dependent photoluminescence signal in both the reverse and forward scan directions. Voltage-dependent PL sweeps in the forward and reverse directions are normalised according to a protocol reported previously by Wagner et al.[6] The difference between the photoluminescence intensity of a point at open circuit $I_{PL}(V_{OC})$ and any voltage below it $I_{PL}(V)$ is related to the charge carrier extraction efficiency ($\Phi_{PL}(V)$) once normalised by $I_{PL}(V_{OC})$.

$$\Phi_{PL}(V) = \frac{I_{PL}(V_{OC}) - I_{PL}(V)}{I_{PL}(V_{OC})} = 1 - \frac{I_{PL}(V)}{I_{PL}(V_{OC})} \approx \frac{J(V)}{J_{gen}}$$

In order to demonstrate this relation, the aim is to relate the current flowing through a solar cell to Δμ which is probed by the photoluminescence. The current flowing through and ideal diode can be written as:

$$J(V) = J_{rec} - J_{gen} = J_0\left(e^{\frac{eV}{k_BT}} - 1\right) - J_{gen}$$

Where $J_{rec}$ is the recombination current, $J_{gen}$ is the generation current from the light-source and $J_0$ is the radiative saturation current and $k_B$ is the Boltzmann constant. This diode equation assumes only radiative recombination, and no transport losses. If we assume that voltage is directly equivalent to Δμ, then this is the relationship needed. However, it has been shown that Δμ is not necessarily equivalent to V in perovskite solar cells, particularly those with low mobility charge transport layers[7,8]. Furthermore, recombination pathways beyond radiative recombination are known to occur in perovskite solar cells. To account for these pathways and to develop an explicit relationship between the current and Δμ, we first begin with a charge carrier recombination rate equation for an intrinsic semiconductor with rate constants for Shockley-Read-Hall recombination ($k_1$), radiative bimolecular ($k_2$) and Auger ($k_3$):

$$\frac{dn}{dt} = -k_1 n - k_2 n^2 - k_3 n^3$$

Writing the recombination rate equation in this form implies an intrinsic semiconductor with equal concentrations of electrons and holes which share non-radiative lifetimes and Auger rate constants. The carrier density can be related to Δμ with the relation $n^2 = n_i^2 \exp\left(\frac{\Delta\mu}{k_BT}\right)$ where $n_i$ is the intrinsic carrier density. Subbing in for the carrier density we obtain:

$$\frac{dn}{dt} = -k_1 n_i e^{\frac{\Delta\mu}{2k_BT}} - k_2 n_i^2 e^{\frac{\Delta\mu}{k_BT}} - k_3 n_i^3 e^{3\Delta\mu/2k_BT}$$

$J_{rec}$ is found by integrating this recombination rate across the active layer of the device. Assuming a thickness invariant Δμ and corresponding recombination rate through a device of thickness d, this gives:

$$J_{rec} = qd\left(-k_1 n_i e^{\frac{\Delta\mu}{2k_BT}} - k_2 n_i^2 e^{\frac{\Delta\mu}{k_BT}} - k_3 n_i^3 e^{\frac{3\Delta\mu}{2k_BT}}\right) \approx J_0\left(e^{\frac{\Delta\mu}{nk_BT}} - 1\right)$$

Where $n$ is the ideality factor which accounts for the dominant recombination mechanism. For an ideal diode undergoing only radiative recombination, n=1, is above 1 for recombination mechanisms involving monomolecular processes such as Shockley-Read-Hall recombination and below one for Auger processes. The ideality factor is typically between 1 and 2 for perovskites[9]. Finally, we obtain an ideal diode like equation relating the current to Δμ:

$$J(\Delta\mu) = J_{rec} - J_{gen} = J_0\left(e^{\frac{eV}{nk_BT}} - 1\right) - J_{gen} \approx J_0\left(e^{\frac{\Delta\mu}{nk_BT}}\right) - J_{gen}$$

Next returning to the original expression for $\Phi_{PL}(V)$, by noting from the ideal Planck law written above that the PL intensity can be written as the product of the absorptance, the photon density of states and an exponential factor depending on Δμ, $\Phi_{PL}(V)$ can be written as:

$$\Phi_{PL}(V) = 1 - \frac{a \cdot \rho \cdot e^{\frac{\Delta\mu_V - E}{k_BT}}}{a \cdot \rho \cdot e^{\frac{\Delta\mu_{OC} - E}{k_BT}}} = 1 - \frac{e^{\frac{\Delta\mu_V}{k_BT}}}{e^{\frac{\Delta\mu_{OC}}{k_BT}}}$$

When the diode equation is at open circuit, no current flows and so we get an expression:

$$e^{\frac{\Delta\mu_{OC}}{k_BT}} = \frac{J_{gen}}{J_0}$$

And so $\Phi_{PL}(V)$ can be written as:

$$\Phi_{PL}(V) = 1 - \frac{J_0}{J_{gen}} \cdot e^{\frac{\Delta\mu_V}{k_BT}}$$

This expression is equal to the current for the diode equation with ideality factor of 1 normalised by the generation current and so:

$$\Phi_{PL}(V) \approx \frac{J(V)}{J_{gen}}$$

Normalising the voltage-dependent photoluminescence signal in this way gives you an approximate value for the ratio of the extracted current at that voltage (J(V)) to the total generated current at that point ($J_{gen}$). This means that if the photoluminescence drops from a finite value at open circuit to 0 at short circuit, all carriers generated by the excitation source are interpreted as being extracted. Any finite luminescence at 0 V implies incomplete charge carrier extraction as carriers are still recombining radiatively within the active layer and there remains a finite quasi-Fermi level splitting. In this normalisation scheme, the radiative recombination acts as a proxy for the total recombination. This relies on the assumption that both the radiative and total recombination scale in the same exponential way with voltage and in this case, an ideality factor of 1. When the ideality factor deviates from one, the expression is no longer exact and so will slightly over or underestimate the extracted current fraction.

To test the assumptions of the model as above, we ran SCAPS simulations[10] and varied an array of parameters to understand under what conditions the model is valid. As above, in the ideal diode case with no non-radiative losses and no transport losses, the relationship between radiative and total recombination is exact. We then simulated an idealised p-i-n perovskite solar cell with no non-radiative recombination losses and low mobility organic transport layers with the following material parameters included in Supplementary Table 1.

**Supplementary Table 1**: Bulk material parameters used in SCAPS drift diffusion simulations.

|  | HTL (PTAA/SAM) | Perovskite | ETL ($C_{60}$) |
| --- | --- | --- | --- |
| Thickness (nm) | 10 | 400 | 20 |
| Bandgap (eV) | 3 | 1.65 | 2.0 |
| Electron Affinity (eV) | 2.5 | 3.9 | 3.9 |
| Dielectric Permittivity (relative) | 3.5 | 22 | 5.0 |
| CB Effective Density of States ($cm^{-3}$) | $1 \times 10^{20}$ | $2.2 \times 10^{18}$ | $1 \times 10^{20}$ |
| VB Effective Density of States ($cm^{-3}$) | $1 \times 10^{20}$ | $2.2 \times 10^{18}$ | $1 \times 10^{20}$ |
| Carrier thermal velocity ($cm\ s^{-1}$) | $1 \times 10^{7}$ | $1 \times 10^{7}$ | $1 \times 10^{7}$ |
| Electron mobility ($cm^2\ V^{-1}\ s^{-1}$) | $1 \times 10^{-5}$ | 20 | $1 \times 10^{-2}$ |
| Hole mobility ($cm^2\ V^{-1}\ s^{-1}$) | $1 \times 10^{-4}$ | 20 | $1 \times 10^{-2}$ |
| Donor density ($cm^{-3}$) | $1 \times 10^{5}$ | $1 \times 10^{10}$ | $1 \times 10^{5}$ |
| Acceptor density ($cm^{-3}$) | $1 \times 10^{5}$ | $1 \times 10^{10}$ | $1 \times 10^{5}$ |
| Radiative recombination coefficient ($cm^3\ s^{-1}$) |  | $6 \times 10^{-11}$ |  |

|  | ITO | Cu |
| --- | --- | --- |
| Work function (eV) | 5.3 | 4.1 |

| Majority carrier barrier height (eV) | 0.2 | 0.2 |
|---|---|---|
| Built in voltage (V) | 1.2 | |

With the above parameters in place we simulated the effects of the addition of parasitic resistance losses on the relationship between radiative recombination and total recombination. For all simulations, solid lines represent the extracted current normalised by the generation current J(V)/J$_{gen}$ while the dashed lines correspond to $\Phi_{PL}$(V). We first varied the impact of series resistance (R$_S$) from 0 to 50 Ω cm$^2$, a wide range from an ideal solar cell to one extremely limited by this loss pathway. The results are plotted in Supplementary Figure 1. There is still an exact overlay between the two metrics over the range of varied parameters showing that series resistance introduces no appreciable errors.

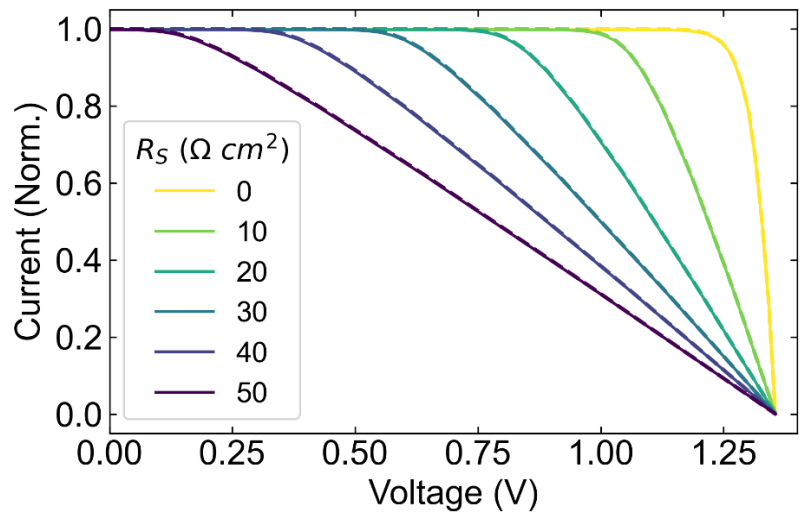

**Supplementary Figure 1.** Impact of series resistance (R$_s$) on relationship between J(V)/J$_{gen}$ (solid line) and $\Phi_{PL}$(V) (dashed line).

By contrast, the effect of low shunt resistance, as shown in Supplementary Figure 2 does introduce a large error. We varied the shunt resistance (R$_{sh}$) from 200-1000 Ω cm$^2$ which corresponds to a significantly shunted device beyond what we see in our own devices to highlight this point. In a case with low shunt resistance, current simply bypasses the diode, meaning that electrical current and radiative recombination currents can become disconnected from one another. In cases where devices exhibit low shunt resistance, it should be noted that the relationship between J(V)/J$_{gen}$ and $\Phi_{PL}$(V) can deviate significantly here. Across the range

$R_{sh}$ simulated, $\Phi_{PL}(V)$ does not appreciably change while the electrical curve changes considerably.

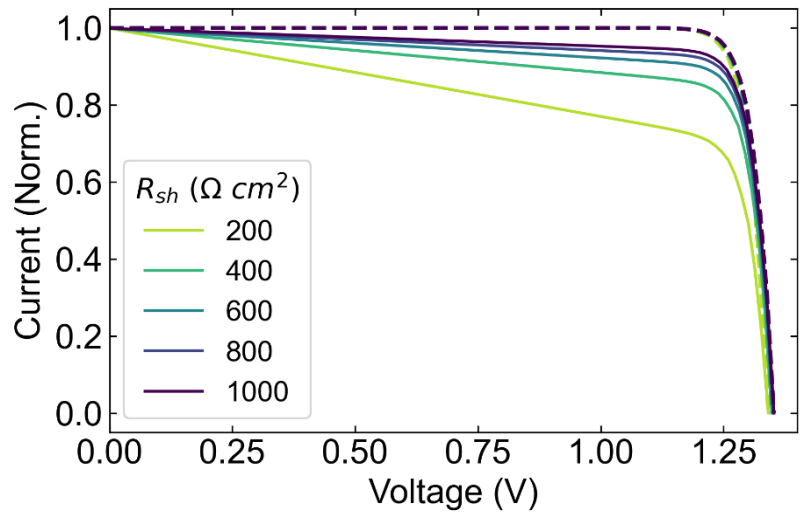

**Supplementary Figure 2.** Impact of shunt resistance ($R_s$) on relationship between $J(V)/J_{gen}$ (solid line) and $\Phi_{PL}(V)$ (dashed line).

Next, the mobility (µ) of charge carriers in the perovskite was also varied to simulate one avenue of poor charge extraction. µ was varied between $1\times10^{-3}$ - 20 cm$^2$ V$^{-1}$ s$^{-1}$ and the results shown in Supplementary Figure 3. The electron and hole mobilities were varied simultaneously so in each simulation, the number listed corresponds to both electron and hole mobilities. While the mobilities at the lower end of the range affected the electrical JV curves significantly, they also affected the radiative recombination in the same way, such that the relationship between optical and electrical JV curves is preserved.

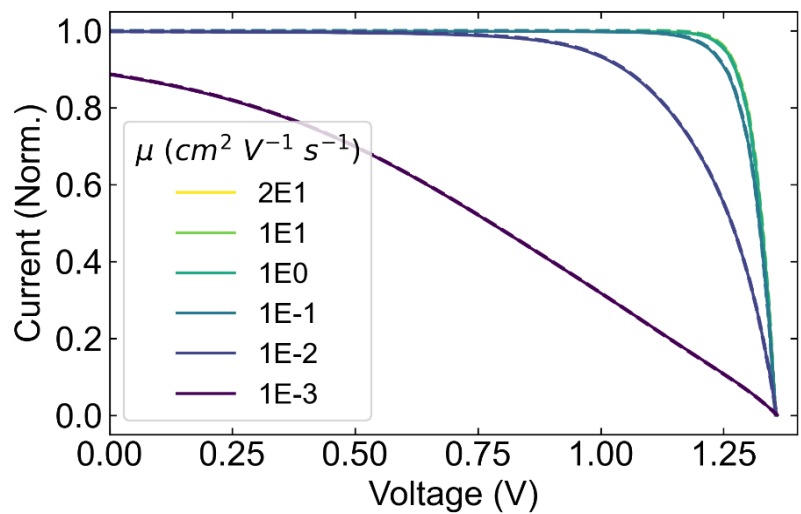

**Supplementary Figure 3.** Impact of carrier mobility (μ) on relationship between J(V)/J$_{gen}$ (solid line) and $\Phi_{PL}$(V) (dashed line).

We then added in bulk non-radiative recombination in the perovskite through a defect state 0.6 eV above the valence band with electron and hole capture cross sections of $1\times10^{-15}$ cm$^2$. The concentration of these defects was varied from $1\times10^{14}$ – $1\times10^{17}$ cm$^{-3}$, which corresponds to a range of non-radiative lifetimes from 1000-1 ns and the results shown in Supplementary Figure 4. In this case, the effective ideality factor of the diode is no longer one and as such, the relationship between J(V)/J$_{gen}$ and $\Phi_{PL}$(V) is no longer exact. In this situation, the optical JV curve can overestimate the fill factor of the device with increasing error at shorter non-radiative lifetimes.

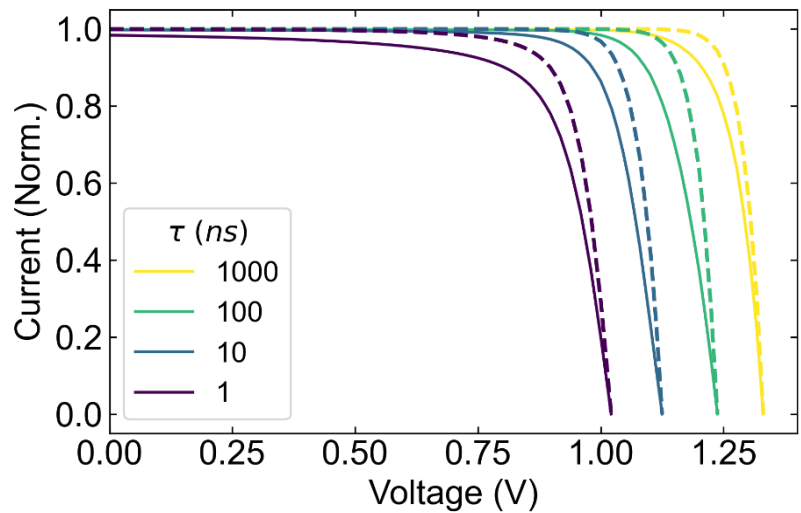

**Supplementary Figure 4.** Impact of non-radiative lifetime (τ) on relationship between J(V)/J$_{gen}$ (solid line) and $\Phi_{PL}$(V) (dashed line).

Next the impact of interfacial recombination was investigated. All other non-radiative recombination pathways were removed and a defect was placed at the interface between the perovskite and C$_{60}$ ETL. The concentration of the defect was varied between $10^7$-$10^{13}$ cm$^{-2}$ which corresponds to surface recombination velocities of $10^{-2}$-$10^4$ cm s$^{-1}$ and results shown in Supplementary Figure 5. Interestingly, while the non-radiative losses at this defect can still be substantial for large surface recombination velocities, the error introduced by this type of loss is actually reduced compared to the bulk non-radiative recombination.

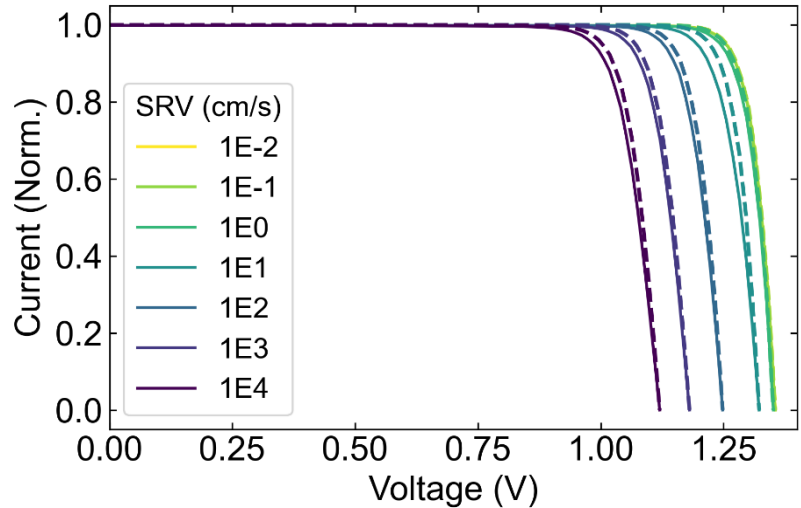

**Supplementary Figure 5.** Impact of surface recombination velocity (SRV) on relationship between J(V)/$J_{gen}$ (solid line) and $\Phi_{PL}$(V) (dashed line).

We simulate a 'realistic' perovskite solar cell, incorporating bulk non-radiative recombination in the perovskite and both transport layers, as well as interfacial recombination losses. The complete list of parameters for bulk and interfacial defects are shown in Supplementary Table 2. These parameters are mirrored in the literature and guided by experimental observations of the bulk Shockley-Read-Hall recombination dominated lifetime of perovskites[11], carrier lifetimes in conjugated organics and fullerenes[12,13], surface recombination velocities at HTL and ETL interfaces[14], paying particular attention to our previous work showing that $C_{60}$ is the more problematic interface[15]. Incorporating all of these recombination losses, we also vary the built in voltage by varying the work function difference between the ITO and copper electrodes from 1-1.2 V. Previous work has shown that for these systems that a built in voltage of at least 1.2 V is required to efficiently extract charges[16], so we wanted to see the effect that dipping below this threshold had on the optical-electrical relationship. The results are shown in Supplementary Figure 6. For the range of parameters over which our solar cells vary, the simulations suggest that the relationship between J(V)/$J_{gen}$ and $\Phi_{PL}$(V) is a good approximation of the true form of the JV curve. For our experimental data, we tend to see a small systematic overestimation of the fill factor from the optical JV data compared to the electrical JV data, suggesting that bulk recombination is still playing an important role as a loss mechanism in our solar cells.

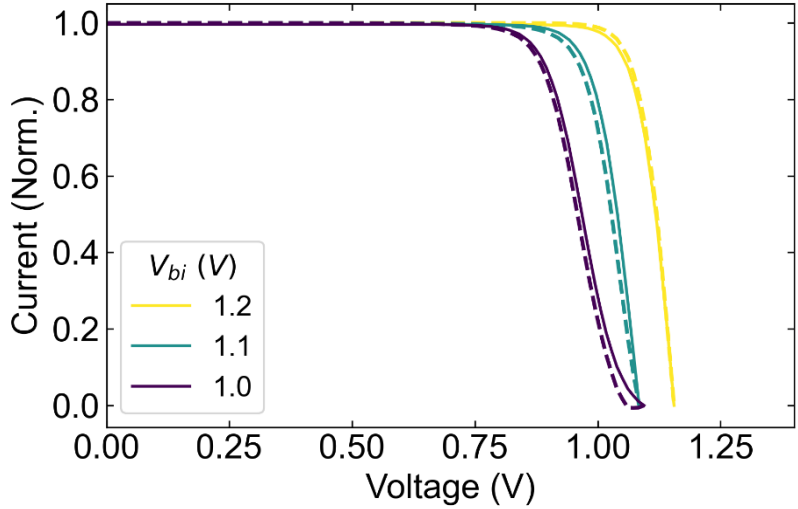

**Supplementary Figure 6.** Impact of built in potential on relationship between J(V)/J$_{gen}$ (solid line) and $\Phi_{PL}$(V) (dashed line) for a 'realistic' perovskite solar cell.

We finally note that for our solar cells, even at 0 V external bias, there is still a large Δμ or internal voltage within the solar cell that can be seen with luminescence as has recently been reported[8]. This has been attributed to the low mobilities and relative permittivities of the charge transport layers in p-i-n perovskite solar cells. In our simulations, plotting the perovskite thickness averaged Δμ versus applied voltage for our 'realistic' perovskite solar cell allows us to reproduce this result well as shown in Supplementary Figure 7 where the internal and external voltage are dramatically different, particularly at voltages close to short circuit. This is a loss mechanism that will need to be addressed for p-i-n perovskite solar cells to approach radiative efficiency limits.

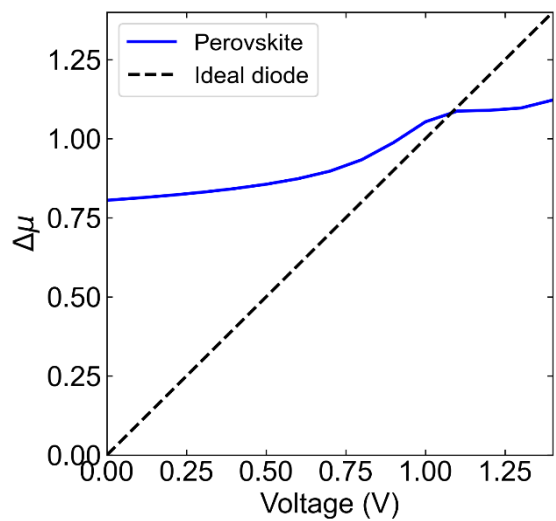

**Supplementary Figure 7:** Δµ versus voltage for ideal diode solar cell (dashed line) and simulated perovskite solar cell (solid line) showing large finite Δµ at 0V.

**Supplementary Table 2**: Parameters for defects used in 'realistic' simulation.

|  | **HTL (PTAA/SAM)** | **Perovskite** | **ETL ($C_{60}$)** |
|---|---|---|---|
| Defect type | Neutral | Neutral | Neutral |
| Electron capture cross section (cm$^2$) | $1\times10^{-15}$ | $1\times10^{-15}$ | $1\times10^{-15}$ |
| Hole capture cross section (cm$^2$) | $1\times10^{-15}$ | $1\times10^{-15}$ | $1\times10^{-15}$ |
| Energetic distribution | Single | Single | Single |
| Energy above valence band (eV) | 0.6 | 0.6 | 0.6 |
| Trap density (cm$^{-3}$) | $1\times10^{17}$ | $2.2\times10^{14}$ | $1\times10^{17}$ |
| Effective non-radiative lifetime (ns) | 1 | 450 | 1 |

|  | **HTL/Perovskite Interface** | **Perovskite/ETL Interface** |
|---|---|---|
| Defect type | Neutral | Neutral |
| Electron capture cross section (cm$^2$) | $1\times10^{-16}$ | $1\times10^{-16}$ |
| Hole capture cross section (cm$^2$) | $1\times10^{-16}$ | $1\times10^{-16}$ |
| Energetic distribution | Single | Single |
| Energy above highest valence band (eV) | 0.6 | 0.6 |
| Trap density (cm$^{-2}$) | $1\times10^{11}$ | $2.2\times10^{12}$ |
| Surface recombination velocity (cm/s) | 100 | 2200 |

After the voltage dependent PL maps have been measured, to process the data, we first employ a principle component analysis as implemented in Hyperspy[17] and then denoised the data by reconstructing the dataset based on the first 10 principle components. In all cases, the first 10 components accounted for the vast majority of the variance in the dataset. Then for the reverse scan, the first image at open circuit was used as the normalisation factor, whereas the last image was used for the forward scan. In some cases, some light soaking resulting in photobrightening could be observed during the measurement in some samples, this results in a shifting of the forward scan up in the forward direction, resulting in an apparent hysteresis in the short circuit current that is not observed in the electrical JV data. The implications of this light soaking on actual device performance are not clear but it is worthy of note.

After the denoising and normalisation, the figures of merit were extracted locally in an analogous way to conventional JV data. $\Phi_{PL}(0)$ is the short circuit current extraction ratio, the macroscopic $V_{OC}$ is implicitly assumed to be uniform across the device in this scheme, the maximum pseudo-power is found by finding the maximum value of the product $\Phi_{PL}(V) \times V$, and the fill factor is obtained by dividing this value by the product of the short circuit current extraction efficiency times the open circuit voltage:

$$FF_{PL} = \frac{\Phi_{PL}(V_{MPP}) \times V_{MPP}}{\Phi_{PL}(0) \times V_{OC}}$$

Doing these calculations at each point produces figure of merit maps across the measured area of the device.

Supplementary Note 3: Image Registration

All samples were initially marked with a small fiducial scratch produced by a diamond tipped scribe on the glass side. This was used as a coarse alignment by initially focusing on the top of the glass substrate to find the region. After finding the scratch, the focus was brought through the glass to the perovskite where a distinctive marking or imperfection was found. The combination of these two protocols allows reliable, rapid returns to the same area of the sample. Image registration between the fresh and aged datasets was performed using Advanced Normalization Tools (ANTs), an open-source and state-of-the-art medical image registration and segmentation toolkit[18]. ANTs provides a flexible and widely-used registration framework, with many different similarity metrics that can be used to tune registration performance[19]. Optical microscopy images were registered using an affine transformation with a squared intensity difference or mutual information similarity metric. nXRF to optical microscopy image registration primarily relied on the mutual information similarity metric for improved performance between different imaging modalities (see Python scripts for details).

Supplementary Note 4: Compositional Correlations

In order to investigate the extent to which local compositional variation correlates with local device performance parameters we performed pixel-by-pixel spatial correlation analysis between the compositional information measured by nXRF and a wide array of optoelectronic figures of merit measured by the optical microscopes across the device space described in the main text. In Supplementary Figure 8, we show 2-dimensional histograms with the Br:Pb ratio on the x-axis and the optoelectronic figure of merit on the y-axis. The color map represents the density of points contained in that particular bin while the dotted black line represent the mean value of the optoelectronic figure of merit for that value of the composition. Superimposed on each histogram is the Spearman's rank correlation coefficient and associated p-value which assesses whether there is a monotonic relationship between the two variables of interest and the strength of said relationship (rather than assuming a linear relationship in the case of Pearson's coefficient). Supplementary Figure 8 shows the case for a pristine double-cation double-halide (DCDH) perovskite fabricated on the 2PACz HTL. Supplementary Figure 8 a shows a strong anticorrelation between open circuit photoluminescence (PL) intensity and Bromine content as we previously qualitatively commented on in the manuscript, while there is a strong positive correlation between the PL centre of mass energy (COM) and Br:Pb. These two opposing trends effectively cancel one another out in the quasi-Fermi level splitting ($\Delta\mu$) which shows effectively no dependence on the Br:Pb, suggesting that a combination of lower trap densities and local carrier funnelling cancel out any voltage penalty induced by the lower bandgap. For the optical Jsc, PCE and fill factor shown in Supplementary Figure 8 b-d, there are all very weak relationships between composition and electrical performance suggesting that for the pristine DCDH sample, the compositional variation has a relatively small effect. The changes that are present are most likely attributable to the wrinkles which we will go into much greater detail in Supplementary Note 5.

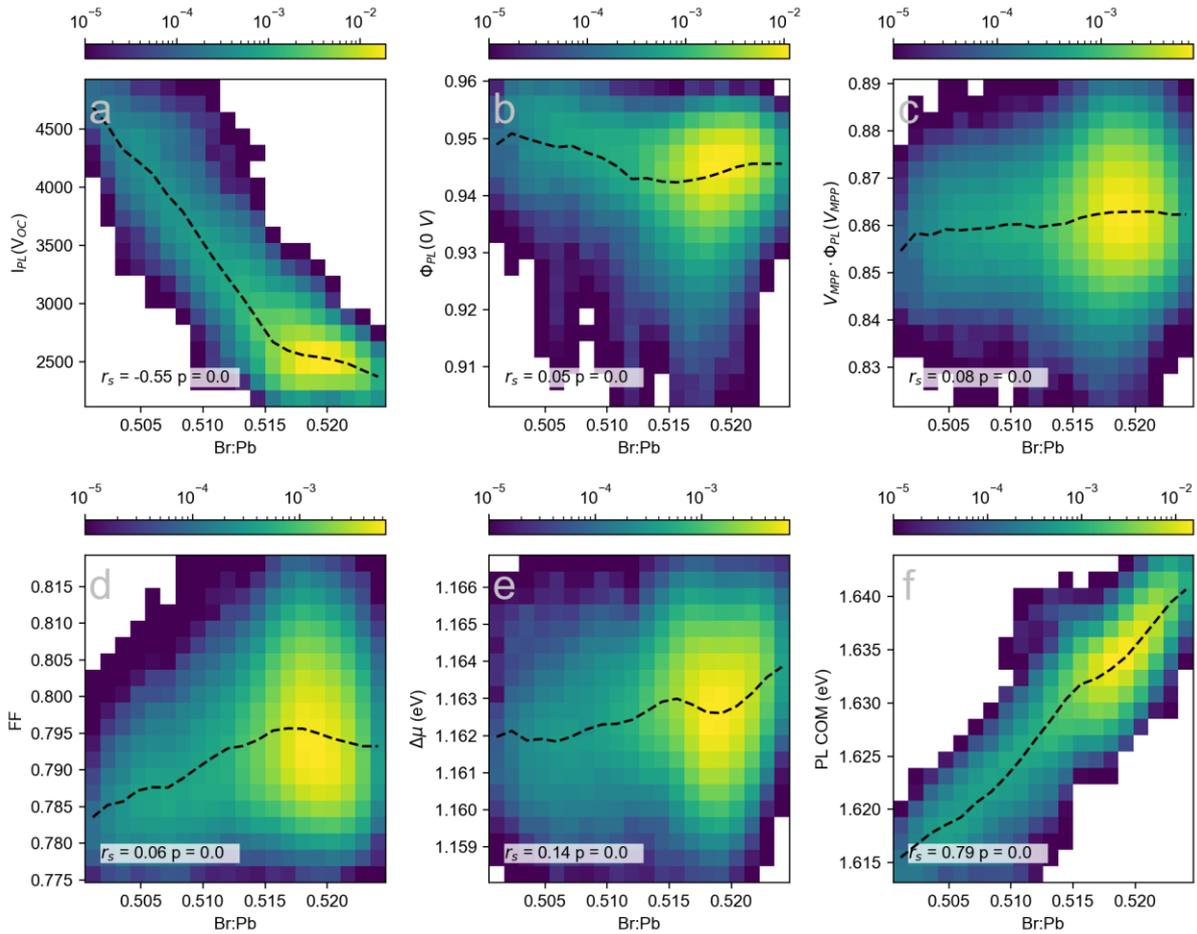

**Supplementary Figure 8:** 2-dimensional histograms of the spatial correlations between the Br:Pb content of a pristine 2PACz/DCDH solar cell and a) open circuit PL intensity ($I_{PL}(V_{OC})$), b) short circuit current extraction efficiency ($\Phi_{PL}(0\,V)$), c) optical power conversion efficiency, d) optical fill factor, e) $\Delta\mu$ and f) PL centre of mass. Overlaid over each map is the Spearman's rank correlation coefficient between the two variables and the related p-value.

By comparison, compositional correlations of a stress tested equivalent of the same sample type (2PACz/DCDH) is included in Supplementary Figure 9. The same correlation/anticorrelation between composition and PL COM/$I_{PL}(V_{OC})$ is still observed. However, there is now a small increase in $\Delta\mu$ with Br:Pb suggesting that the low Br regions may have degraded slightly more than the other areas and this is also reflected in the optoelectronic figures of merit. However, the effect size is extremely small in $\Delta\mu$ and in the optical PCE where there is a few percent change across the Br:Pb range. In Supplementary Figure 37, one can see that in the optical PCE maps of the pristine sample, the wrinkles are quite visible, but in the operated samples, they are less evident as larger scale variations dominate the contrast. However, the behaviour of the 2PACz/DCDH is not entirely general across the sample space.

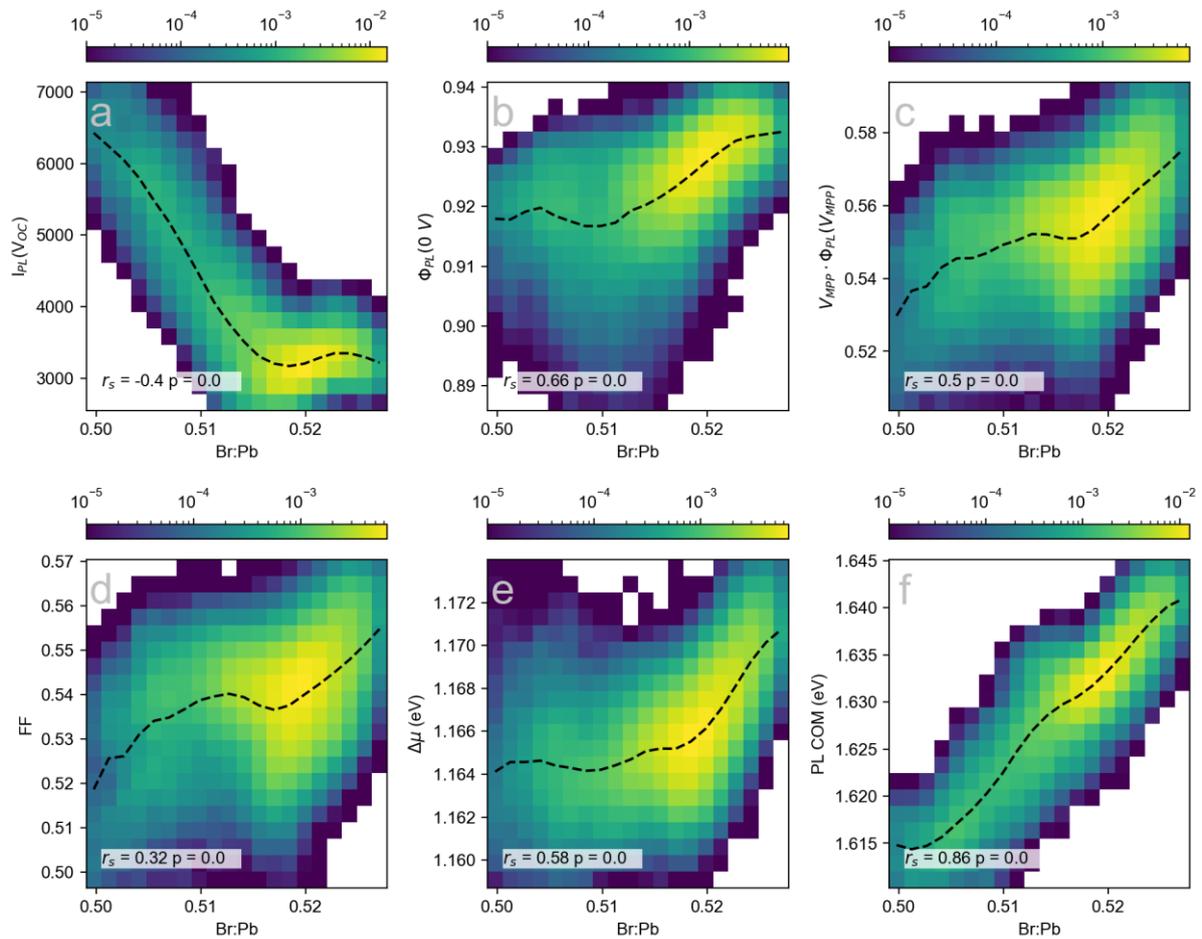

**Supplementary Figure 9:** 2-dimensional histograms of the spatial correlations between the Br:Pb content of a operationally stress tested 2PACz/DCDH solar cell and a) open circuit PL intensity ($I_{PL}(V_{OC})$), b) short circuit current extraction efficiency ($\Phi_{PL}(0\ V)$), c) optical power conversion efficiency, d) optical fill factor, e) $\Delta\mu$ and f) PL centre of mass. Overlaid over each map is the Spearman's rank correlation coefficient between the two variables and the related p-value.

The most extreme case of phase segregation (and therefore largest range of Br:Pb ratios observable which are most likely to detect the compositional limits at which losses may occur) is seen in the 2PACz/DCTH samples, particularly those after operation. This can be seen in Figure 3 and Supplementary Figures 45, 47 and 48. . In the analysis of the fresh DCTH sample (Supplementary Figure 10), it is evident that there is now a significant correlation between the Br:Pb content and the optoelectronic figures of merit (in particular charge extraction efficiency and optical PCE) where there was not before, as is also qualitatively observable in

Supplementary Figure S45 and 47. This implies that the low Br regions/wrinkles are negative for performance, in particular charge extraction, in this device architecture. However, in the operationally stressed devices as shown in Supplementary Figure 11, there is the emergence of highly Br deficient, phase segregated regions with markedly different recombination and transport behaviour and the resulting spread of composition is much larger than in other sample types. Previously, the distribution of PL COM and the associated correlation with Br:Pb was effectively unimodal, however, the PL COM of these samples indicate two distinct populations with low Br: wrinkles and segregation. The segregated regions have an ever more pronounced impact on the PL COM (they are the lower band in the plot), and actually appear to be positive for the charge extraction relative to the surroundings. The overall statistical result suggests that above a threshold value of Br:Pb, there is very little effect of composition on optical PCE, however below this value, low Br actually improves the charge transport and therefore the performance.

Taken together, these results suggest that in pristine samples, there often appears to be a statistically significant positive correlation with small effect size between composition and optical PCE. However, there is additionally a Br threshold relative to its surroundings, below which, the sample is sufficiently segregated to have a large, positive effect on the charge extraction. This suggests that the phase segregated low bandgap regions are not the cause of current loss in wide bandgap solar cells, but it may indeed be the unsegregated or wide gap regions left behind.

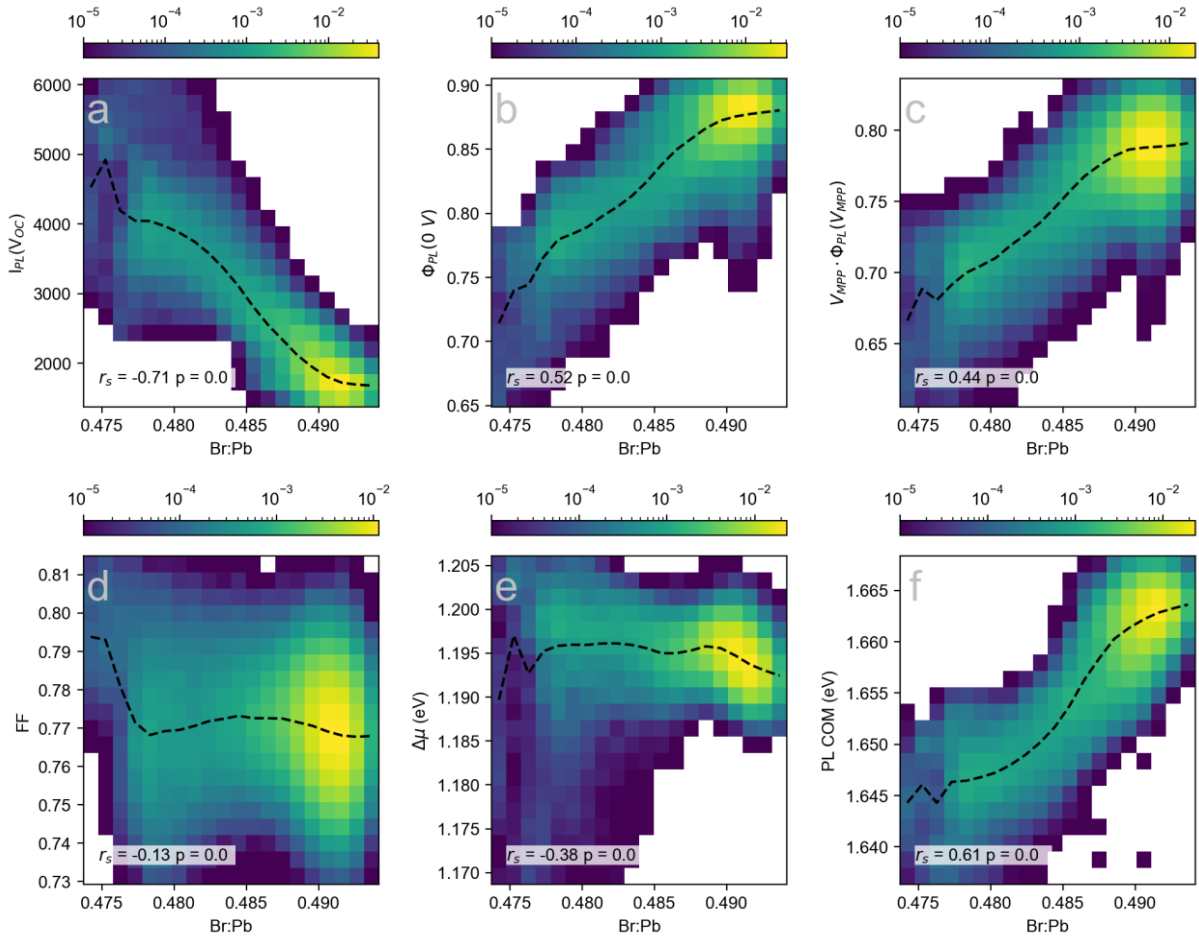

**Supplementary Figure 10:** 2-dimensional histograms of the spatial correlations between the Br:Pb content of a pristine 2PACz/DCTH solar cell and a) open circuit PL intensity ($I_{PL}(V_{OC})$), b) short circuit current extraction efficiency ($\Phi_{PL}(0\,V)$), c) optical power conversion efficiency, d) optical fill factor, e) $\Delta\mu$ and f) PL centre of mass. Overlaid over each map is the Spearman's rank correlation coefficient between the two variables and the related p-value.

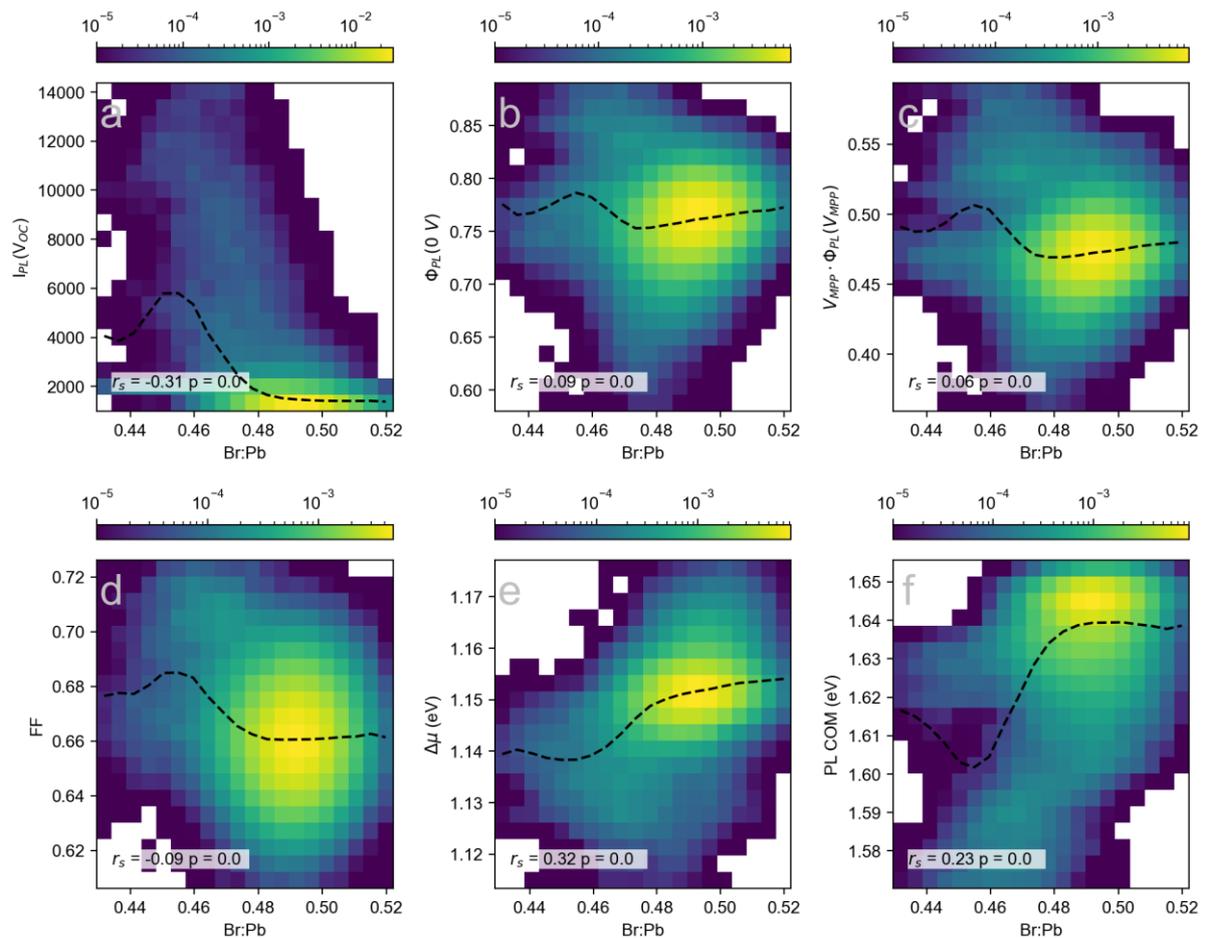

**Supplementary Figure 11:** 2-dimensional histograms of the spatial correlations between the Br:Pb content of an operationally stress tested 2PACz/DCTH solar cell and a) open circuit PL intensity ($I_{PL}(V_{OC})$), b) short circuit current extraction efficiency ($\Phi_{PL}(0\,V)$), c) optical power conversion efficiency, d) optical fill factor, e) $\Delta\mu$ and f) PL centre of mass. Overlaid over each map is the Spearman's rank correlation coefficient between the two variables and the related p-value.

Supplementary Note 5: Wrinkle Correlations

Much like compositional variation, a feature present throughout all of the perovskite films and devices we measured in this study is the presence of wrinkles that cover a significant fraction of the perovskite's area. The field has been aware of the presence of these wrinkles for some time[20], with the microscopy approach demonstrated here, we have the ability to comment directly on influence of wrinkles on local device figures of merit and stability.

The wrinkles are very apparent in bright-field optical reflectance images so we used these as the basis for a thresholding tool to produce a mask specifically for the wrinkles. We used the python package Pyclesperanto, a tool traditionally used for quantitative cell segmentation and analysis in a biological context. After a series of thresholding, filtering and segmentation steps, a mask can be reliably produced with statistical information about the size, shape and spatial distribution of each wrinkle, enabling us to perform a detailed quantitative analysis. An example of a bright field reflectance image of a 2PACz/DCDH device is shown in Supplementary Figure 12 a. The wrinkles very clearly appear as regions of low intensity in the reflected image. Their morphology may prevent coherent thin-film reflection effects such as interference and may waveguide light laterally into the device to reduce reflectance. The masked wrinkles from the reflectance image are shown in Supplementary Figure 12 b. Each individual wrinkle object is marked by a colour distinct from its neighbours. The distribution of different wrinkle areas, the total number of wrinkles and the total surface coverage of wrinkles across the area of interest is shown in Supplementary Figure 12 c. This sample is in line with the most wrinkled of the devices in this study showing a ~20% surface coverage depending on the location of measurement. The filtered wrinkle mask was either used as a positive or negative mask to filter only the regions at or away from the wrinkles. Equivalent masks and distributions were generated for all the measured devices.

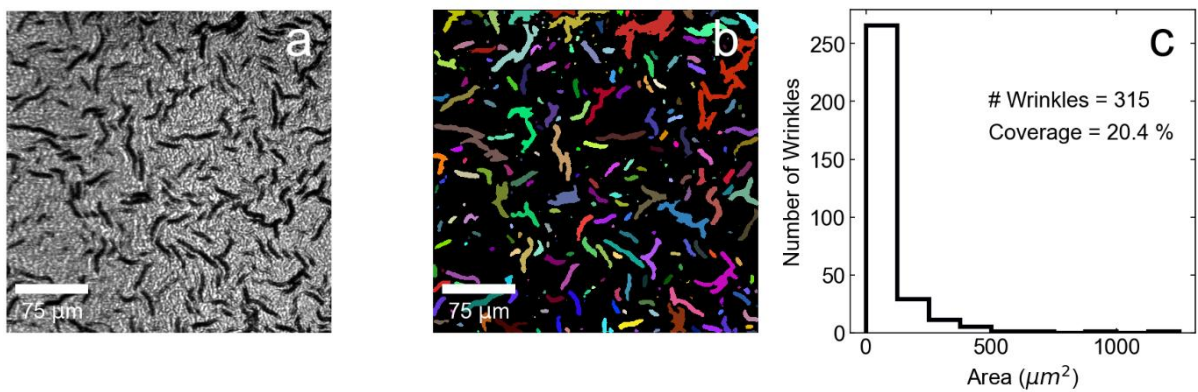

**Supplementary Figure 12:** a) Bright field reflectance image of 2PACz/DCDH solar cell. b) Filtered wrinkle mask from the reflectance image in a. c) Histogram of the size distribution of wrinkles in panel b.

Having established a robust protocol for masking regions where the wrinkles are present versus not present, we then apply them as a mask for the optoelectronic measurements. Supplementary Figure 13 a and b show the mean PL spectra and optical JV curves before and after the operational stress test on and off the wrinkled areas. The wrinkled areas show higher intensity, red shifted PL compared to the wrinkle free areas but both areas show effectively equivalent Δμ. After operational stress, both areas have increased in intensity, with the spectral differences between both regions increasingly similar. The optical JV curves in this sample show effectively no difference in either the forward or reverse direction before or after operational stress – showing that while the wrinkles to mediate the emission spectrum, the charge extraction and stability is close to unaffected by the presence of the wrinkles in these samples. To show this more explicitly, we calculated the distributions of a range of device relevant optoelectronic parameters in the wrinkled versus non wrinkled cases. These are shown in Supplementary Figure 14. The open circuit PL follows a predictable trend and PL COM follows a similar trend to the compositional correlations above where the wrinkled areas have higher intensity, red-shifted emission. The other optoelectronic figures of merit follow suit, showing very close agreement both before and after operational stress testing, further solidifying our claim that the wrinkles are not seeds of high performance or degradation. One caveat to this dataset is the degradation was dominated by an edge effect which may blur out the effects we might otherwise with the wrinkles (see Supplementary Figures 35 and 36). We therefore include further analysis on regions and devices where this is not the case.

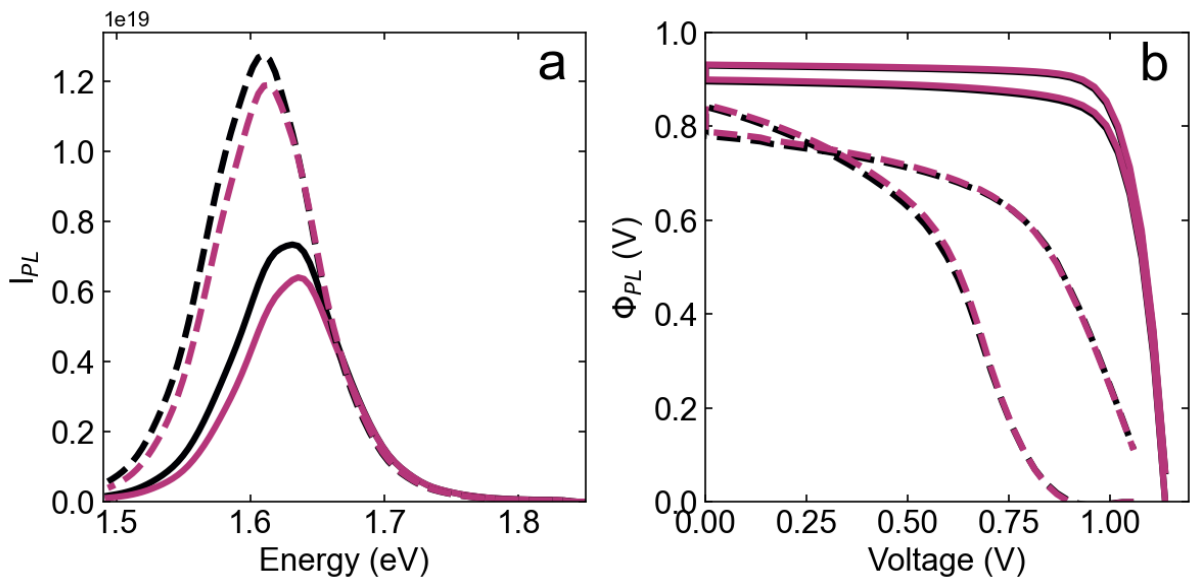

**Supplementary Figure 13**: a) Average PL spectra of non-wrinkled (red) and wrinkled (black) regions in a 2PACz/DCDH device before (solid lines) and after (dashed lines) operational stress. b) Mean optical JV curves of the same device and masked regions before and after operational stress.

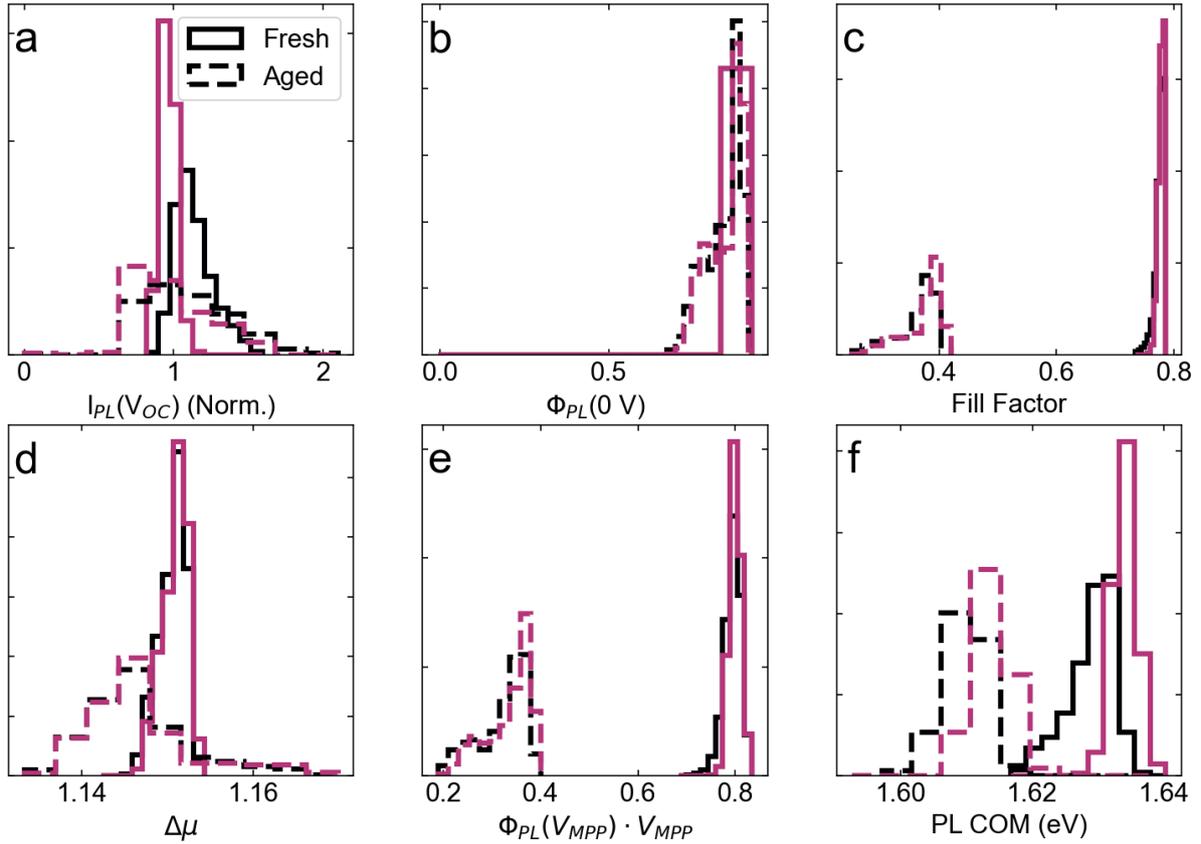

**Supplementary Figure 14**: Distributions of optoelectronic parameters on non-wrinkled (red) and wrinkled (black) regions before (solid) and after (dashed) operational stress of a 2PACz/DCDH device. a) open circuit PL intensity, b) short circuit optical charge extraction efficiency, c) optical fill factor, d) quasi-Fermi level splitting, e) optical power conversion efficiency and f) PL centre of mass.

We perform the same masking analysis on the complete range of devices measured beginning with 2PACz/DCTH. The size distribution and the surface coverage in these samples as shown in Supplementary Figure 15 is similar to the DCDH case as we previously claimed without quantitative evidence. The mean PL spectra before and after operational stress in Supplementary Figure 16 a show a consistent PL intensity increase and red shift in the wrinkled areas. However, after ageing both regions display quite a stark red shift and this is indicative of the fact that the highly phase segregated regions appear to form on both the wrinkled and non-wrinkled regions – as opposed to their growth being preferentially seeded by one versus another. The optical JV curves in the pristine sample show that the wrinkles exhibit worse optical charge extraction efficiency and fill factor, but this variation disappears after operational stress where the performance of the wrinkles has homogenised with the rest of the sample (Supplementary Figure 16 b), and the only relevant deviation from the norm is the highly segregated regions which show reduced Δμ and improved optical extraction efficiency. This is further highlighted in Supplementary Figure 17 which show this homogenisation of the properties of the wrinkled and non-wrinkled areas after operational stress.

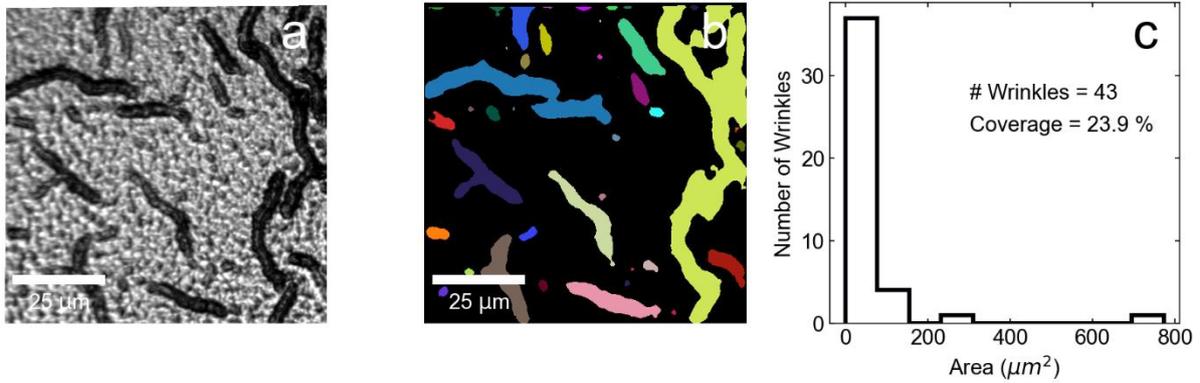

**Supplementary Figure 15:** a) Bright field reflectance image of 2PACz/DCTH solar cell. b) Filtered wrinkle mask from the reflectance image in a. c) Histogram of the size distribution of wrinkles in panel b.

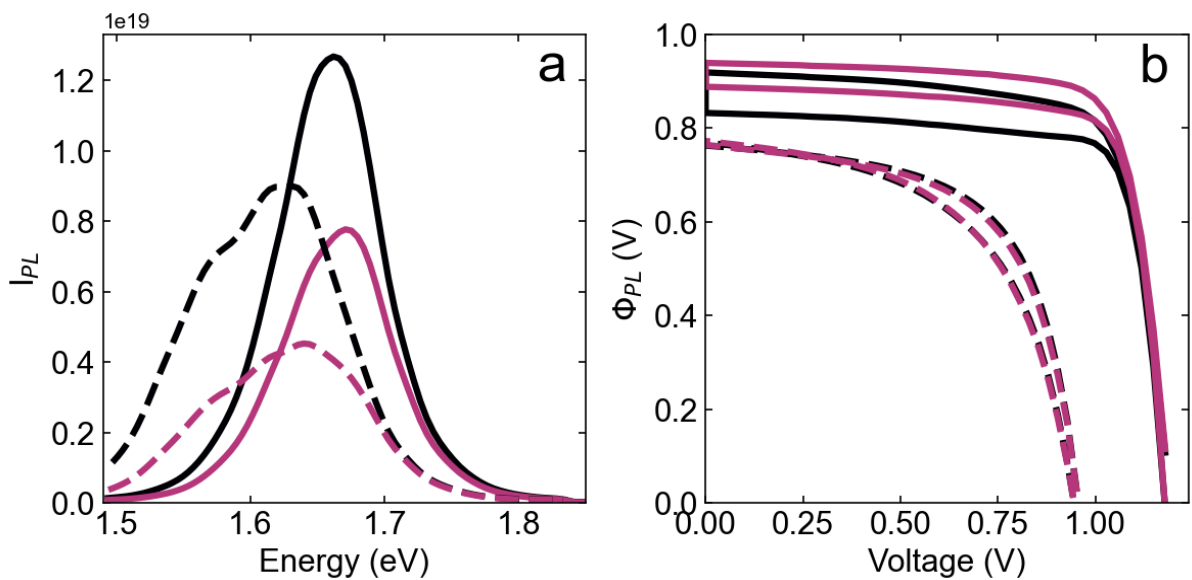

**Supplementary Figure 16**: a) Average PL spectra of non-wrinkled (red) and wrinkled (black) regions in a 2PACz/DCTH device before (solid lines) and after (dashed lines) operational stress. b) Mean optical JV curves of the same device and masked regions before and after operational stress.

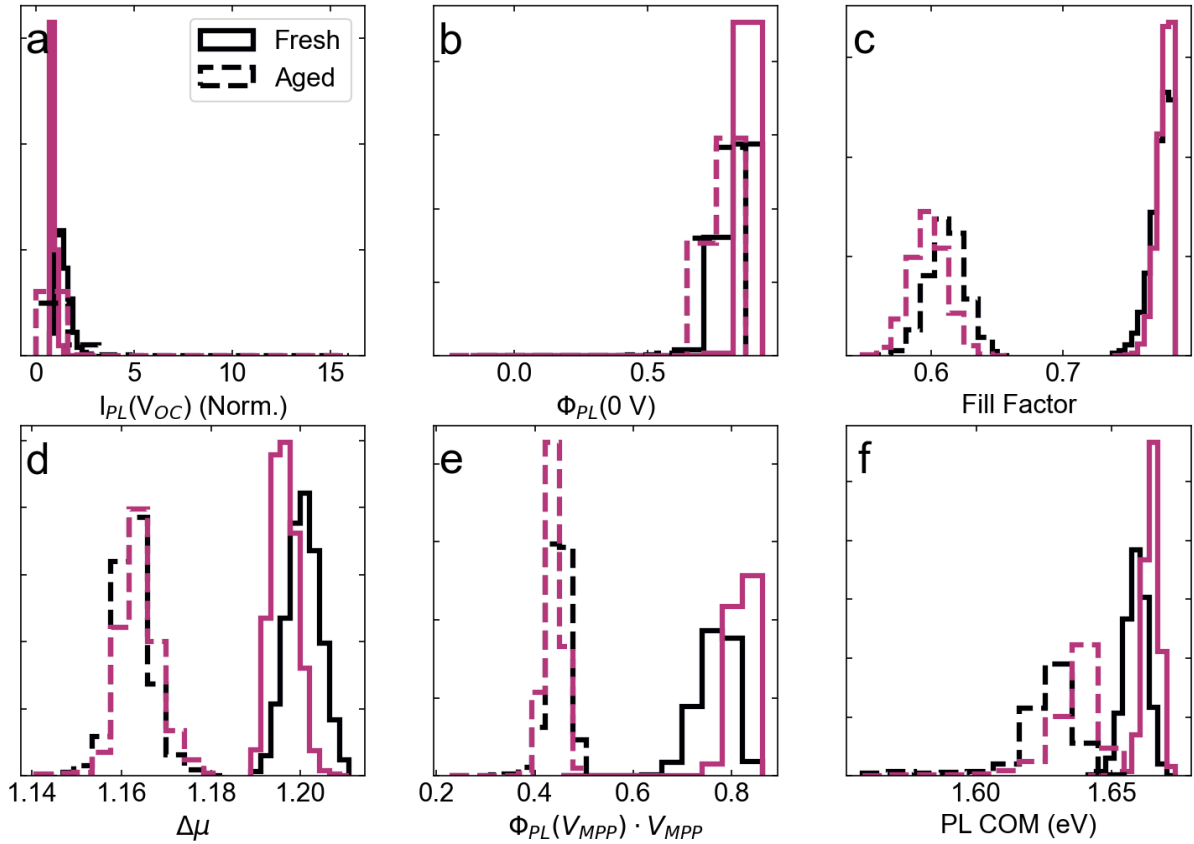

**Supplementary Figure 17**: Distributions of optoelectronic parameters on non-wrinkled (red) and wrinkled (black) regions before (solid) and after (dashed) operational stress of a 2PACz/DCTH device. a) open circuit PL intensity, b) short circuit optical charge extraction efficiency, c) optical fill factor, d) quasi-Fermi level splitting, e) optical power conversion efficiency and f) PL centre of mass.

We perform the same analysis on the 2PACz/TCTH which we qualitatively observed to have a much lower density of wrinkles than the DCDH and TCTH and plot the distribution in Supplementary Figure 18 where it is evident that both the number of wrinkles per unit area and the total surface coverage of wrinkles in this sample is much lower. This was also observed in equivalent samples with the same TCTH perovskite fabricated on other hole transporting layers (Me-4PACz and MeO-2PACz). Supplementary Figures 19 and 20 show a similar story to the case of the DCTH sample where the wrinkled areas show lower performance in pristine samples but this homogenises after operational stress testing.

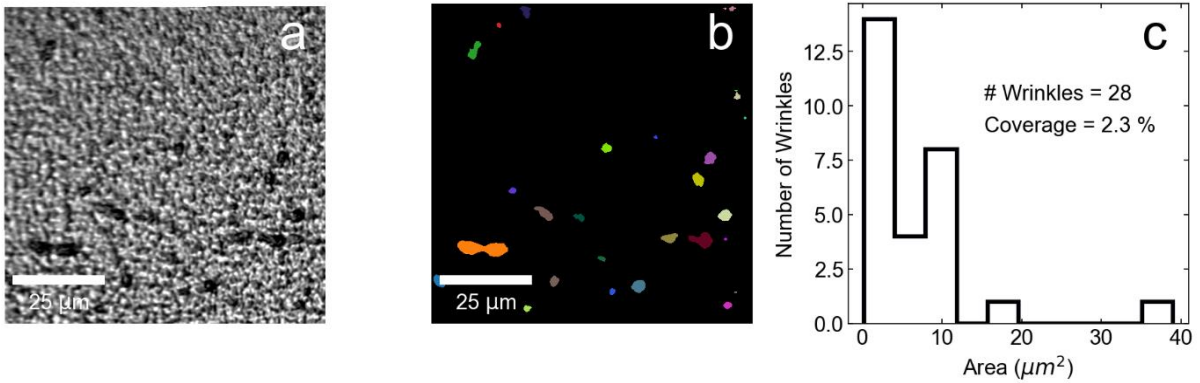

**Supplementary Figure 18:** a) Bright field reflectance image of 2PACz/TCTH solar cell. b) Filtered wrinkle mask from the reflectance image in a. c) Histogram of the size distribution of wrinkles in panel b.

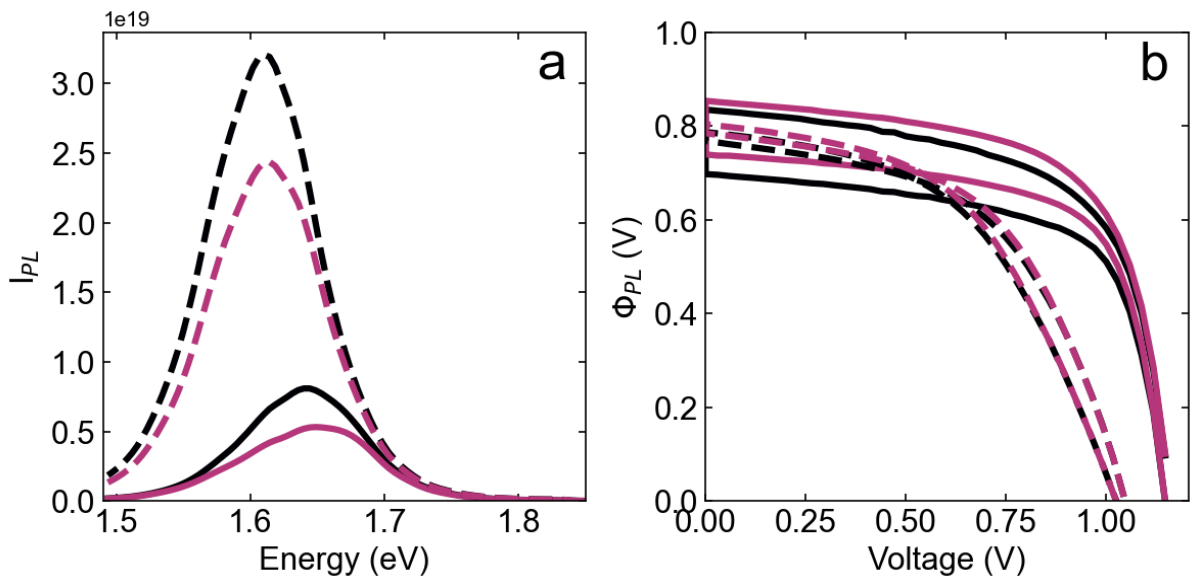

**Supplementary Figure 19**: a) Average PL spectra of non-wrinkled (red) and wrinkled (black) regions in a 2PACz/TCTH device before (solid lines) and after (dashed lines) operational stress. b) Mean optical JV curves of the same device and masked regions before and after operational stress.

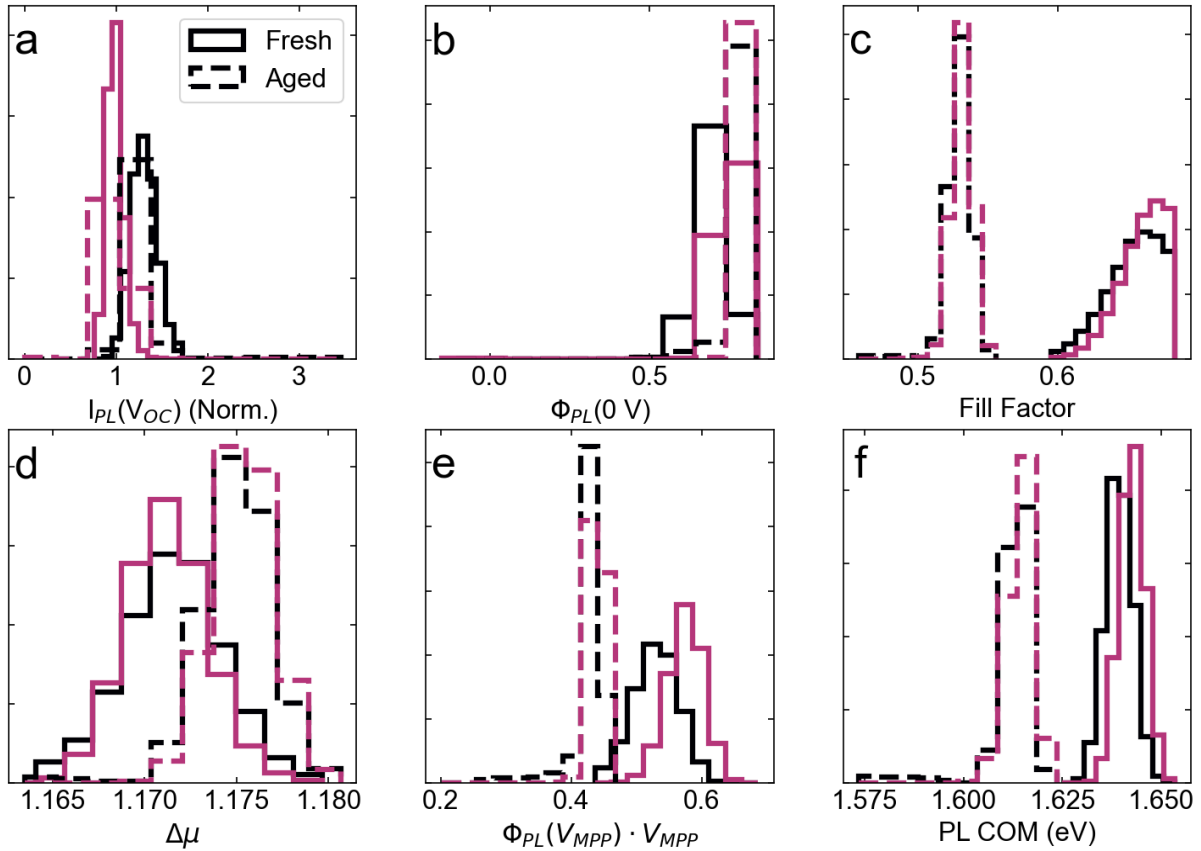

**Supplementary Figure 20**: Distributions of optoelectronic parameters on non-wrinkled (red) and wrinkled (black) regions before (solid) and after (dashed) operational stress of a 2PACz/TCTH device. a) open circuit PL intensity, b) short circuit optical charge extraction efficiency, c) optical fill factor, d) quasi-Fermi level splitting, e) optical power conversion efficiency and f) PL centre of mass.

One point that is important to mention is that while within batches, the distribution of wrinkling for a given perovskite/device is consistent, this is not always true between batches. Solvent drying, glovebox atmosphere and crystallisation kinetics play a very important role in the formation (or absence) of wrinkling, and the composition of the solution and substrate will certainly play a role, other factors are important to consider including the temperature, atmosphere and solvent vapor pressure in the glovebox at the time of fabrication. However, we find interestingly that even large changes to the wrinkling distribution have little impact on the performance or stability of these devices. To demonstrate this, we show a 2PACz/TCTH device from a separate batch of devices fabricated several months after the above reported samples which has a much higher surface coverage of wrinkles (Supplementary Figure 21), comparable to the distributions in the DCDH and DCTH samples. The optoelectronic and stability behaviour of this device (Supplementary Figure 22 and 23) is almost identical to the much less wrinkled equivalent shown above. The wrinkles have a small negative impact on the charge extraction efficiency in the pristine device which disappears after operation.

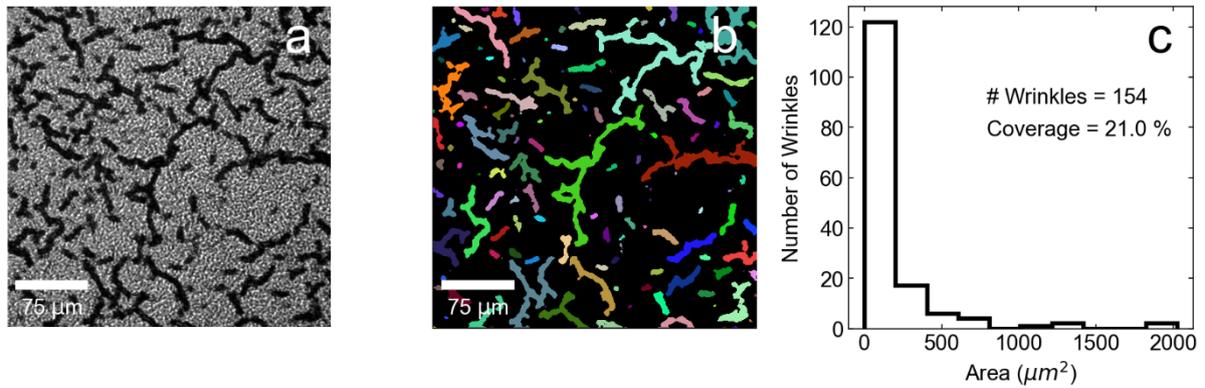

**Supplementary Figure 21:** a) Bright field reflectance image of 2PACz/TCTH solar cell. b) Filtered wrinkle mask from the reflectance image in a. c) Histogram of the size distribution of wrinkles in panel b.

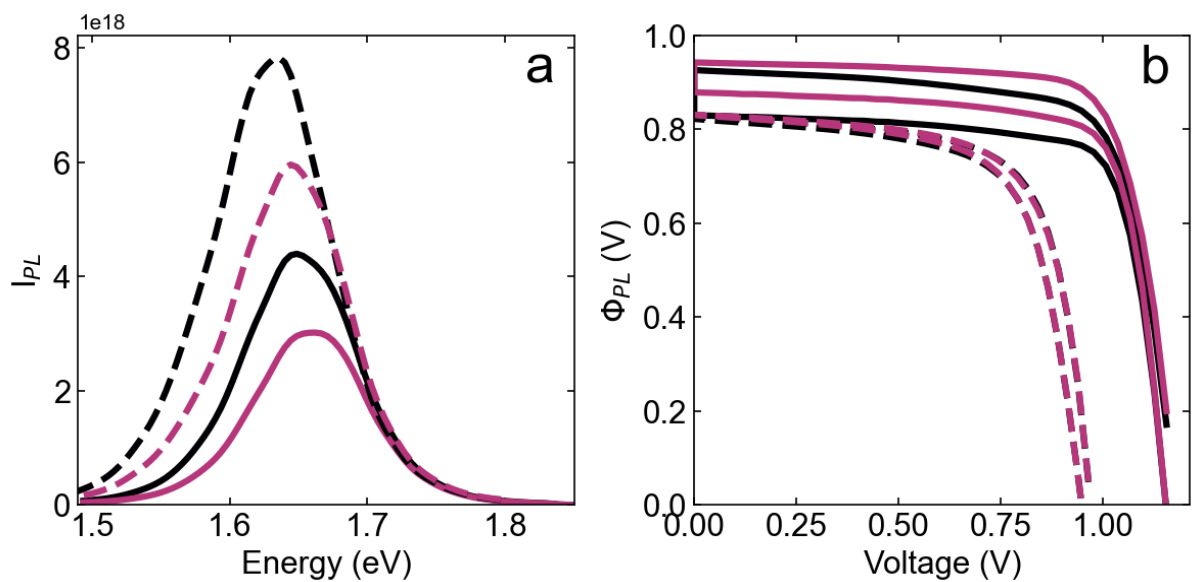

**Supplementary Figure 22**: a) Average PL spectra of non-wrinkled (red) and wrinkled (black) regions in a 2PACz/TCTH device before (solid lines) and after (dashed lines) operational stress. b) Mean optical JV curves of the same device and masked regions before and after operational stress.

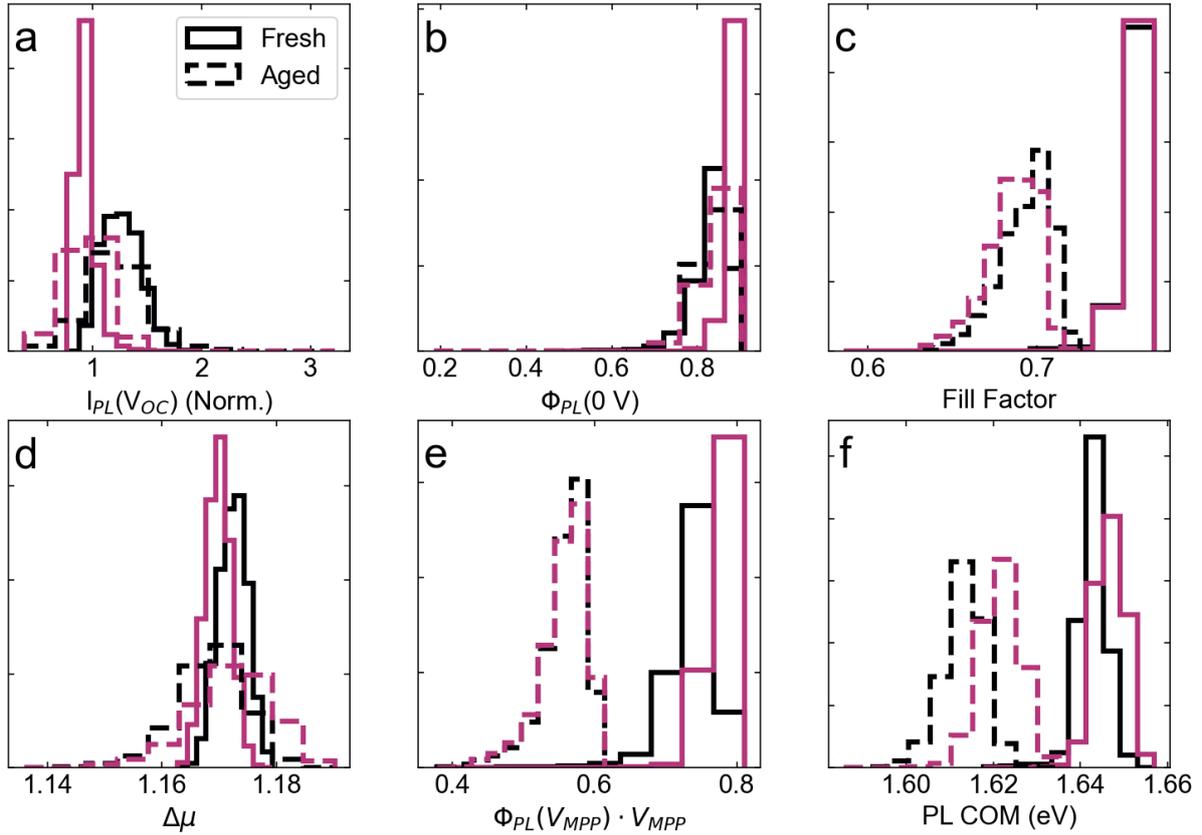

**Supplementary Figure 23**: Distributions of optoelectronic parameters on non-wrinkled (red) and wrinkled (black) regions before (solid) and after (dashed) operational stress of a 2PACz/TCTH device. a) open circuit PL intensity, b) short circuit optical charge extraction efficiency, c) optical fill factor, d) quasi-Fermi level splitting, e) optical power conversion efficiency and f) PL centre of mass.

To confirm that our optical measurements were not somehow missing a proportion of the wrinkling or light outcoupling was causing distortions skewing our view of their size and distribution, we used another morphology sensitive imaging technique, scanning electron microscopy (SEM) in order to image the wrinkles and grain structure across all of the device stacks studied. The images of the large scales wrinkle structures and smaller scale apparent grain morphologies are shown in Supplementary Figure 24. The relatively sparse coverage of the wrinkles and their interesting shape and features across each individual wrinkle matches well with our optical measurements, giving us further confidence that our above analysis is valid. To show that these features are not just 2-dimensional texture in the film and to get a closer look at the structure of the wrinkles, we performed some tilted SEM images with the devices tilted at 40° relative to normal incidence which are shown in Supplementary Figure 25. Several observations are immediately apparent from these images. The first simple observation is that the wrinkles do indeed have an appreciable height change relative to their surroundings as expected. The second is that rather than each wrinkles having a single 'peak', each wrinkle has two peaks, one on each side surrounded by a valley in the centre. This structure helps us explain some observations observed in the optical microscopy data where often each wrinkle would appear as one entity in bright-field reflection mode, but as two distinct features in PL (see Figure 1h, Figure S15 for examples). This unusual morphology clearly has a defining role in this effect. The final, more unexpected observation is that the morphological grains on the wrinkles are substantially larger on average than the surrounding flat film. The red shift of the PL from these domains may therefore also be influenced by the larger grain size, in addition to the thickness change and compositional variation.

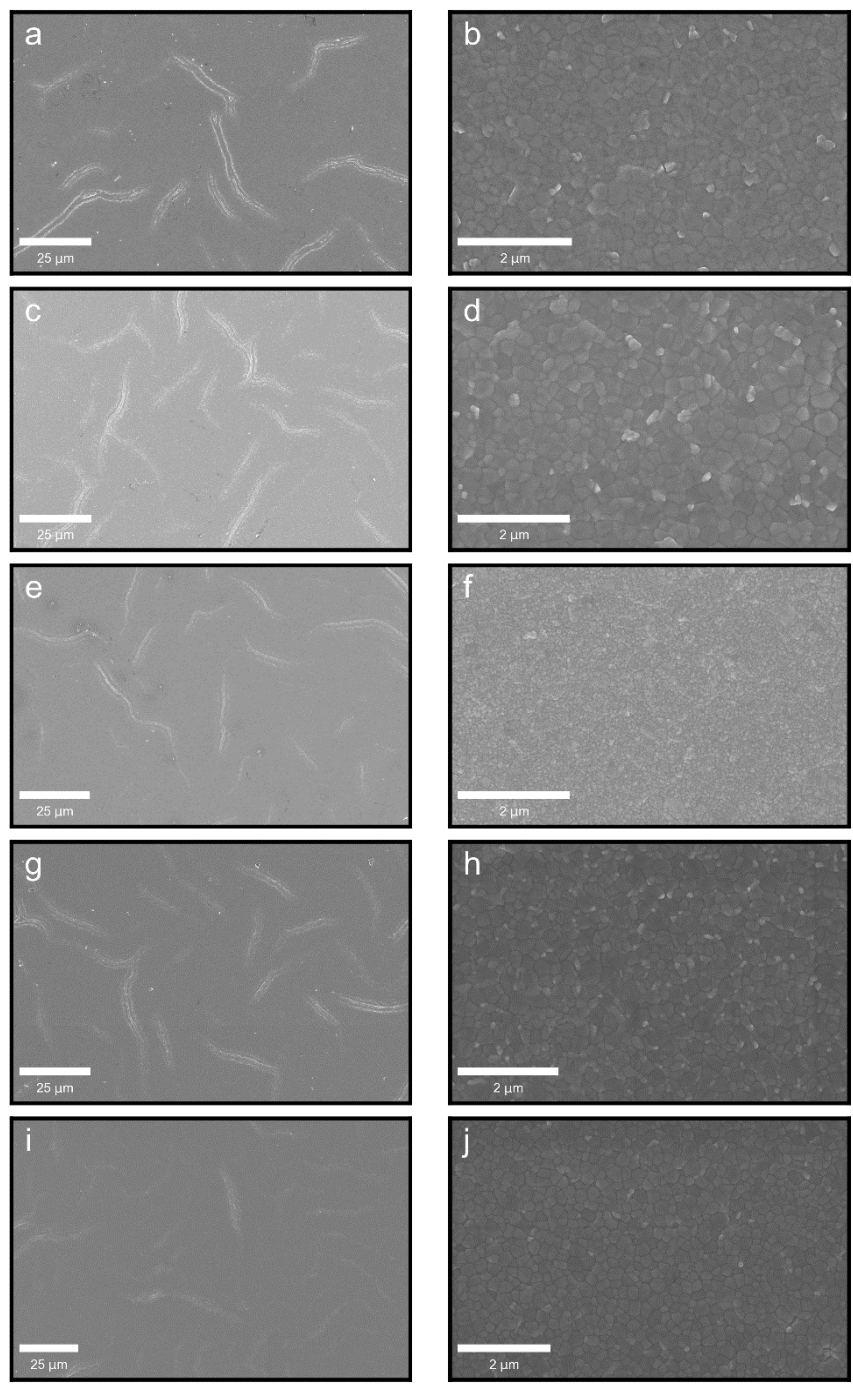

**Supplementary Figure 24:** Low mag images showing the larger scale wrinkle structure and small-scale grain structure respectively of a), b) Me-4PACz/TCTH, c), d), MeO-2PACz/TCTH, e), f), 2PACz/TCTH, g),h), 2PACz/DCTH and i), j) 2PACz/DCDH.

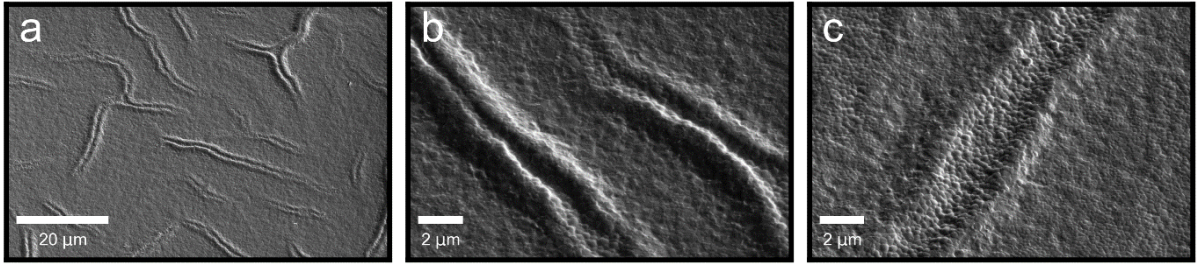

**Supplementary Figure 25:** SEM image of a) and b) 2PACz/DCTH and c) 2PACz/DCDH wrinkled perovskite devices taken at 40° tilt angle.

Taken together, these results suggest that the wrinkles are quite benign to device stability as they do not appear to be the seeds of spreading degradation or degrade any more than their surroundings. This is somewhat surprising as we might naively assume that these large-scale heterogeneities may invite degradation, but this appears not to be the case.

Supplementary Note 6: Hysteresis Spatial Variations

Here we investigate whether the local emission properties of the perovskite ($\Delta\mu$, open circuit PL intensity and COM) correlate more strongly with optoelectronic properties measured from voltage dependent PL mapping measured in either the forward or reverse voltage scan direction. To investigate this further, we took 2PACz/DCDH dataset and first performed autocorrelation plots between figures of merit extracted from the reverse and forward scans. If these are highly correlated, it means that the optical figures of merit ($\Delta\mu$ and COM) will have a similar correlation to both scan directions. In Supplementary Figure 26, we plot the autocorrelation of the optical extraction efficiency, fill factor and PCE in the pristine 2PACz/DCDH device. Each column in the figure shows maps of the reverse and forward scan figures of merit, the histograms of their respective distributions before a 2D histogram to show the autocorrelations. Both qualitative inspection and the statistical analysis from the Spearman's coefficient indicate a strong spatial correlation between the forward and reverse direction figures of merit in the pristine sample, implying the optical figures of merit will be correlated equally with the figures of merit from both scan directions. We confirm this in Supplementary Figure 27 where once again, both the qualitative appearance and statistical analysis of the data of both $\Delta\mu$ and COM correlations are very similar in both forward and reverse scans.

The situation is more complex after the operational stress test as there is a particularly large disparity between forward and reverse scans in this device. We plot the autocorrelation of the stress tested figures of merit in Supplementary Figure 28. In this device, the optical extraction efficiency is still relatively homogeneous in the forward and reverse directions, however, strikingly we find a strong gradient in performance across the sample from one corner to the other which particularly manifests in the fill factor and the resulting optical PCE. However, the gradients are of opposite signs in the forward and reverse scans, strongly suggesting that the gradient observed is due to a gradient in mobile ionic concentrations. This means that there is a positive autocorrelation in the extraction efficiency but a negative correlation in both FF and optical PCE. The gradient in the fill factor in the forward scan qualitatively matches the gradient of the optical figures of merit, in particular $\Delta\mu$. The correlations are shown in Supplementary Figure 29. The COM and $\Delta\mu$ values positively correlate with forward direction figures of merit and anticorrelate with the reverse scan figures, suggesting that the 'true' optoelectronic performance of the perovskite itself is exhibited in the forward scan, before being somewhat masked by the ionic effects. However, we caveat this statement that given these values are particularly affected by the large lateral variation caused by the edge effects, it is worth considering a region that is less affected by these effects. We perform the same correlation analysis on a different region of the same stress tested device that is less hysteretic (Supplementary Figure 30). Here we find that the optical figures of merit correlate positively with the reverse scan, while there is a more complex, non-monotonic relationship with the forward scan. We assert that in a complete device, simply measuring the microscopic properties of the perovskite ($\Delta\mu$ and COM) in isolation is not sufficient to determine local performance, particularly in cases with large concentrations of mobile ionic species and hysteresis.

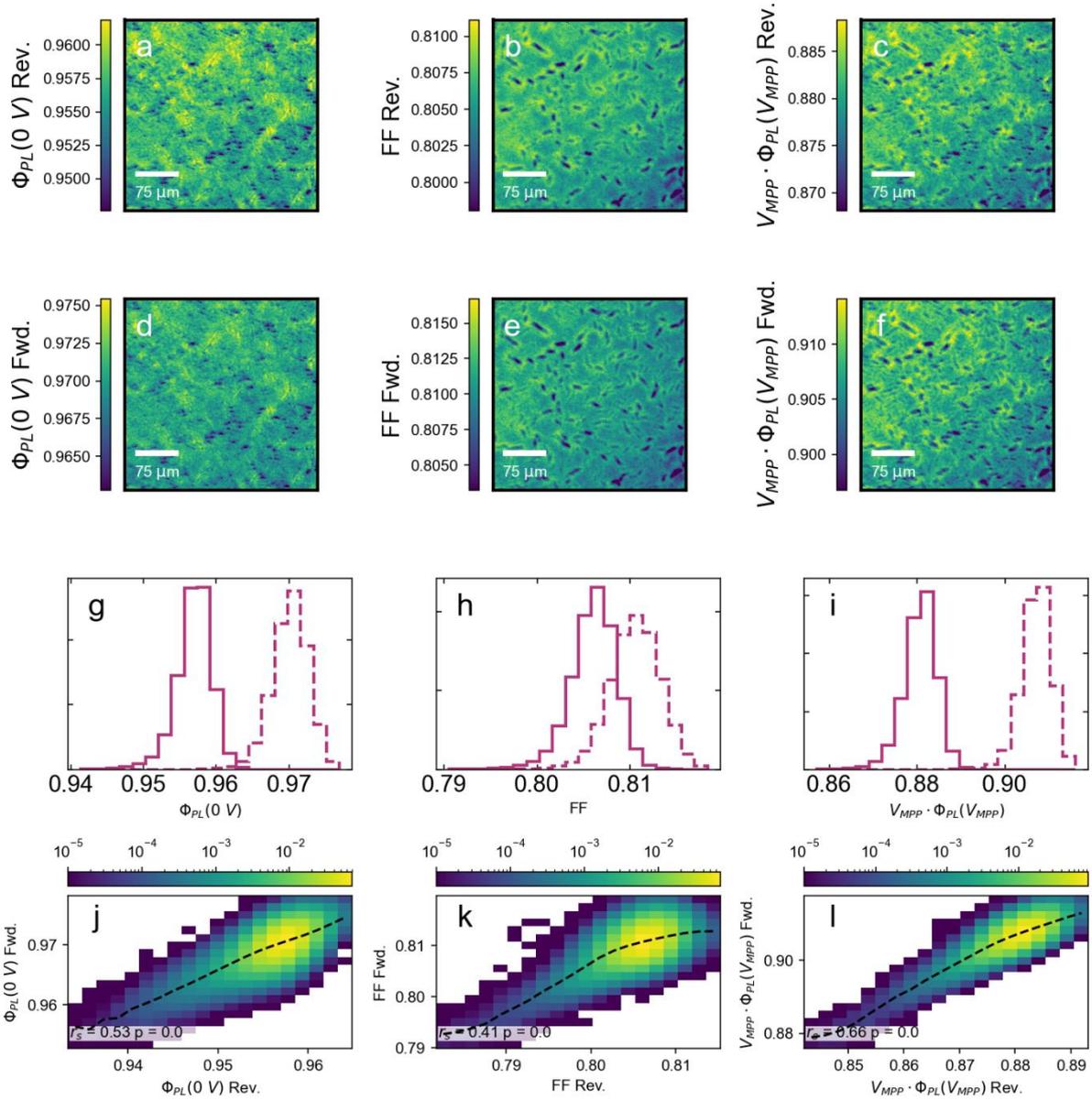

**Supplementary Figure 26**: Optical JV parameters extracted in both the reverse and forward directions from a pristine 2PACz/DCDH solar cell. Maps of the reverse scan a) optical extraction efficiency, b) fill factor and c) optical PCE. Maps of the forward scan d) optical extraction efficiency, e) fill factor and f) optical PCE. Histograms comparing the distributions of g) optical extraction efficiency, h) FF and i) optical PCE extracted in reverse (solid line) and forward (dashed line) figures of merit. 2D histograms showing the correlation with j) optical extraction efficiency, k) fill factor and i) optical PCE in the reverse and forward direction.

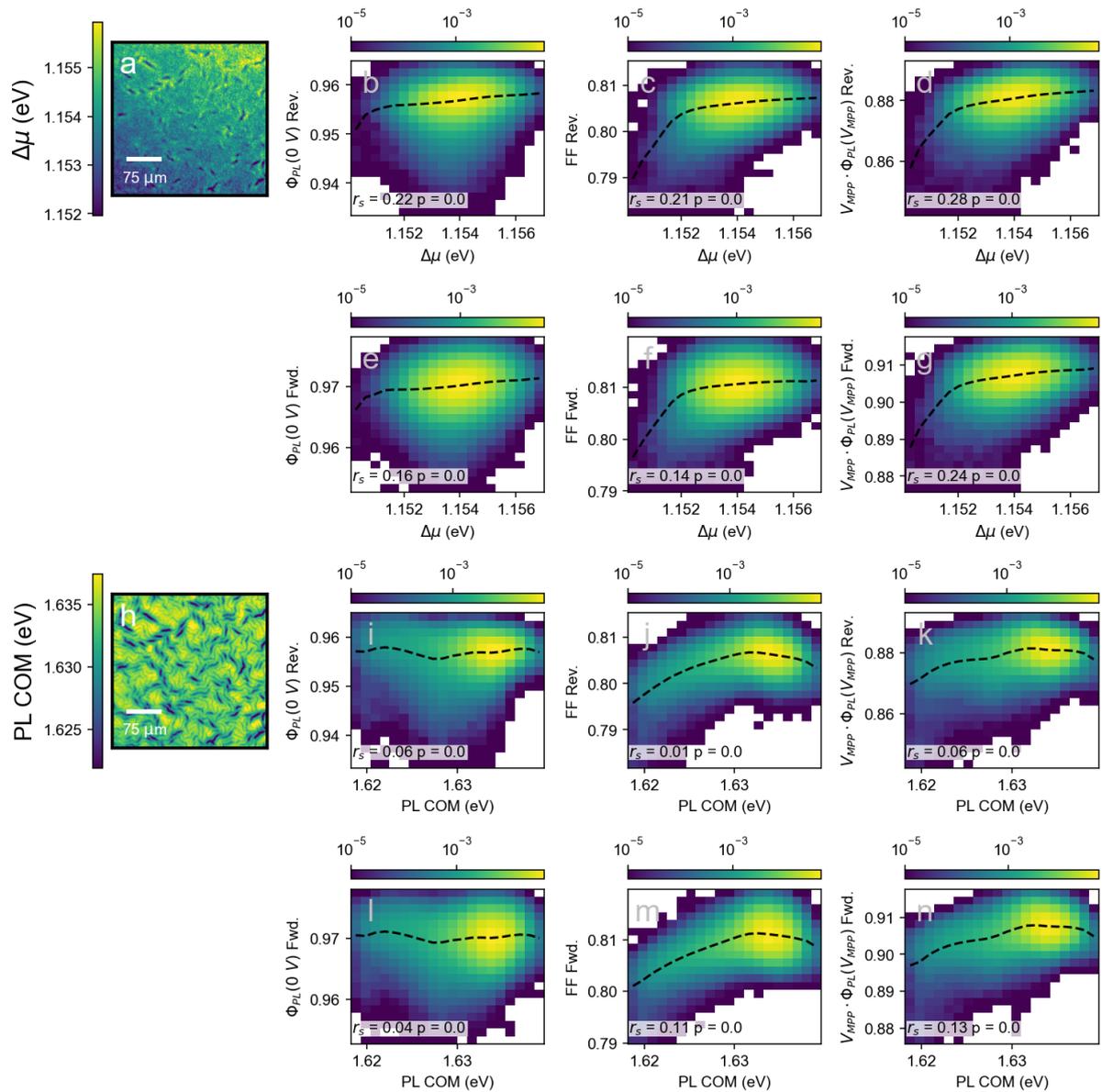

**Supplementary Figure 27:** Correlations between optical figures of merit and optoelectronic figures of merit extracted from forward and reverse scans in pristine 2PACz/DCDH solar cell. a) Map of Δμ. 2D histograms showing correlation between Δμ and b), e) optical extraction efficiency, c) f) fill factor and d), g) optical PCE in the reverse and forward directions respectively. h) Map of COM. 2D histograms showing correlation between COM and i), l) optical extraction efficiency, j) m) fill factor and k), n) optical PCE in the reverse and forward directions respectively.

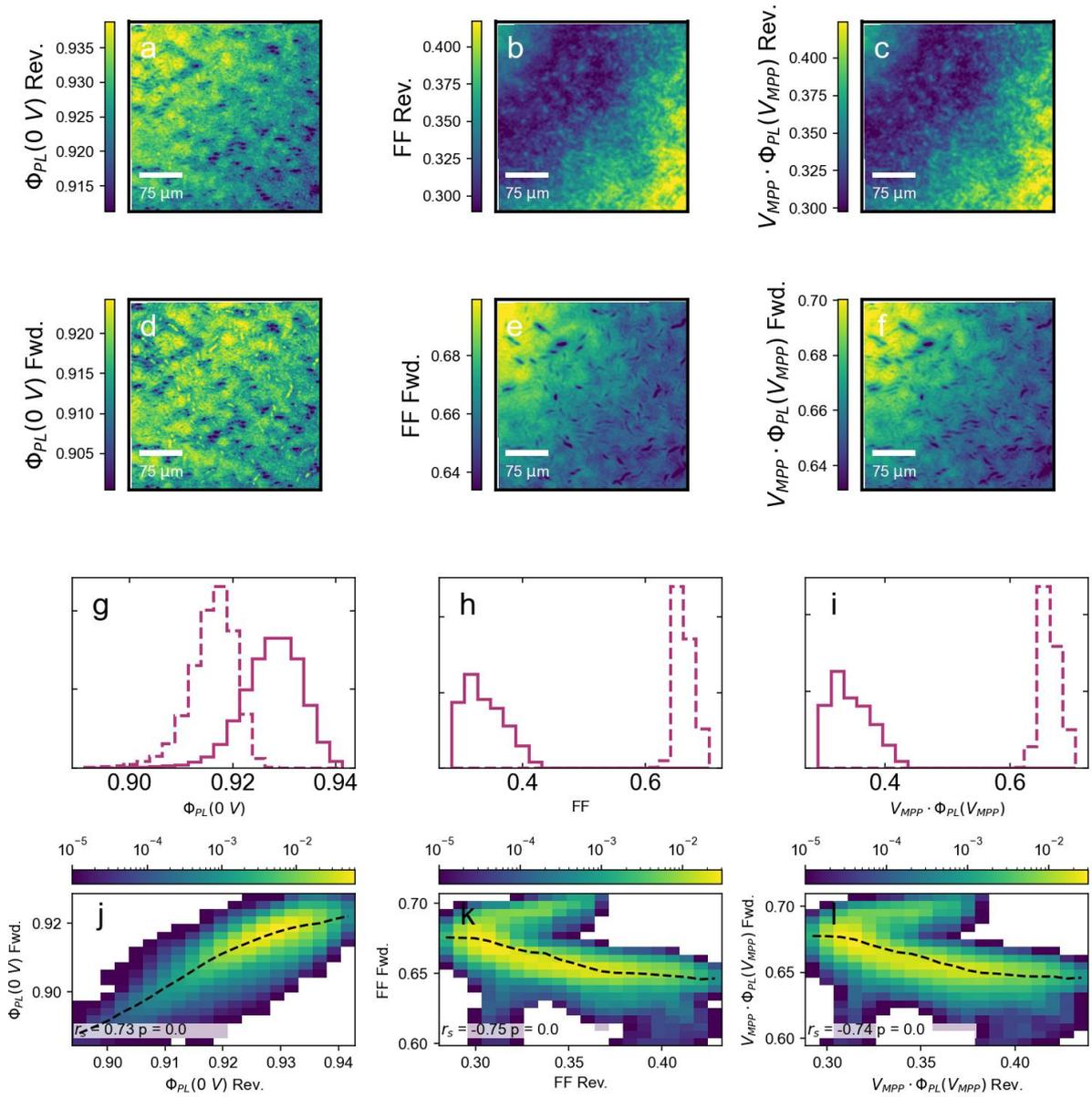

**Supplementary Figure 28**: Optical JV parameters extracted in both the reverse and forward directions from an operationally stress tested 2PACz/DCDH solar cell. Maps of the reverse scan a) optical extraction efficiency, b) fill factor and c) optical PCE. Maps of the forward scan d) optical extraction efficiency, e) fill factor and f) optical PCE. Histograms comparing the distributions of g) optical extraction efficiency, h) FF and i) optical PCE extracted in reverse (solid line) and forward (dashed line) figures of merit. 2D histograms showing the correlation with j) optical extraction efficiency, k) fill factor and i) optical PCE in the reverse and forward direction.

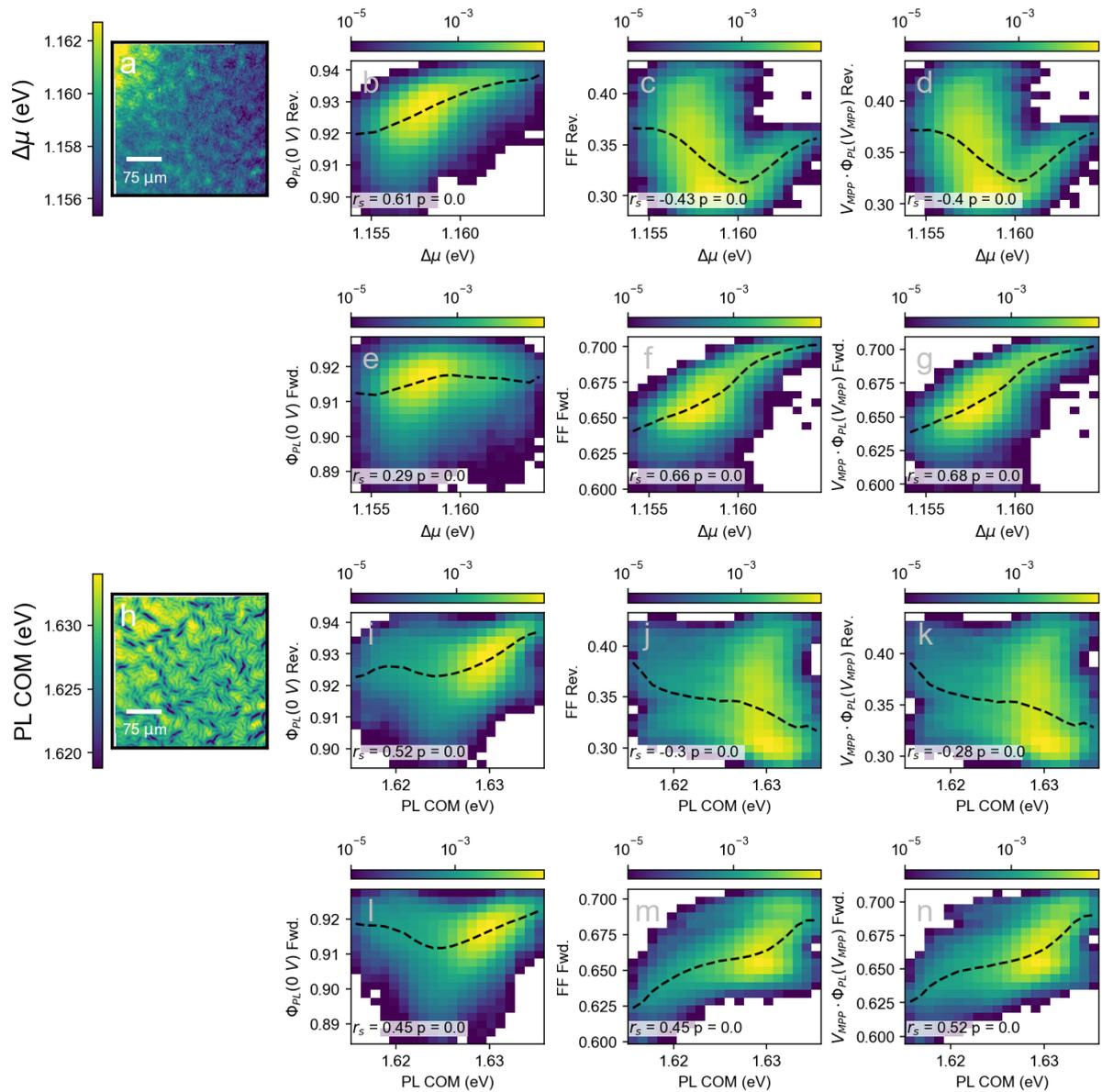

**Supplementary Figure 29:** Correlations between optical figures of merit and optoelectronic figures of merit extracted from forward and reverse scans in operationally stress tested 2PACz/DCDH solar cell. a) Map of $\Delta\mu$. 2D histograms showing correlation between $\Delta\mu$ and b), e) optical extraction efficiency, c) f) fill factor and d), g) optical PCE in the reverse and forward directions respectively. h) Map of COM. 2D histograms showing correlation between COM and i), l) optical extraction efficiency, j) m) fill factor and k), n) optical PCE in the reverse and forward directions respectively.

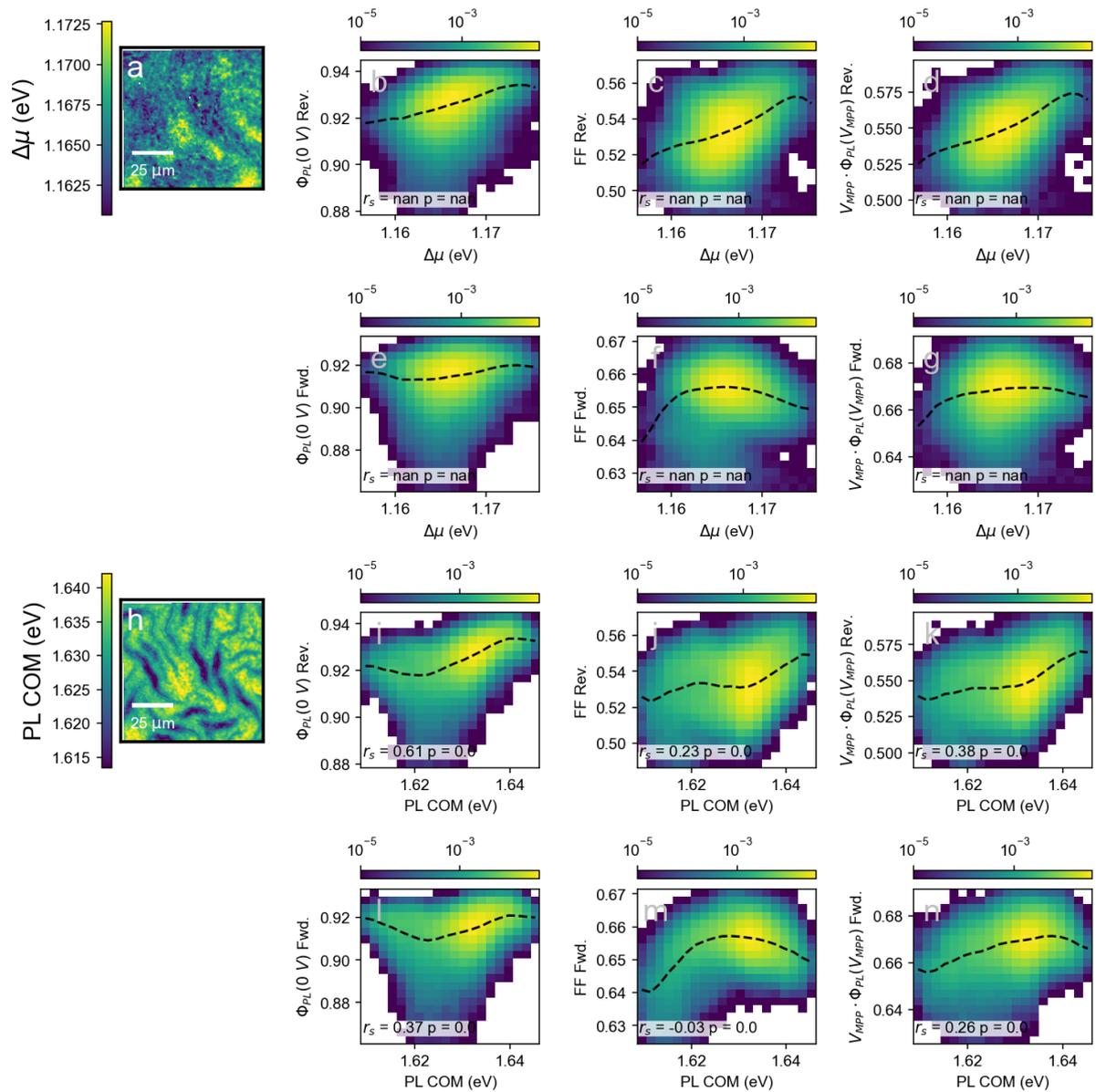

**Supplementary Figure 30:** Correlations between optical figures of merit and optoelectronic figures of merit extracted from forward and reverse scans in operationally stress tested 2PACz/DCDH solar cell. a) Map of Δμ. 2D histograms showing correlation between Δμ and b), e) optical extraction efficiency, c) f) fill factor and d), g) optical PCE in the reverse and forward directions respectively. h) Map of COM. 2D histograms showing correlation between COM and i), l) optical extraction efficiency, j) m) fill factor and k), n) optical PCE in the reverse and forward directions respectively.

**Supplementary Figures**

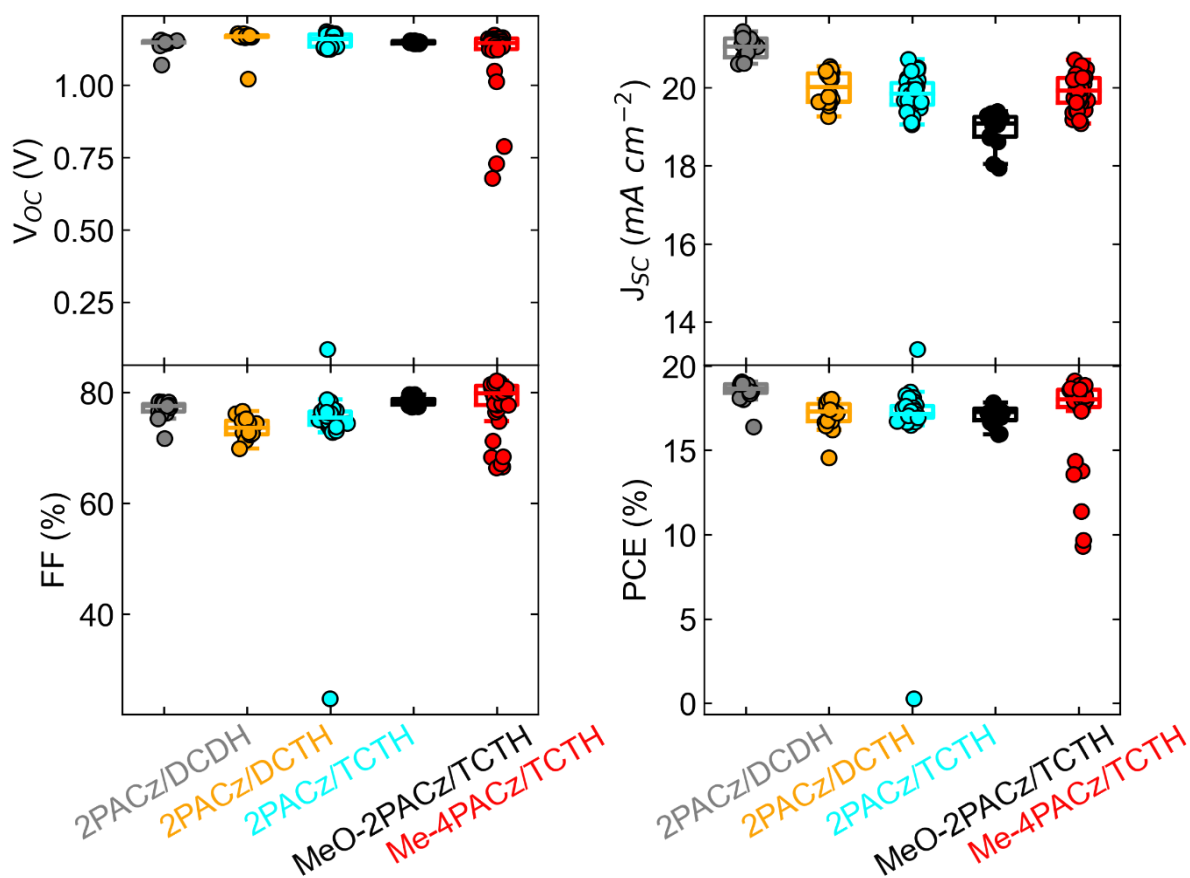

**Supplementary Figure 31: Electrically measured JV figures of merit.** Box plots of the 4 main solar cell figures of merit across the compositional and HTL phase space described in the main text. Data shown for 116 devices. Boxplot horizontal lines represent the median, the box represents the 1st and 3rd quartile. Whiskers extend to the last data point less/greater than the upper/lower quartile plus 1.5 times the interquartile range.

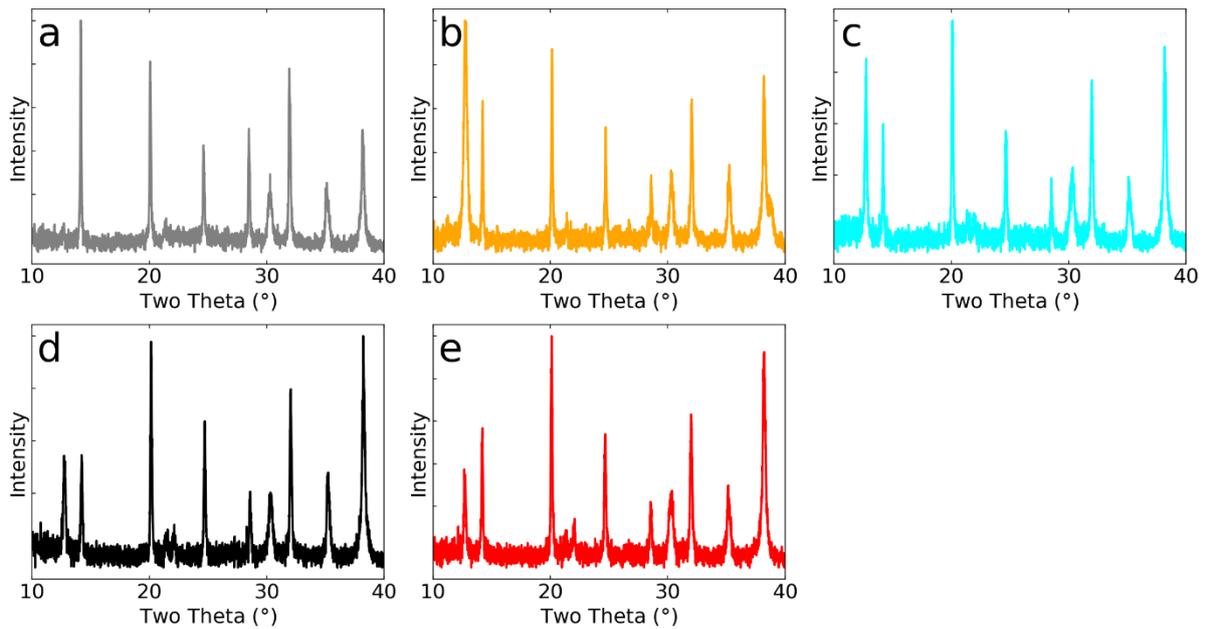

**Supplementary Figure 32:** XRD patterns of the 5 main device stacks in the study. a) 2PACz/DCDH, b) 2PACz/DCTH, c) 2PACz/TCTH, d) MeO-2PACz/TCTH, e) Me-4PACz/TCTH.

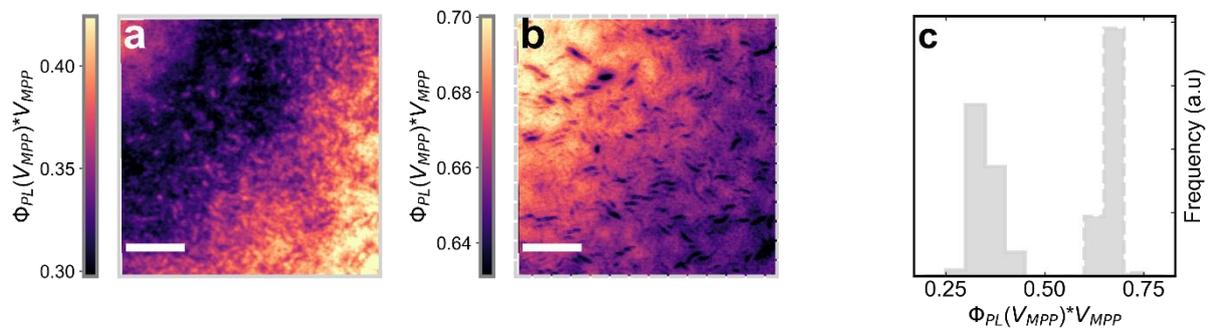

**Supplementary Figure 33. Spatial distribution of hysteresis in DCDH solar cell after operational stress test.** Optical PCE maps of DCDH solar cells after operational stress extracted from the a) reverse scan (as shown in Figure 2) and b) the forward voltage sweep. Scalebars are 75 µm. c) Histogram of optical PCE values from the maps in a and b (dashed lines represent forward voltage sweep values).

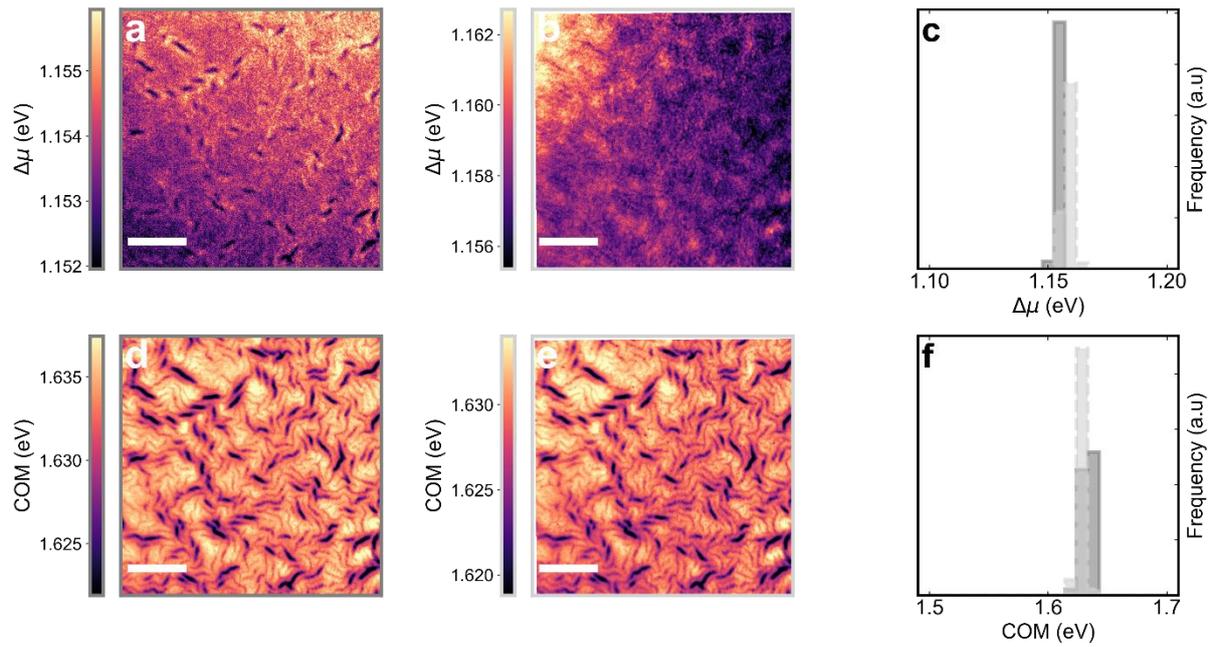

**Supplementary Figure 34. Δμ and COM of DCDH solar cell before after accelerated ageing.** Δμ a) before and b) after accelerated ageing, distributions summarised in panel c). Solid bars are before ageing, dashed bars are after ageing. COM d) before and e) after accelerated ageing, distributions summarised in panel f). Dark gray is before and light gray is after the accelerated ageing protocol. The scan area is the same as for the voltage dependent measurement shown in Figure 2. Scalebars are 75 µm.

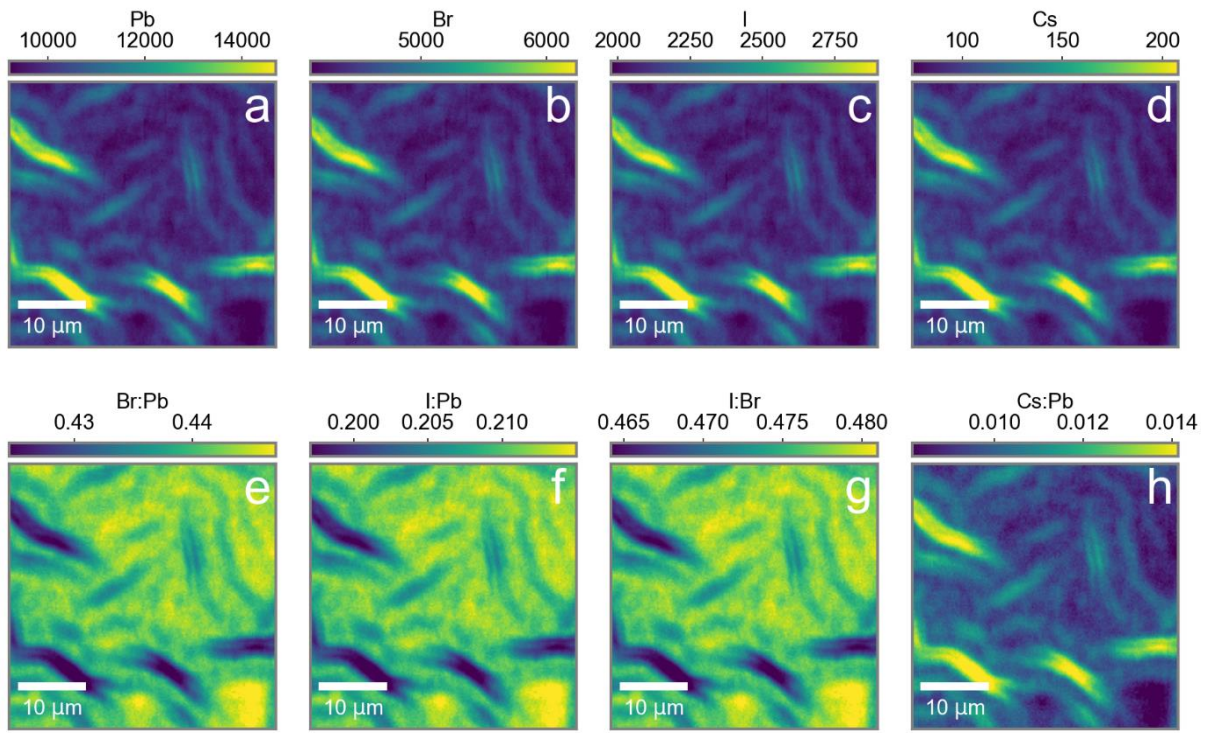

**Supplementary Figure 35:** XRF maps of region 1 from Figure 2 in the main text showing the a) Pb L, b) Br K, c) I L and d) Cs L lines. Ratios of the e) Br:Pb lines, f) I:Pb lines, g) I:Br lines and h) Cs:Pb lines.

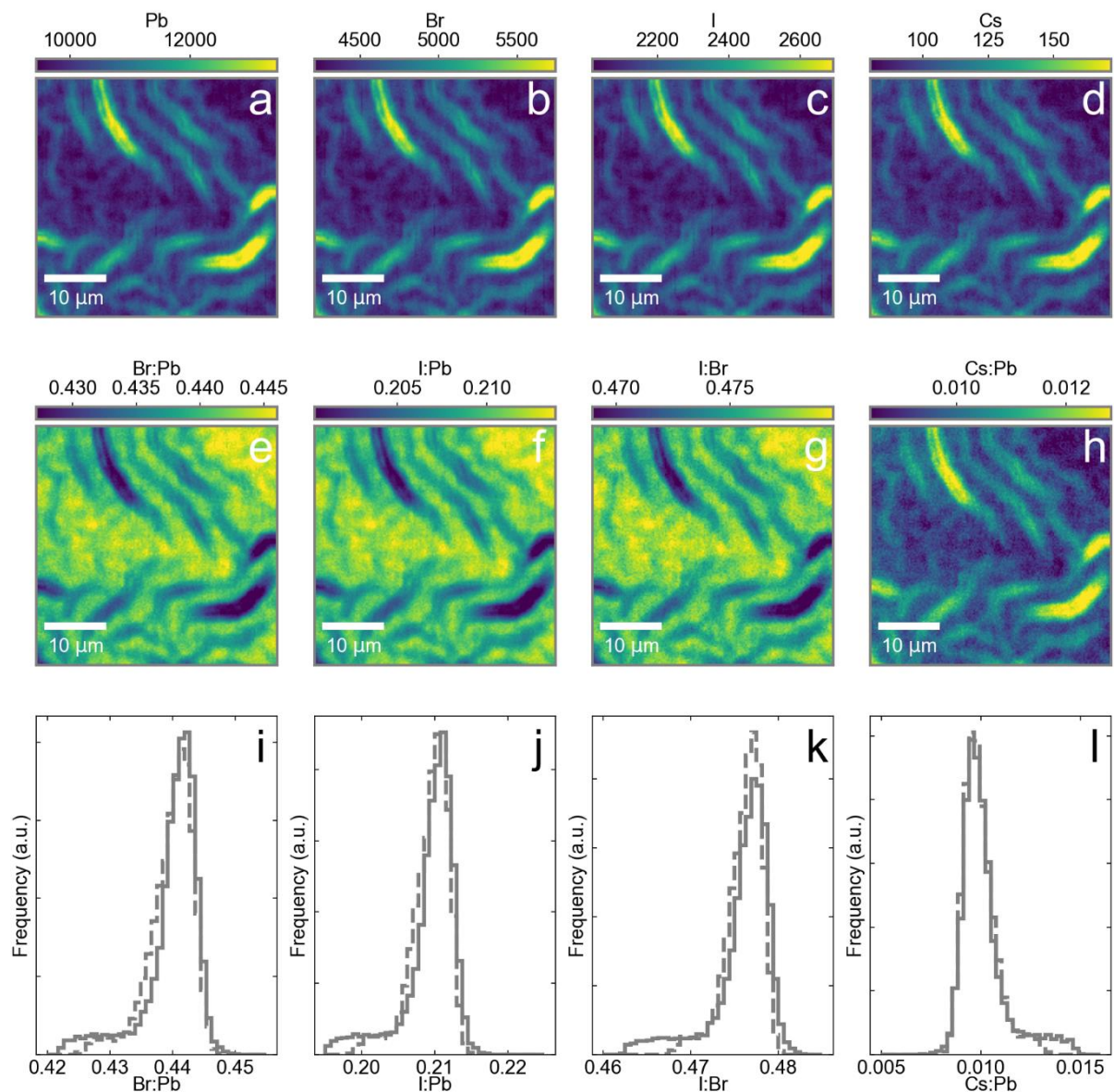

**Supplementary Figure 36:** XRF maps of region 2 from Figure 2 in the main text showing the a) Pb L, b) Br K, c) I L and d) Cs L lines. Ratios of the e) Br:Pb lines, f) I:Pb lines, g) I:Br lines and h) Cs:Pb lines. Histograms comparing the XRF intensities between region 1 (solid line) and region 2 (dashed lines) for the i) Br:Pb ratio, j) I:Pb ratio, k) I:Br and l) Cs:Pb lines.

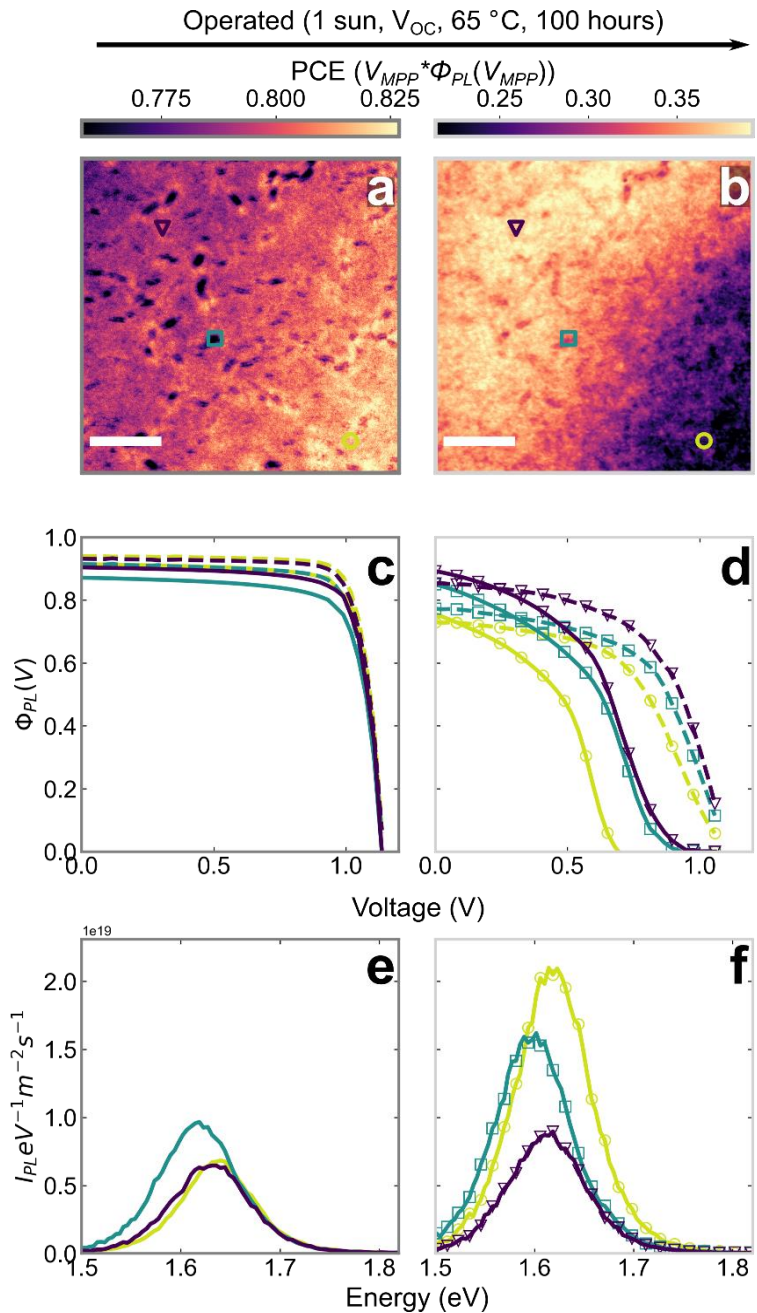

**Supplementary Figure 37. Optical and electronic spectra of DCDH before and after operation.** Optical PCE maps of the same area of a fresh a) and operated b) DCDH solar cell after 100 hours at $V_{OC}$, 65 °C and 1 sun illumination. c) and d) Optical JV curves before and after ageing from the points marked in panels a and b. Solid lines are reverse scans, dashed lines are forward scans. e) and f) PL spectra from the same marked areas before and after ageing respectively. Scalebars in a and b are 75 µm.

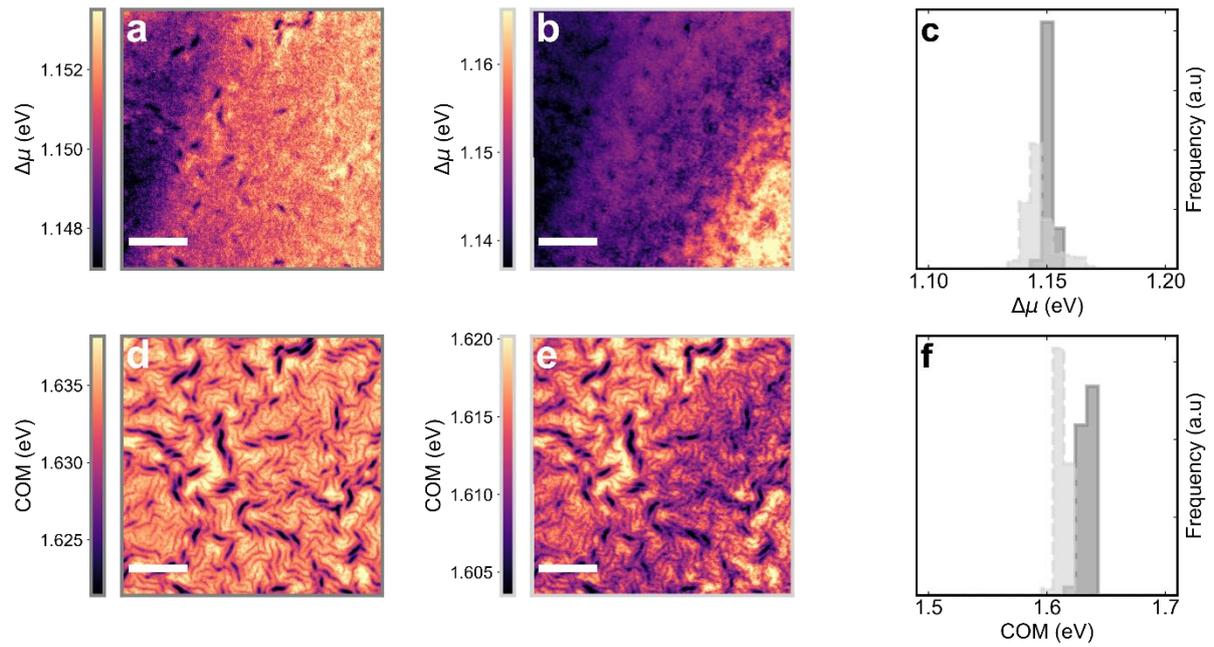

**Supplementary Figure 38. Δµ and COM of DCDH solar cell before after accelerated ageing.** Δµ a) before and b) after accelerated ageing, distributions summarised in panel c). Solid bars are before ageing, dashed bars are after ageing. COM d) before and e) after accelerated ageing, distributions summarised in panel f). Dark gray is before and light gray is after the accelerated ageing protocol. The scan area is the same as for the voltage dependent measurement shown in Supplementary Figure 13. Scalebars are 75 µm.

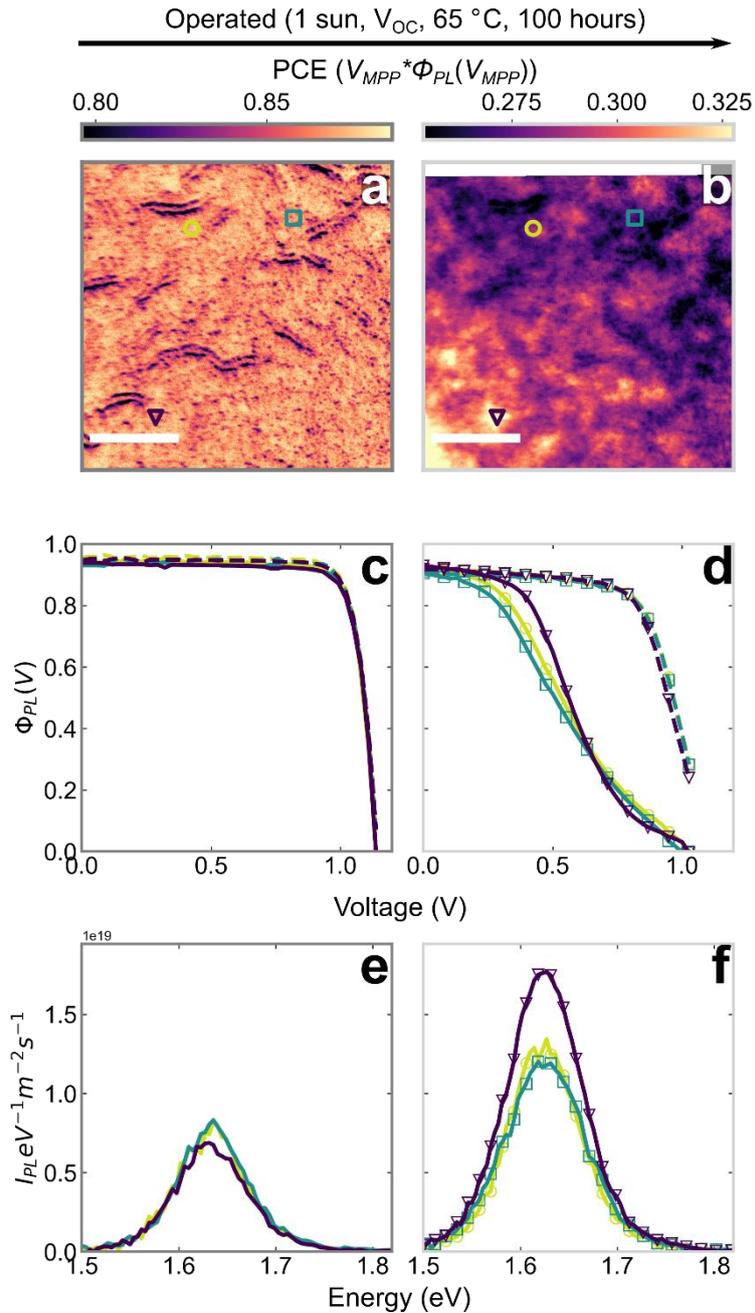

**Supplementary Figure 39. Optical and electronic spectra of DCDH before and after operation away from an edge.** Optical PCE maps of the same area of a fresh a) and operated b) DCDH solar cell after 100 hours at $V_{OC}$, 65 °C and 1 sun illumination. c) and d) Optical JV curves before and after ageing from the points marked in panels a and b. Solid lines are reverse scans, dashed lines are forward scans. e) and f) PL spectra from the same marked areas before and after ageing respectively. Scalebars in a and b are 25 µm.

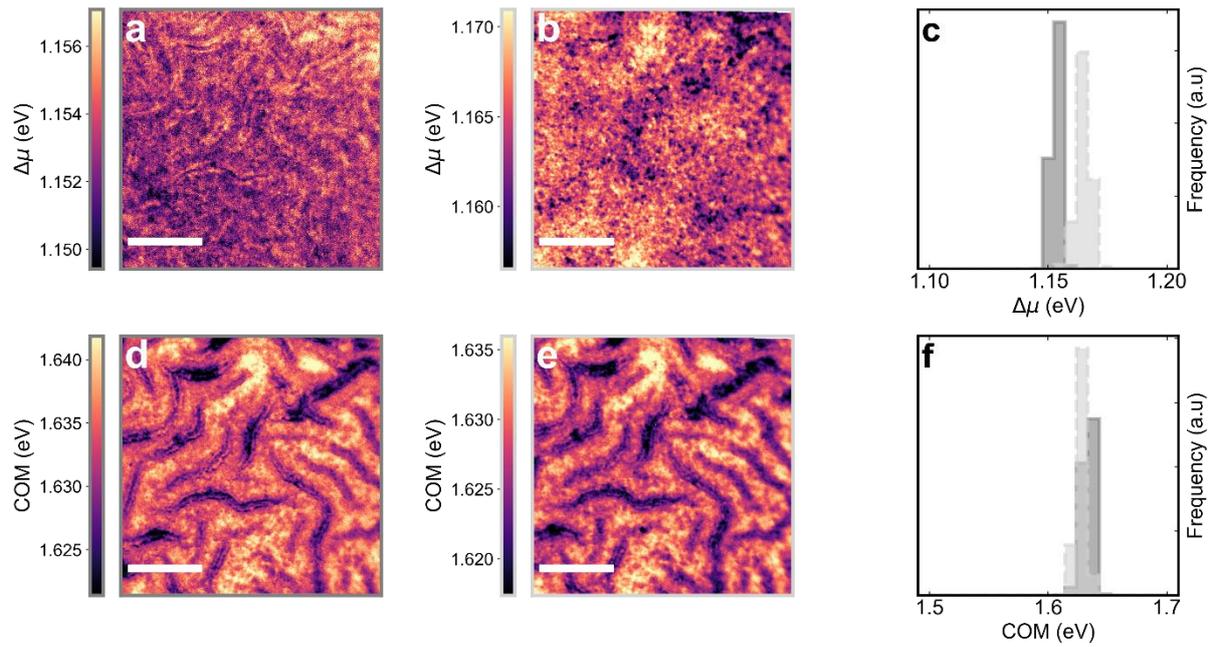

**Supplementary Figure 40. Δμ and COM of DCDH solar cell before after accelerated ageing.** Δμ a) before and b) after accelerated ageing, distributions summarised in panel c). Solid bars are before ageing, dashed bars are after ageing. COM d) before and e) after accelerated ageing, distributions summarised in panel f). Dark gray is before and light gray is after the accelerated ageing protocol. The scan area is the same as for the voltage dependent measurement shown in Supplementary Figure 15. Scalebars are 25 µm.

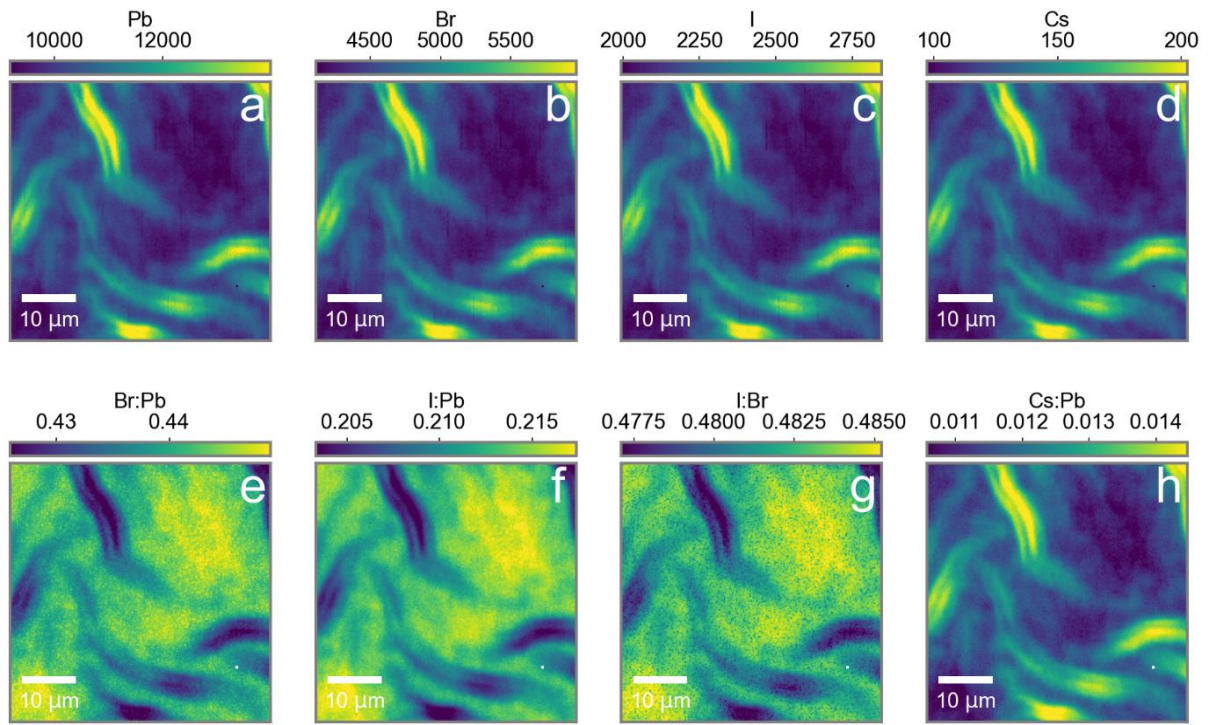

**Supplementary Figure 41:** XRF maps of a DCDH solar cell showing the a) Pb L, b) Br K, c) I L and d) Cs L lines. Ratios of the e) Br:Pb lines, f) I:Pb lines, g) I:Br lines and h) Cs:Pb lines.

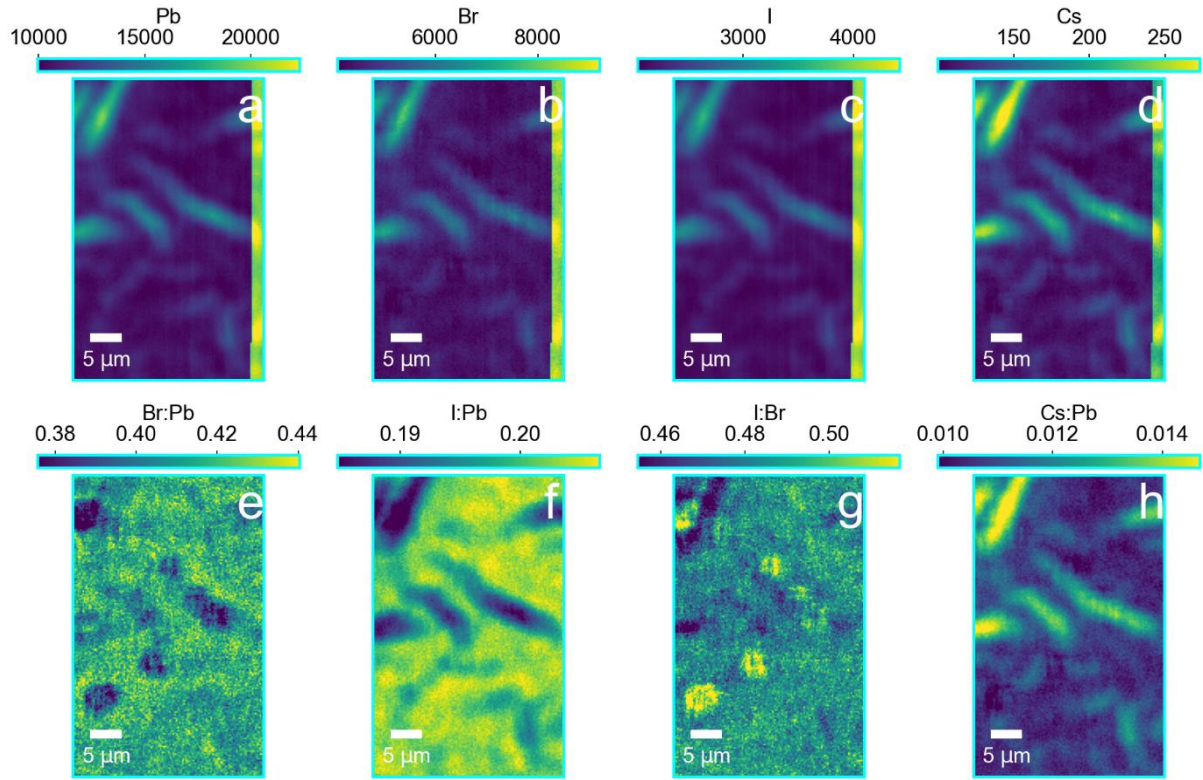

**Supplementary Figure 42:** XRF maps of a DCTH solar cell showing the a) Pb L, b) Br K, c) I L and d) Cs L lines. Ratios of the e) Br:Pb lines, f) I:Pb lines, g) I:Br lines and h) Cs:Pb lines.

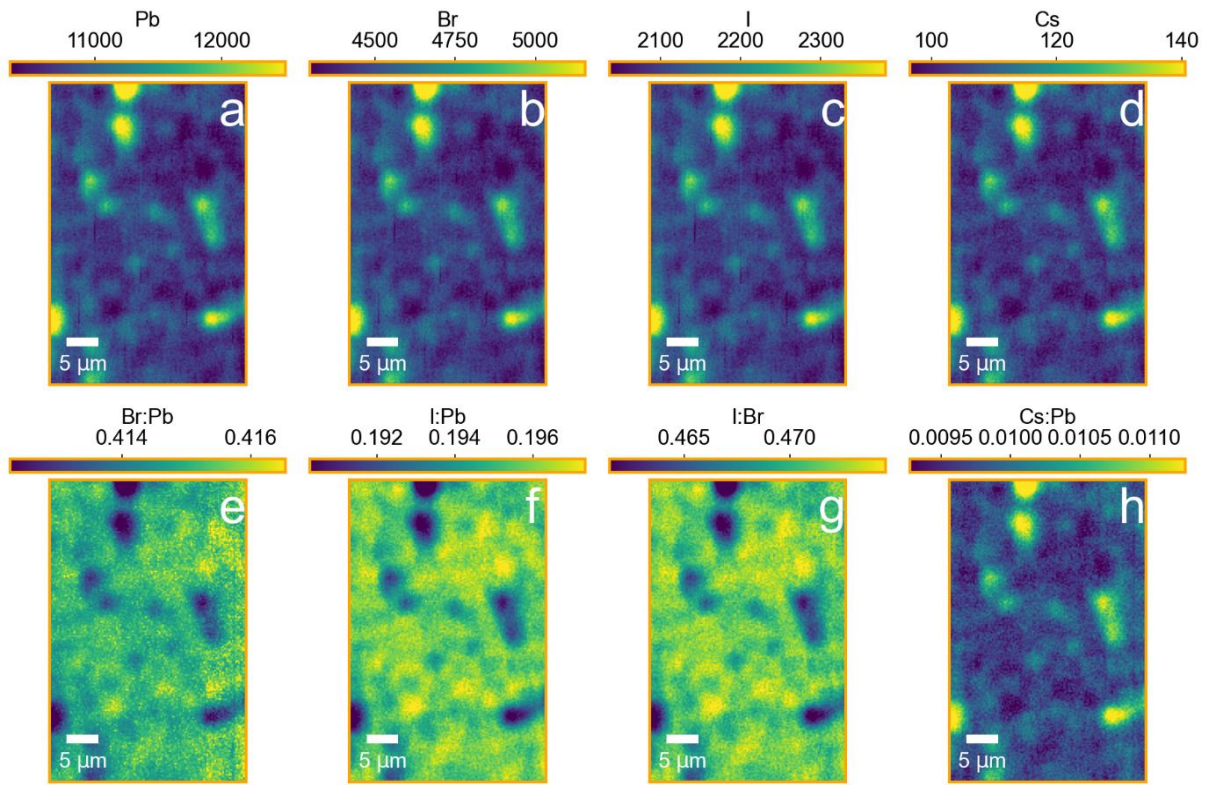

**Supplementary Figure 43:** XRF maps of a TCTH solar cell showing the a) Pb L, b) Br K, c) I L and d) Cs L lines. Ratios of the e) Br:Pb lines, f) I:Pb lines, g) I:Br lines and h) Cs:Pb lines.

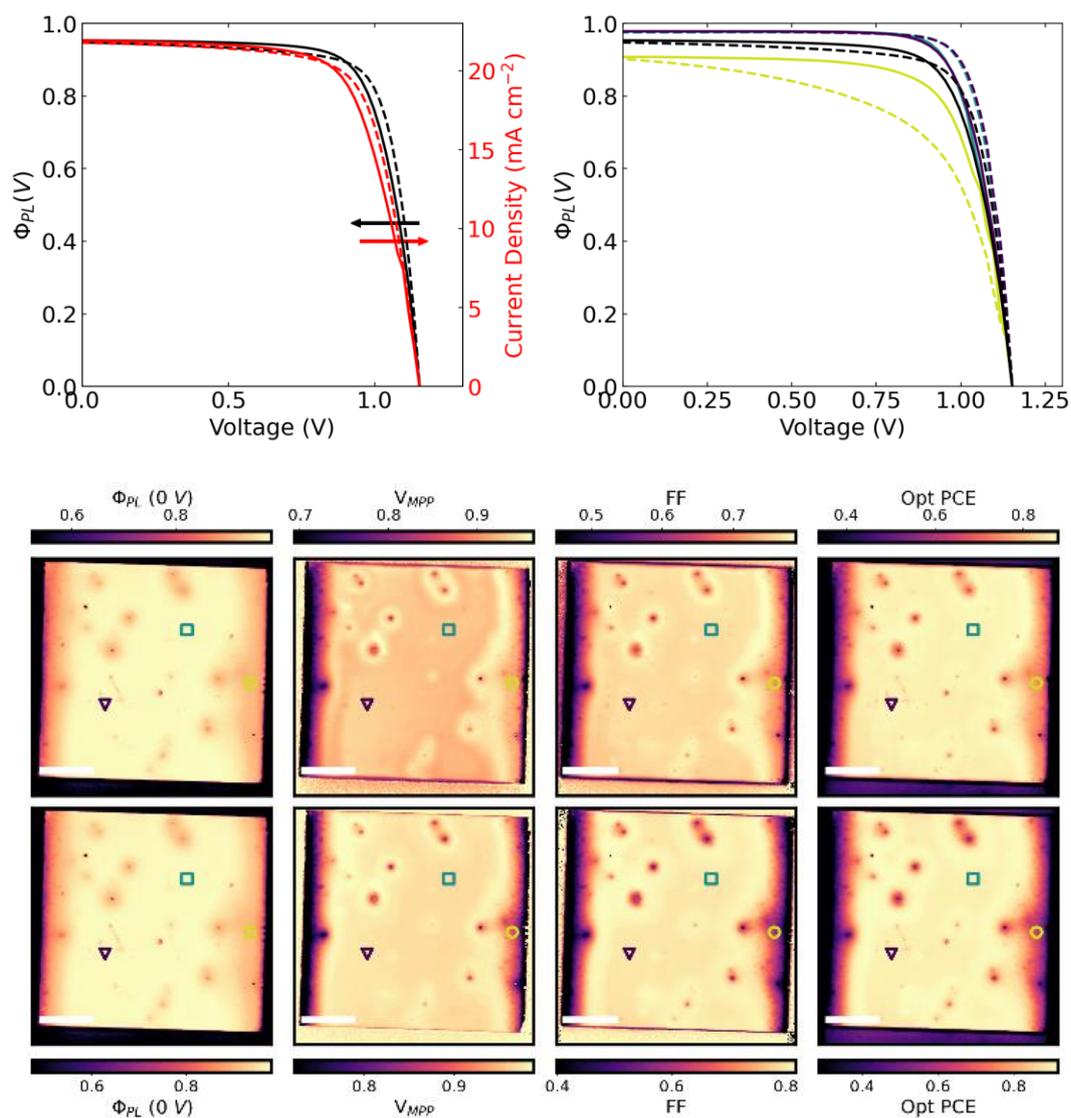

**Supplementary Figure 44**: Large scale optical JV measurements of pristine 2PACz/DCDH perovskite solar cell. a) mean optical (black) and electrical (red) JV measurements of the device. b) Optical JV curves extracted from the areas marked in panels c-j. Maps of the optical JV figures of merit c) optical extraction efficiency, d) max power point voltage, e) fill factor, f) optical PCE extracted from the reverse scan. g) optical extraction efficiency, h) max power point voltage, i) fill factor, j optical PCE extracted from the forward scan. Scalebars are 1 mm.

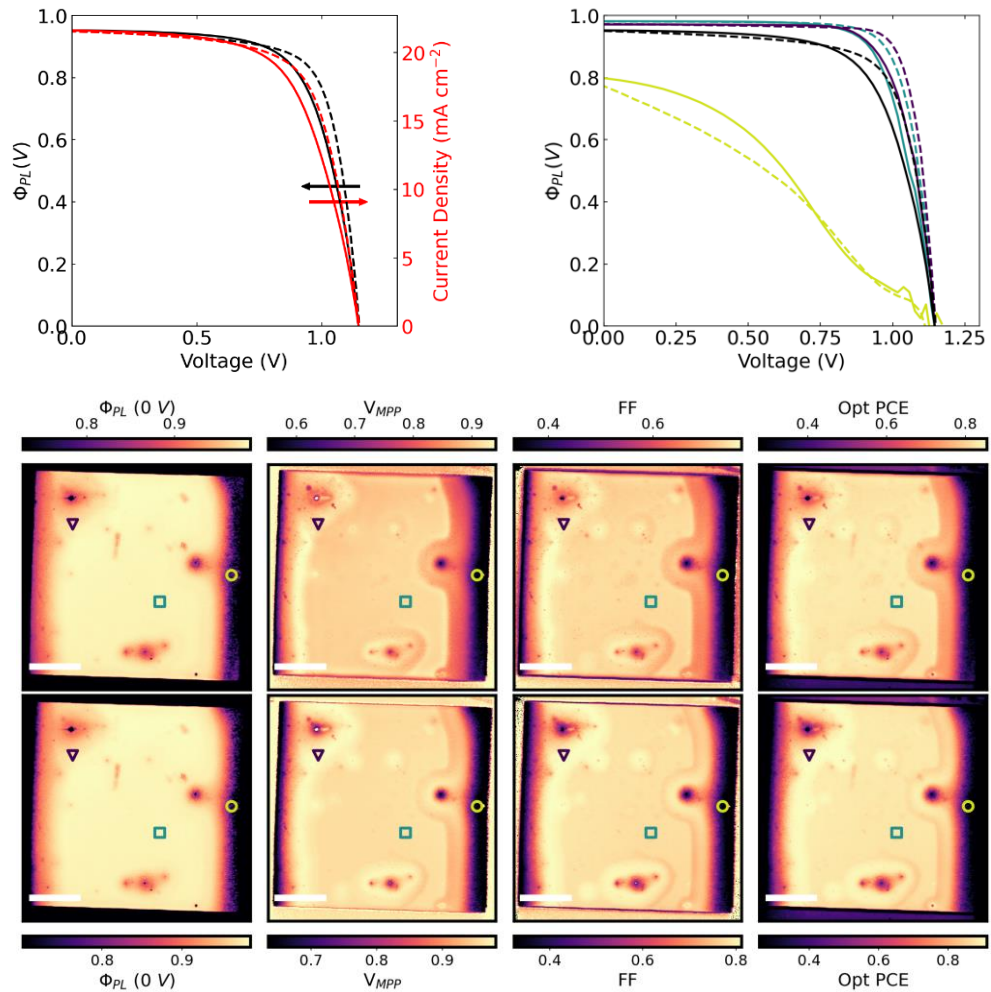

**Supplementary Figure 45** : Large scale optical JV measurements of pristine 2PACz/DCDH perovskite solar cell. a) mean optical (black) and electrical (red) JV measurements of the device. b) Optical JV curves extracted from the areas marked in panels c-j. Maps of the optical JV figures of merit c) optical extraction efficiency, d) max power point voltage, e) fill factor, f) optical PCE extracted from the reverse scan. g) optical extraction efficiency, h) max power point voltage, i) fill factor, j optical PCE extracted from the forward scan. Scalebars are 1 mm.

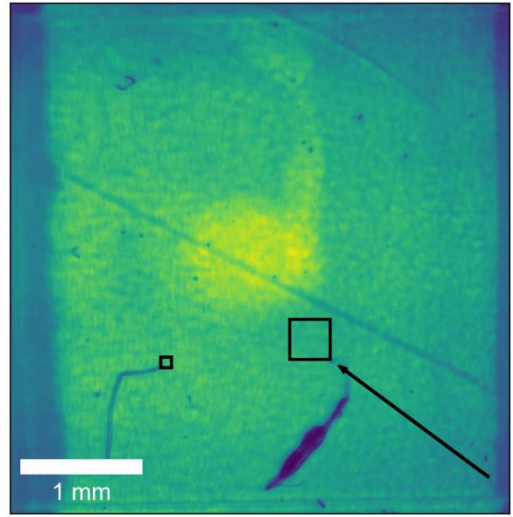

**Supplementary Figure 46:** Reflection image of the DCDH device shown in Figure 2 with the large hysteresis. The large black square shows the area with pronounced hysteresis shown in the main text, while the small area shows the area with lesser hysteresis further away from the degradation front depicted by the arrow.

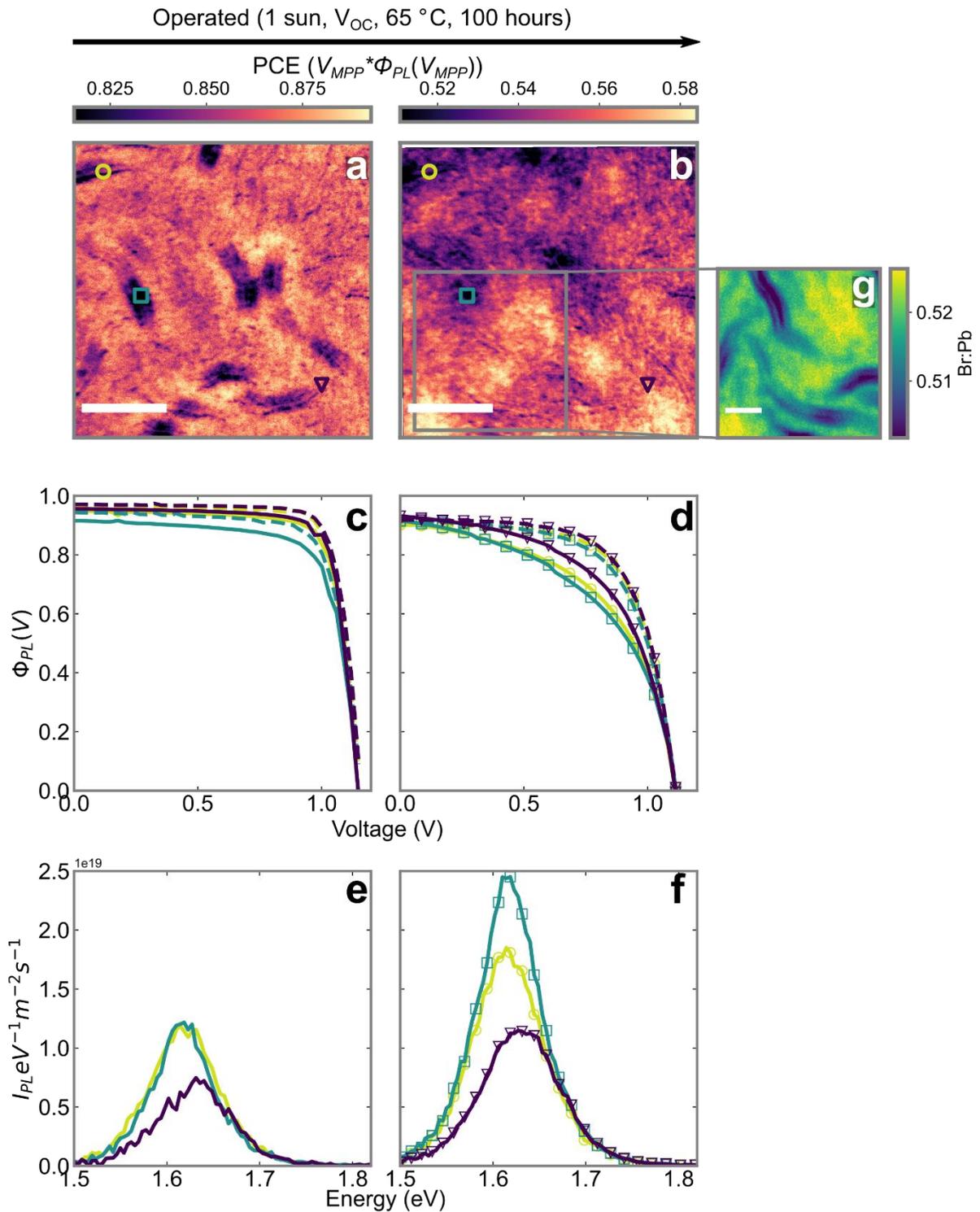

**Supplementary Figure 47. Optical and electronic spectra of DCDH before and after operation away from an edge.** Optical PCE maps of the same area of a fresh a) and operated b) DCDH solar cell after 100 hours at $V_{OC}$, 65 °C and 1 sun illumination. c) and d) Optical JV curves before and after ageing from the points marked in panels a and b. Solid lines are reverse scans, dashed lines are forward scans. e) and f) PL spectra from the same marked areas before and after ageing respectively. g) Br:Pb map of region marked in panel b consistent with relatively unchanged perovskite chemical character post accelerated ageing. Scalebars in a and b are 25 µm, scalebar in g is 10 µm.

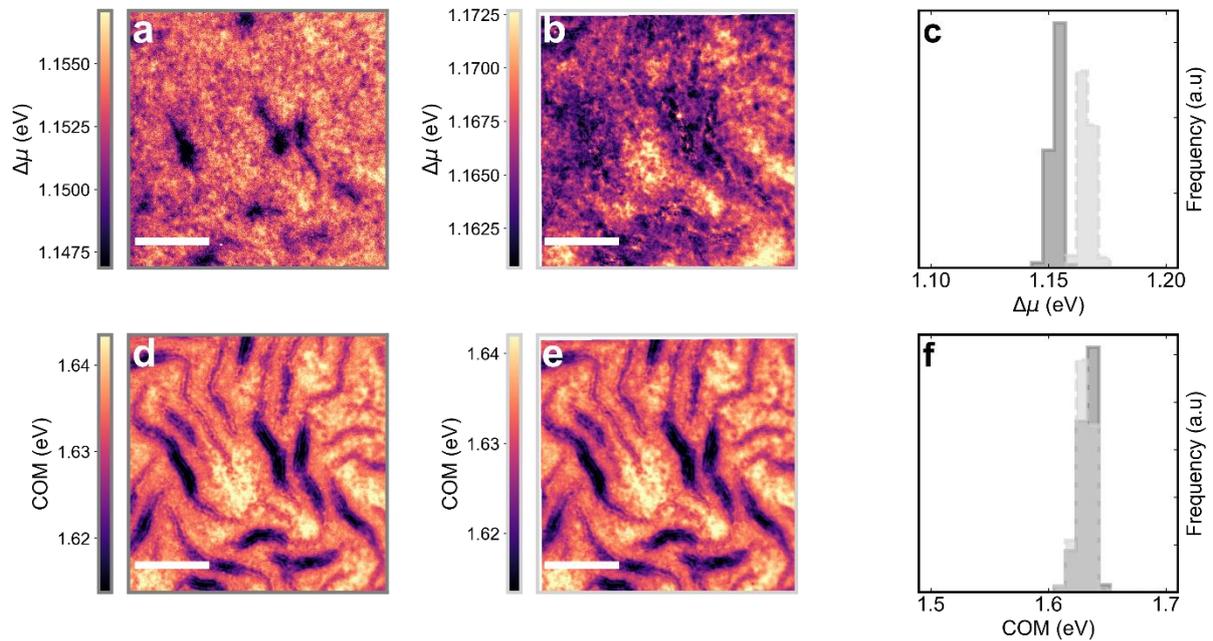

**Supplementary Figure 48. Δµ and COM of DCDH solar cell before after accelerated ageing away from an edge.** Δµ a) before and b) after accelerated ageing, distributions summarised in panel c). COM d) before and e) after accelerated ageing, distributions summarised in panel f). The scan area is the same as for the voltage dependent measurement shown in Supplementary Figure 17. Scalebars are 25 µm.

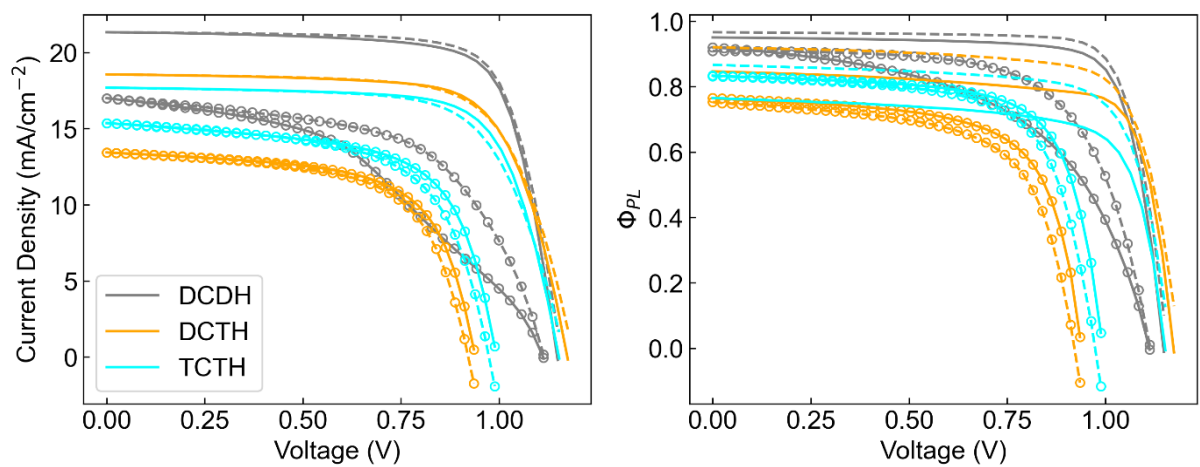

**Supplementary Figure 49. Averaged electrical and optoelectronic properties of perovskite compositional series before and after operation.** a) Shows the average electronic JV curves for the perovskite compositional series before (no markers) and after (markers)

operation. b) Shows the spatially averaged optical JV curves for the perovskite compositional series before (no markers) and after (markers) operation. c) Shows the average open circuit PL spectra for the perovskite compositional series before (no markers) and after (markers) operation.

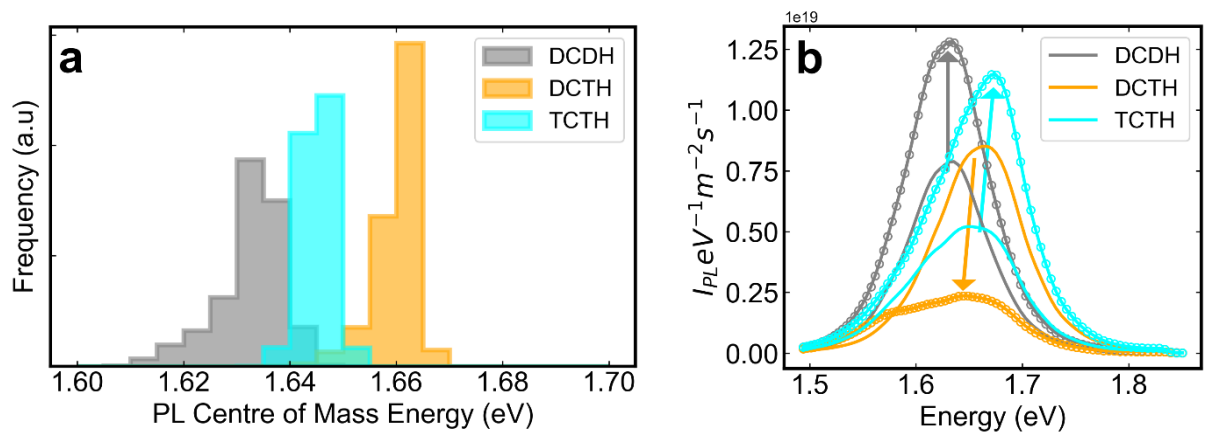

**Supplementary Figure 50. PL spectral characteristics of perovskite compositional series.** a) Histograms of PL COM energies extracted at each spatial point for DCDH (grey), DCTH (orange) and TCTH (cyan) perovskite solar cells. b) Spatially averaged PL spectra of devices before (solid line no markers) and after accelerated ageing (solid line with markers).

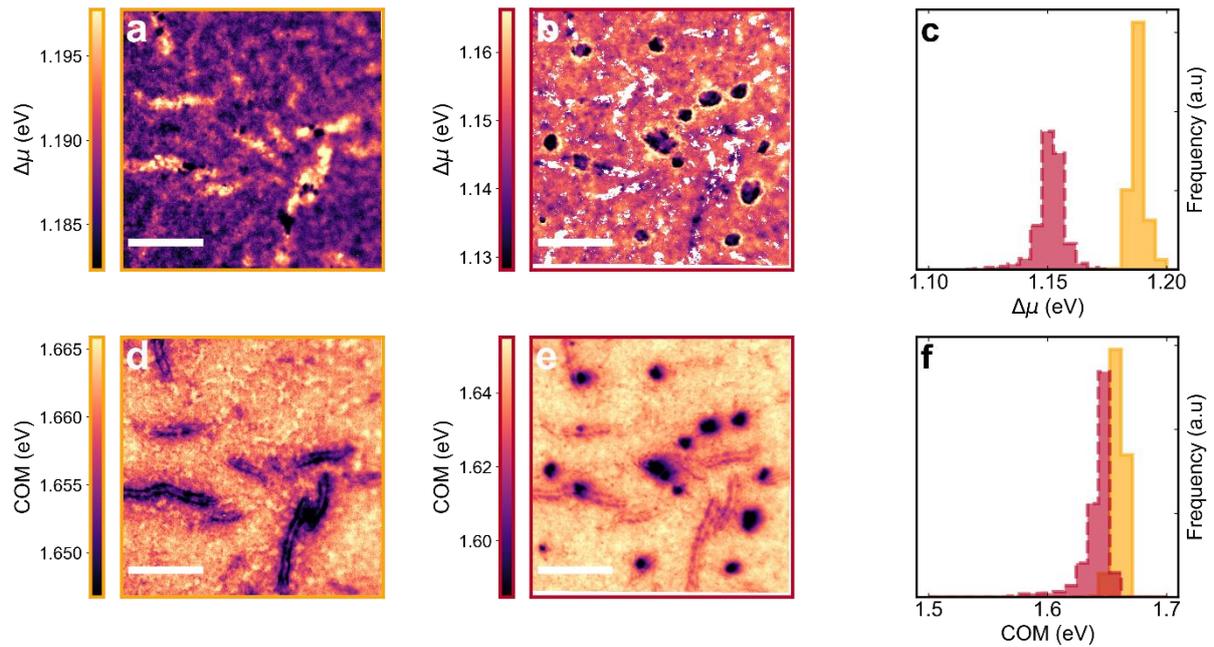

**Supplementary Figure 51. Δμ and COM of DCTH solar cell before after accelerated ageing away from an edge.** Δμ a) before and b) after accelerated ageing, distributions summarised in panel c). COM d) before and e) after accelerated ageing, distributions summarised in panel f). The scan area is the same as for the voltage dependent measurement shown in Figure 3. Scalebars are 25 µm.

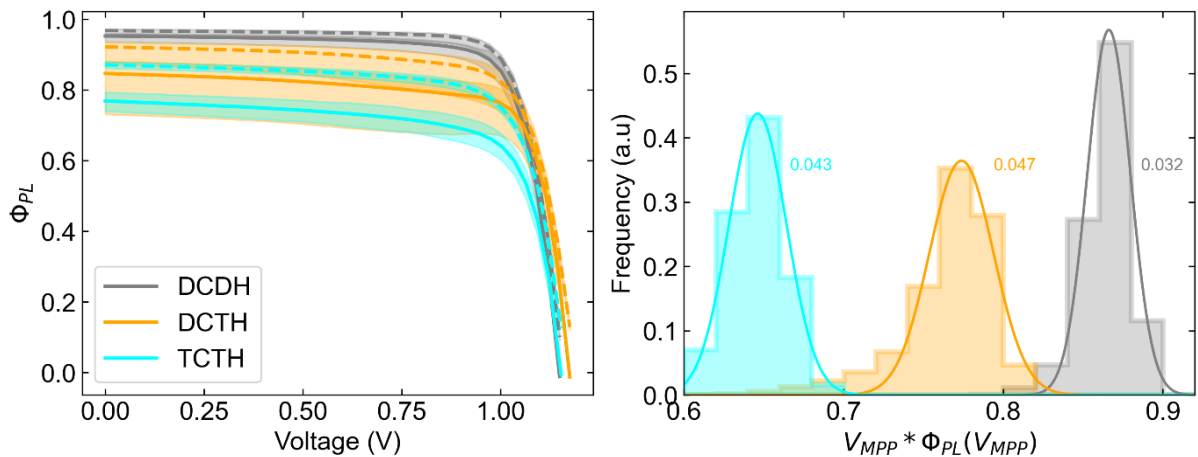

**Supplementary Figure 52: Spatial PCE heterogeneity as a function of perovskite composition.** a) Reverse (solid lines) and forward (dotted lines) spatially averaged optical JV curves. Shaded areas indicate the 5th and 95th percentile optical PCE area. b) Histograms of optical PCE distributions for a range of perovskite compositions.

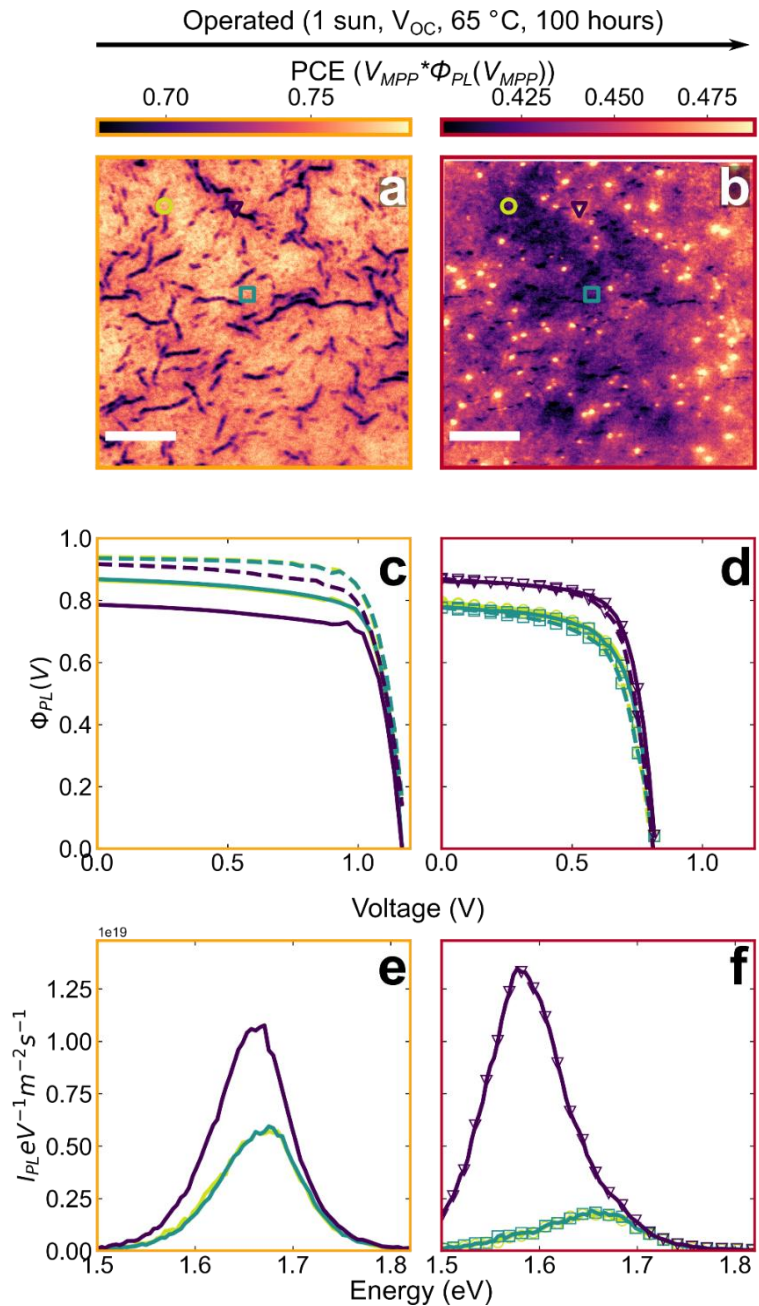

**Supplementary Figure 53: Optical and electronic spectra of DCTH solar cell before and after operation.** Optical PCE maps of the same area of a fresh a) and operated b) DCTH solar cell after 100 hours at $V_{OC}$, 65 °C and 1 sun illumination. c) and d) Optical JV curves before and after ageing from the points marked in panels a and b. Solid lines are reverse scans, dashed lines are forward scans. e) and f) PL spectra from the same marked areas before and after ageing respectively. g) Br:Pb map of region marked in panel b consistent with relatively unchanged perovskite chemical character post accelerated ageing. Scalebars in a and b are 75 µm.

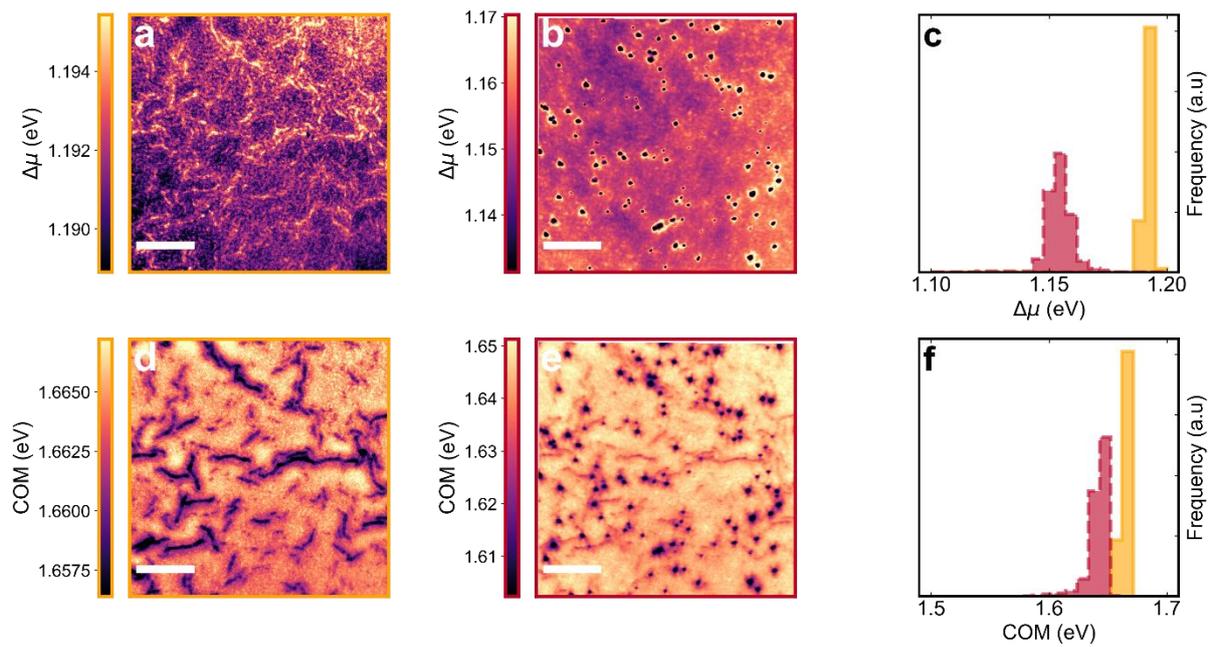

**Supplementary Figure 54. Δμ and COM of DCTH solar cell before after accelerated ageing.** Δμ a) before and b) after accelerated ageing, distributions summarised in panel c). COM d) before and e) after accelerated ageing, distributions summarised in panel f). The scan area is the same as for the voltage dependent measurement shown in Supplementary Figure 21. Scalebars are 75 µm.

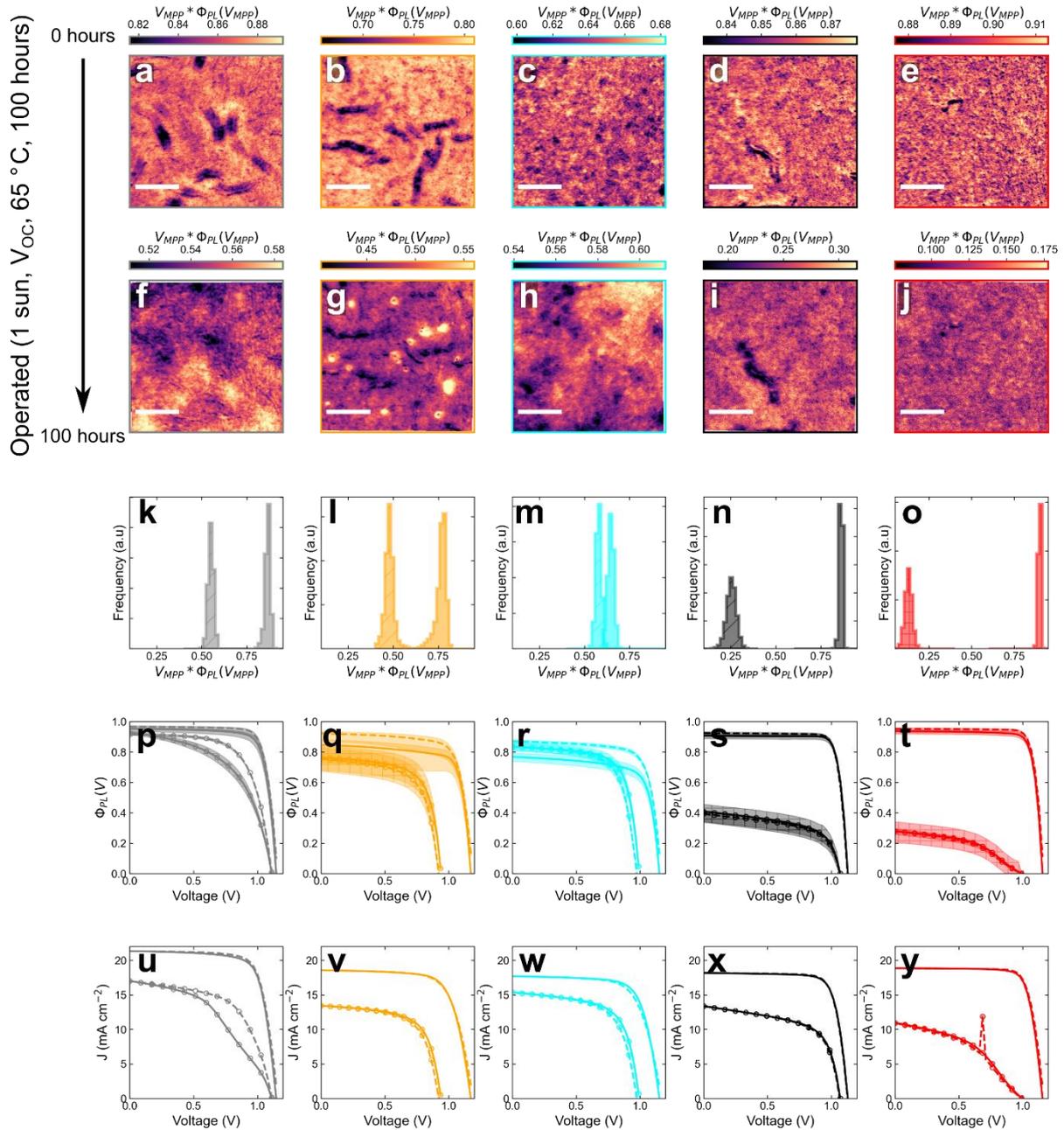

**Supplementary Figure 55. Summary of operational degradation observation across the device phase space.** Optical PCE maps of a) 2PACz/DCDH, b) 2PACz/DCTH, c) 2PACz/TCTH, d) MeO-2PACz/TCTH and e) Me-4PACz/TCTH before accelerated stress test. Optical PCE maps of f) 2PACz/DCDH, g) 2PACz/DCTH, h) 2PACz/TCTH, i) MeO-2PACz/TCTH and j) Me-4PACz/TCTH after accelerated stress test. Optical PCE histograms of k) 2PACz/DCDH, l) 2PACz/DCTH, m) 2PACz/TCTH, n) MeO-2PACz/TCTH and o) Me-4PACz/TCTH before and after (cross-hatched) the accelerated stress test. Optical JV curves of p) 2PACz/DCDH, q) 2PACz/DCTH, r) 2PACz/TCTH, s) MeO-2PACz/TCTH and t) Me-4PACz/TCTH before (no markers) and after (markers) the accelerated stress test. The intervals indicate the spatial variation in the optical JV curves across the regions measured. Electrical JV curves of u) 2PACz/DCDH, v) 2PACz/DCTH, w) 2PACz/TCTH, x) MeO-2PACz/TCTH and y) Me-4PACz/TCTH before (no markers) and after (markers) the accelerated stress test.

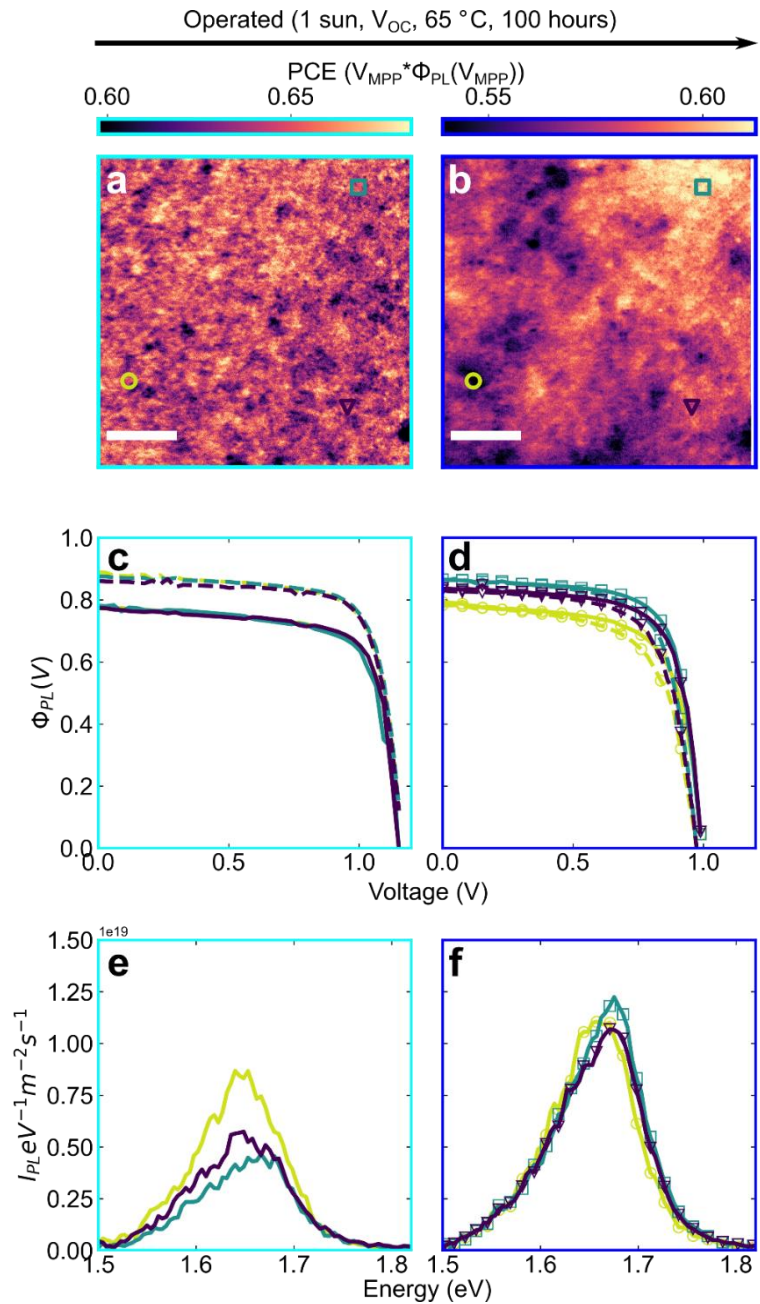

**Supplementary Figure 56. Optical and electronic spectra of TCTH solar cell at high magnification before and after operation away from an edge.** Optical PCE maps of the same area of a fresh a) and operated b) TCTH solar cell after 100 hours at $V_{OC}$, 65 °C and 1

sun illumination. c) and d) Optical JV curves before and after ageing from the points marked in panels a and b. Solid lines are reverse scans, dashed lines are forward scans. e) and f) PL spectra from the same marked areas before and after ageing respectively. Scalebars are 25 µm.

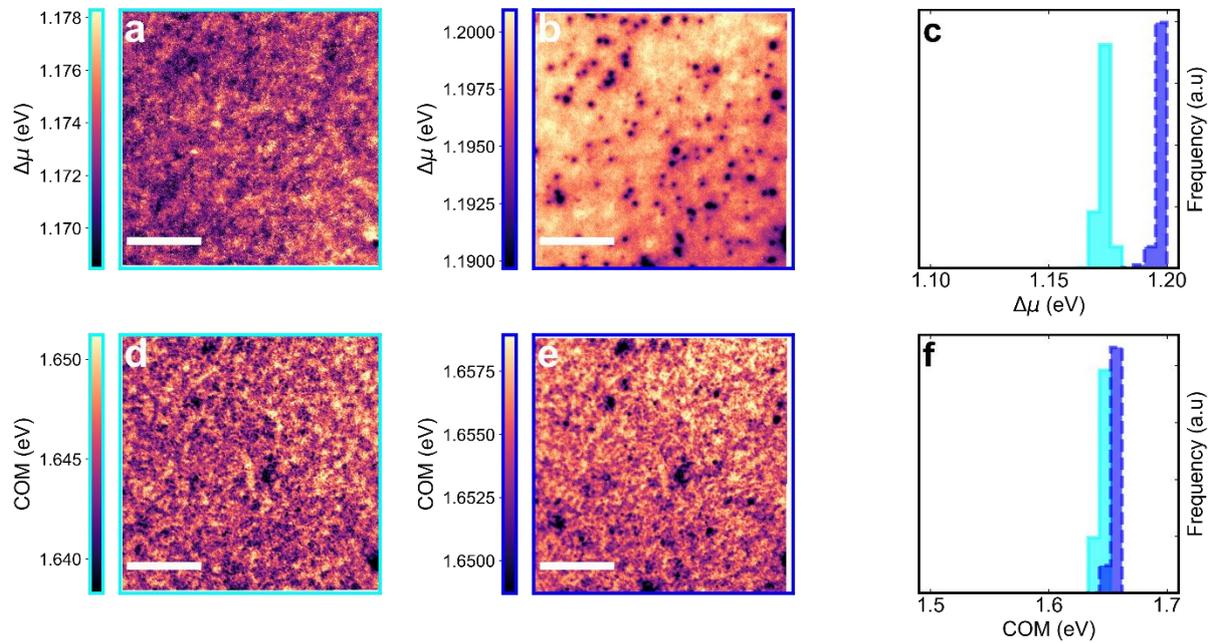

**Supplementary Figure 57. Δµ and COM of TCTH solar cell before after accelerated ageing.** Δµ a) before and b) after accelerated ageing, distributions summarised in panel c). COM d) before and e) after accelerated ageing, distributions summarised in panel f). The scan area is the same as for the voltage dependent measurement shown in Supplementary Figure 26. Scalebars are 75 µm.

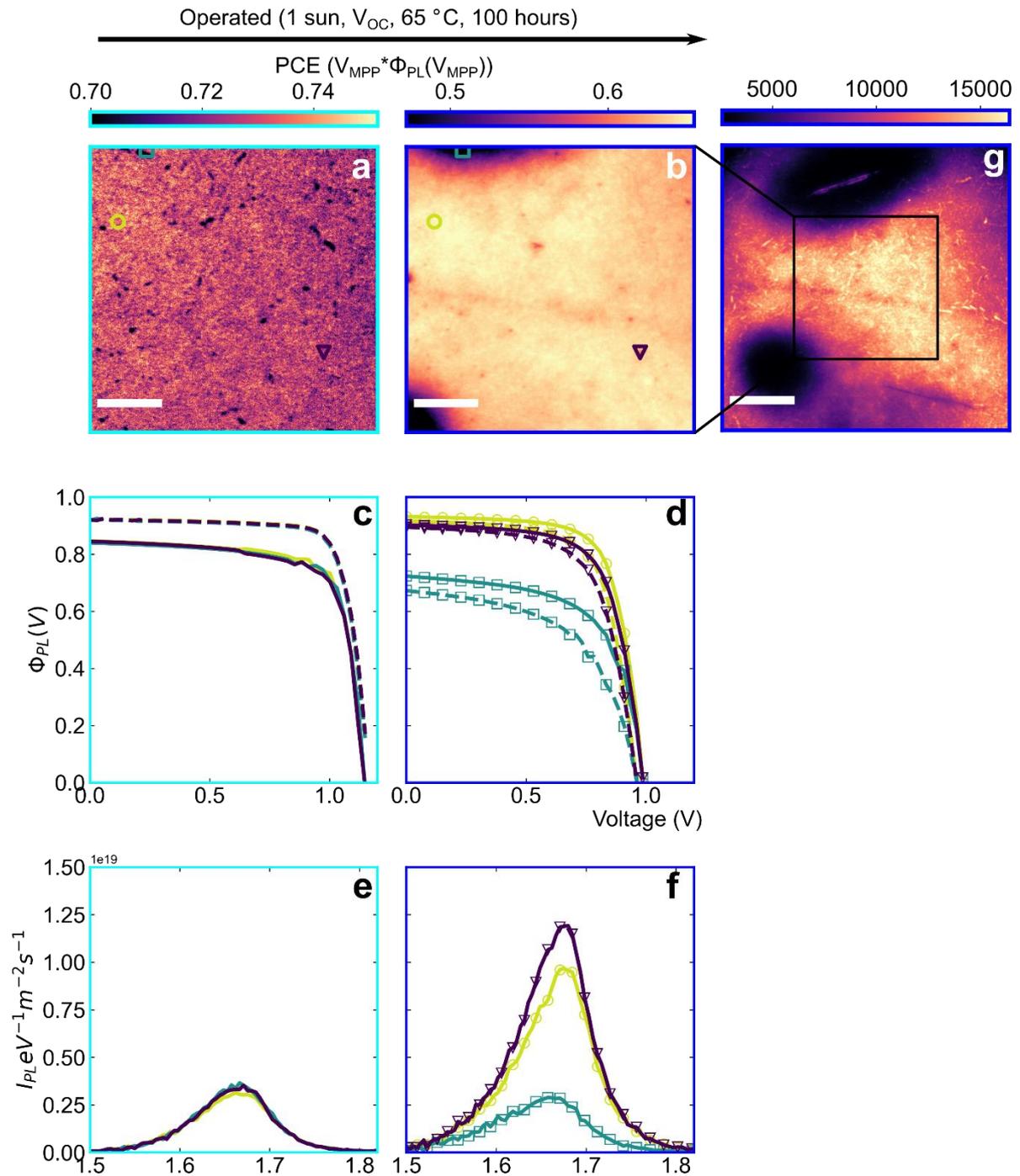

**Supplementary Figure 58. Optical and electronic spectra of TCTH solar cell at low magnification before and after operation away from an edge.** Optical PCE maps of the same area of a fresh a) and operated b) TCTH solar cell after 100 hours at $V_{OC}$, 65 °C and 1 sun illumination. c) and d) Optical JV curves before and after ageing from the points marked in panels a and b. Solid lines are reverse scans, dashed lines are forward scans. e) and f) PL spectra from the same marked areas before and after ageing respectively. g) Large area broadband PL map at open circuit containing the region shown in panels a and b highlighting

the large scale degradation features forming. Scalebars in a and b are 75 µm, scalebar in g is 150 µm.

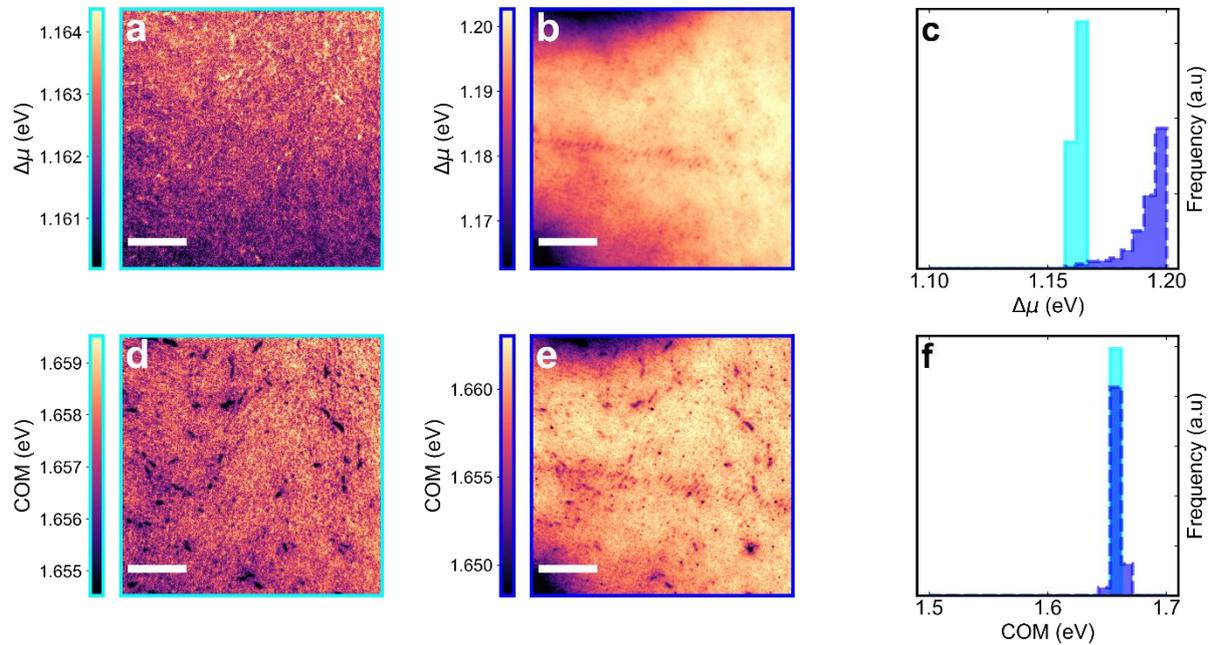

**Supplementary Figure 59. Δμ and COM of TCTH solar cell before after accelerated ageing.** Δμ a) before and b) after accelerated ageing, distributions summarised in panel c). COM d) before and e) after accelerated ageing, distributions summarised in panel f). The scan area is the same as for the voltage dependent measurement shown in Supplementary Figure 28. Scalebars are 75 µm.

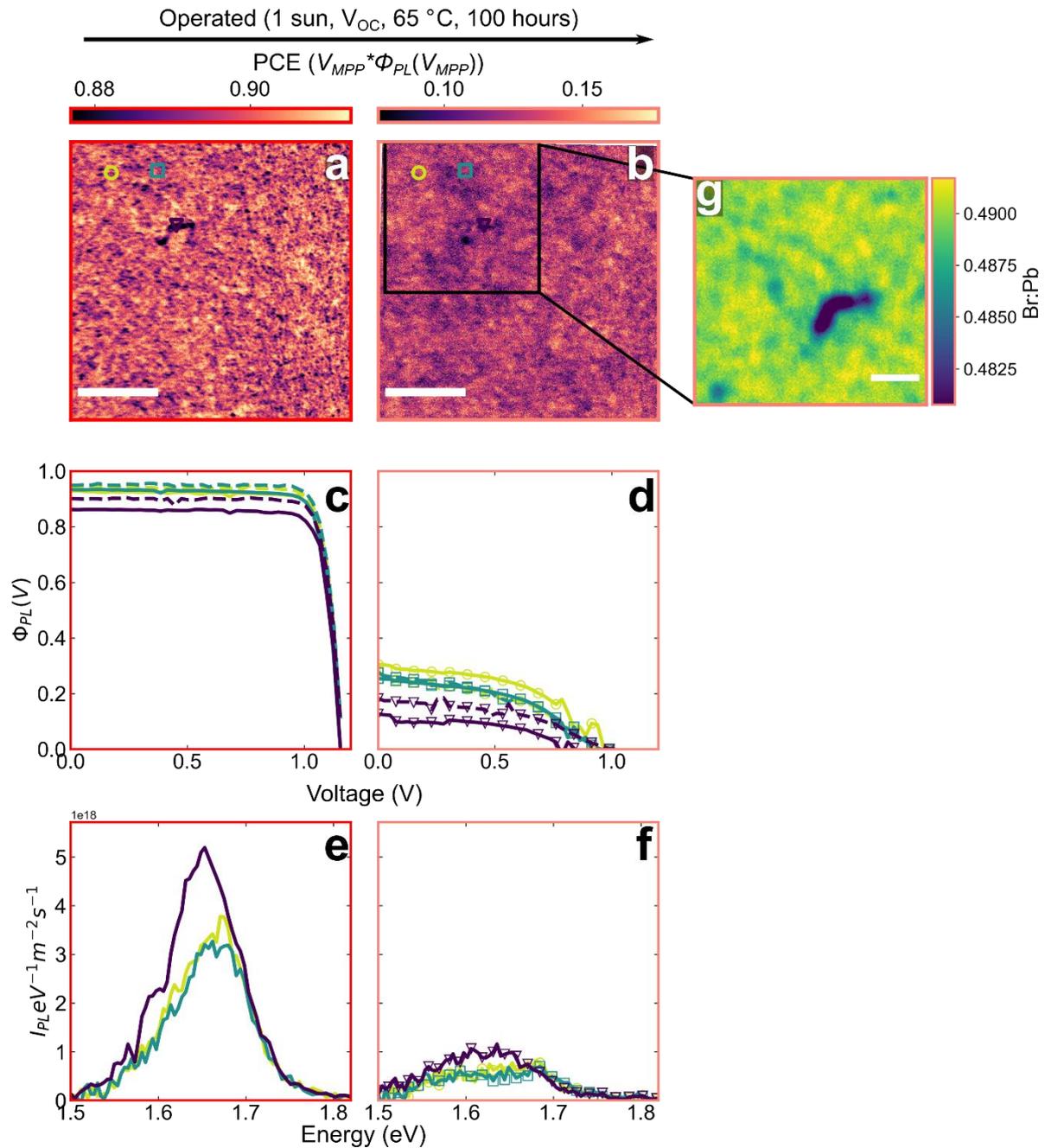

**Supplementary Figure 60. Optical and electronic spectra of TCTH solar cell on Me-4PACz before and after operation away from an edge.** Optical PCE maps of the same area of a fresh a) and operated b) TCTH solar cell after 100 hours at $V_{OC}$, 65 °C and 1 sun illumination. c) and d) Optical JV curves before and after ageing from the points marked in panels a and b. Solid lines are reverse scans, dashed lines are forward scans. e) and f) PL spectra from the same marked areas before and after ageing respectively. g) Br:Pb map of region marked in panel b. Scalebars in a and b are 25 µm, scalebar in g is 10 µm.

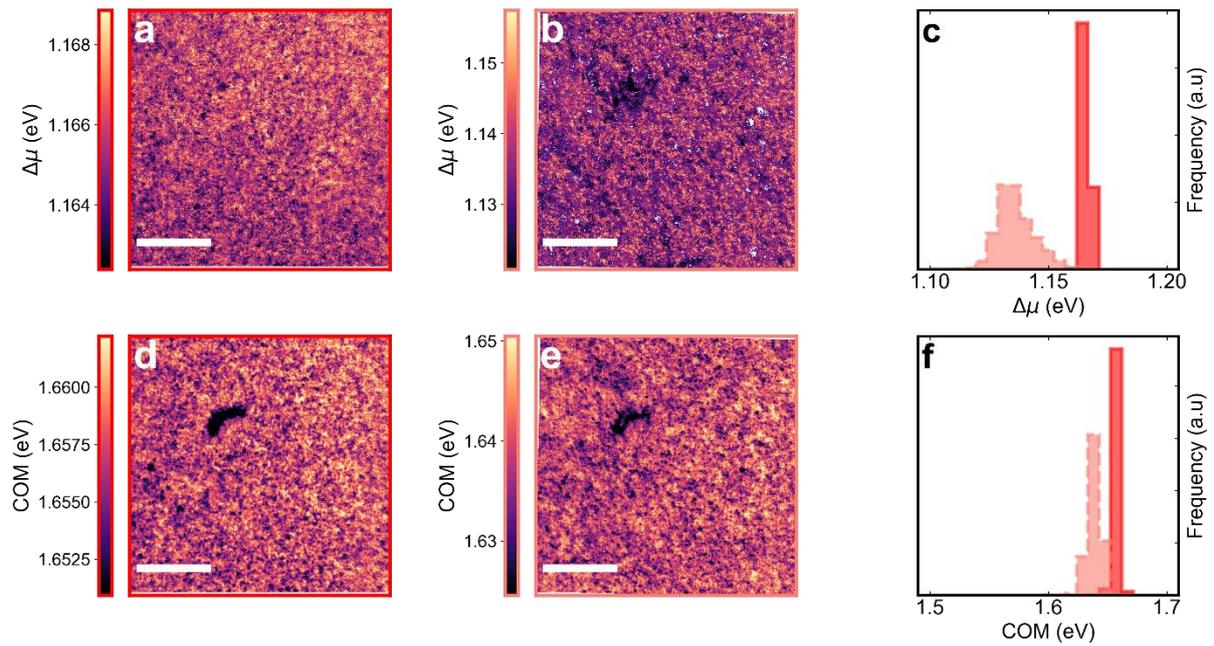

**Supplementary Figure 61. Δμ and COM of TCTH on Me-4PACz solar cell before after accelerated ageing.** Δμ a) before and b) after accelerated ageing, distributions summarised in panel c). COM d) before and e) after accelerated ageing, distributions summarised in panel f). The scan area is the same as for the voltage dependent measurement shown in Supplementary Figure 34. Scalebars are 25 µm.

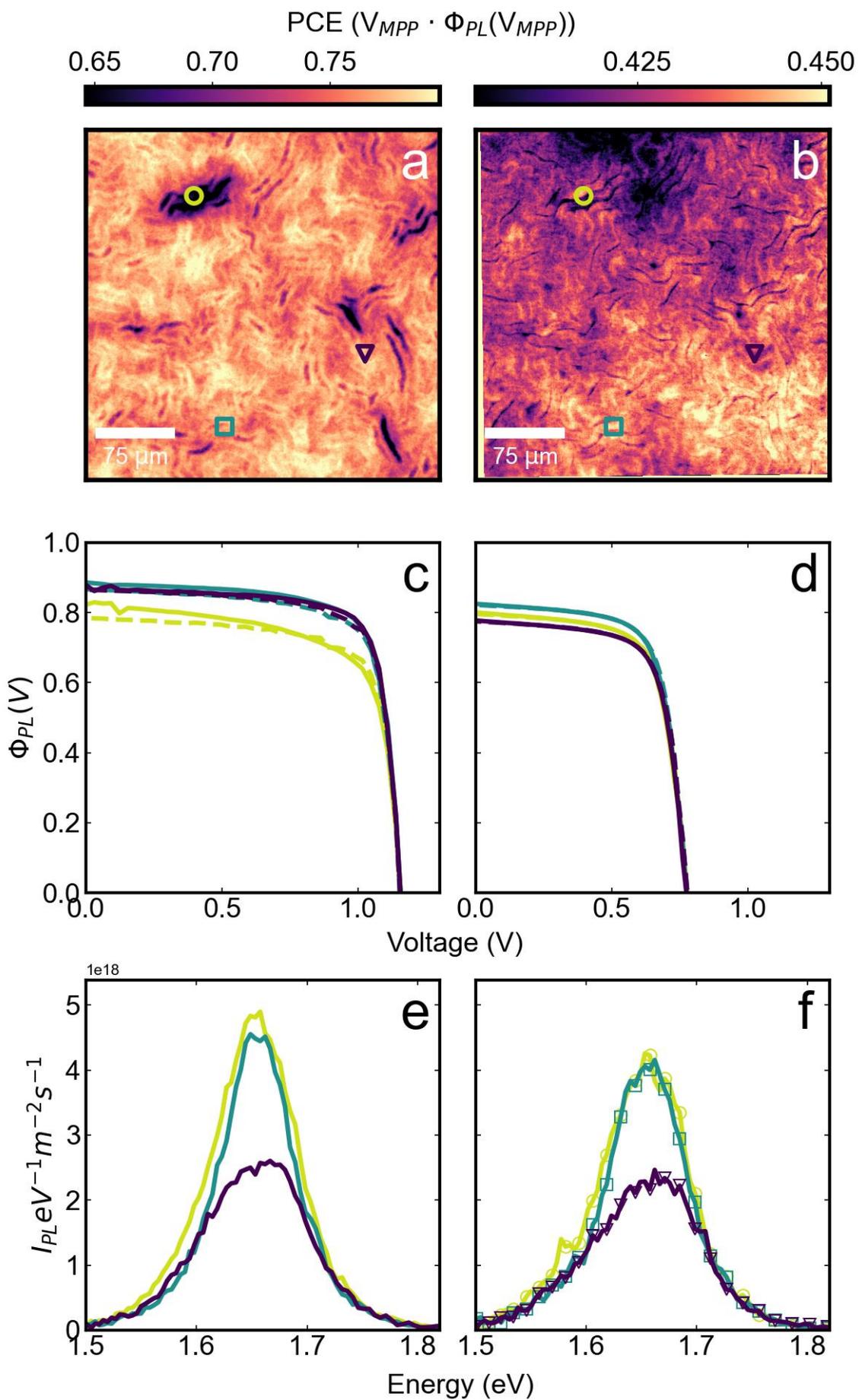

**Supplementary Figure 62. Optical and electronic spectra of wrinkled 2PACz/TCTH solar cell.** Optical PCE maps of the same area of a fresh a) and operated b) PI passivated TCTH solar cell after 100 hours at $V_{OC}$, 65 °C and 1 sun illumination. c) and d) Optical JV curves before and after ageing from the points marked in panels a and b. Solid lines are reverse scans, dashed lines are forward scans. e) and f) PL spectra from the same marked areas before and after ageing respectively.

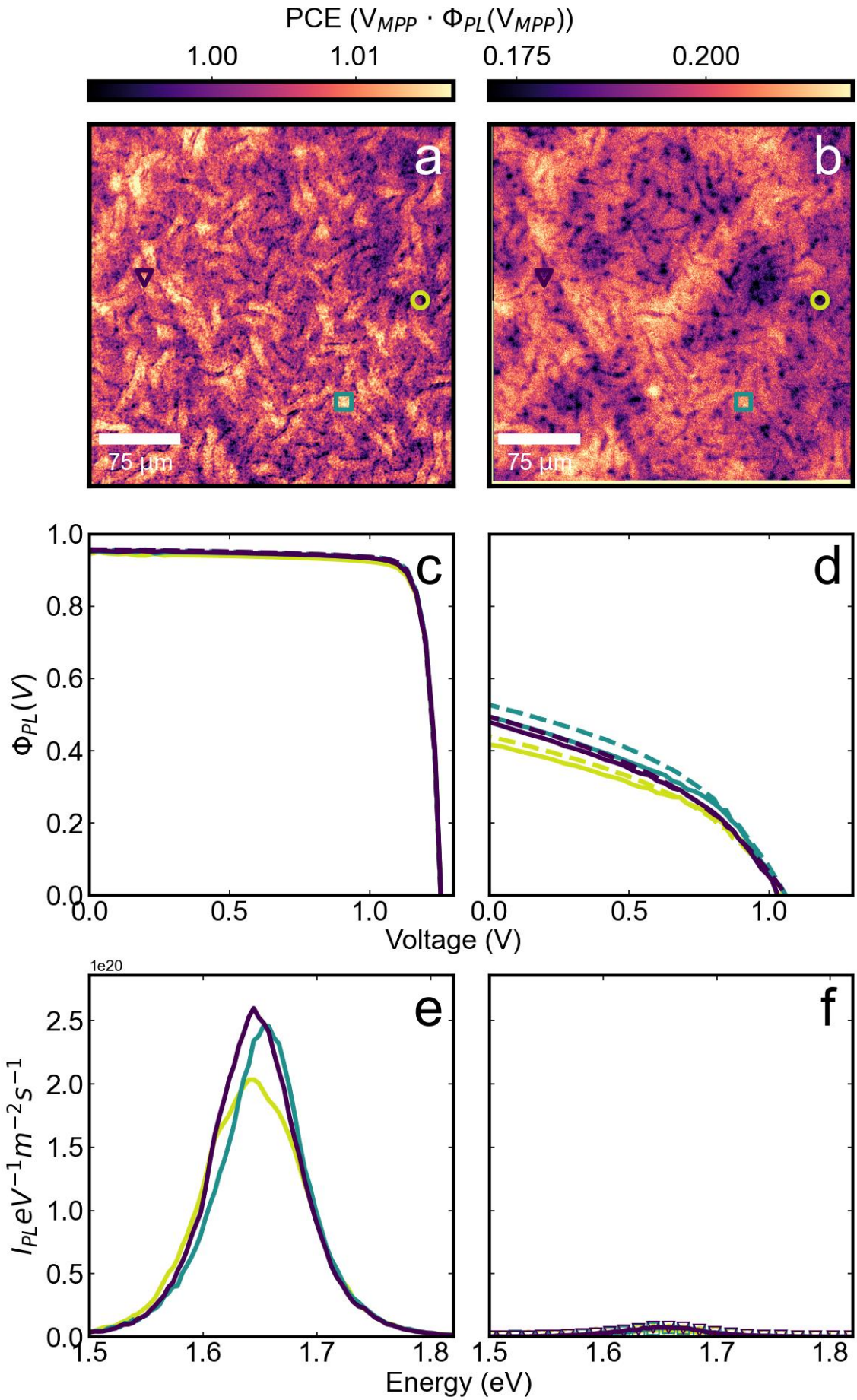

**Supplementary Figure 63. Optical and electronic spectra of TCTH solar cell 2PACz passivated with PI before and after.** Optical PCE maps of the same area of a fresh a) and operated b) PI passivated TCTH solar cell after 100 hours at $V_{OC}$, 65 °C and 1 sun illumination. c) and d) Optical JV curves before and after ageing from the points marked in panels a and b. Solid lines are reverse scans, dashed lines are forward scans. e) and f) PL spectra from the same marked areas before and after ageing respectively.

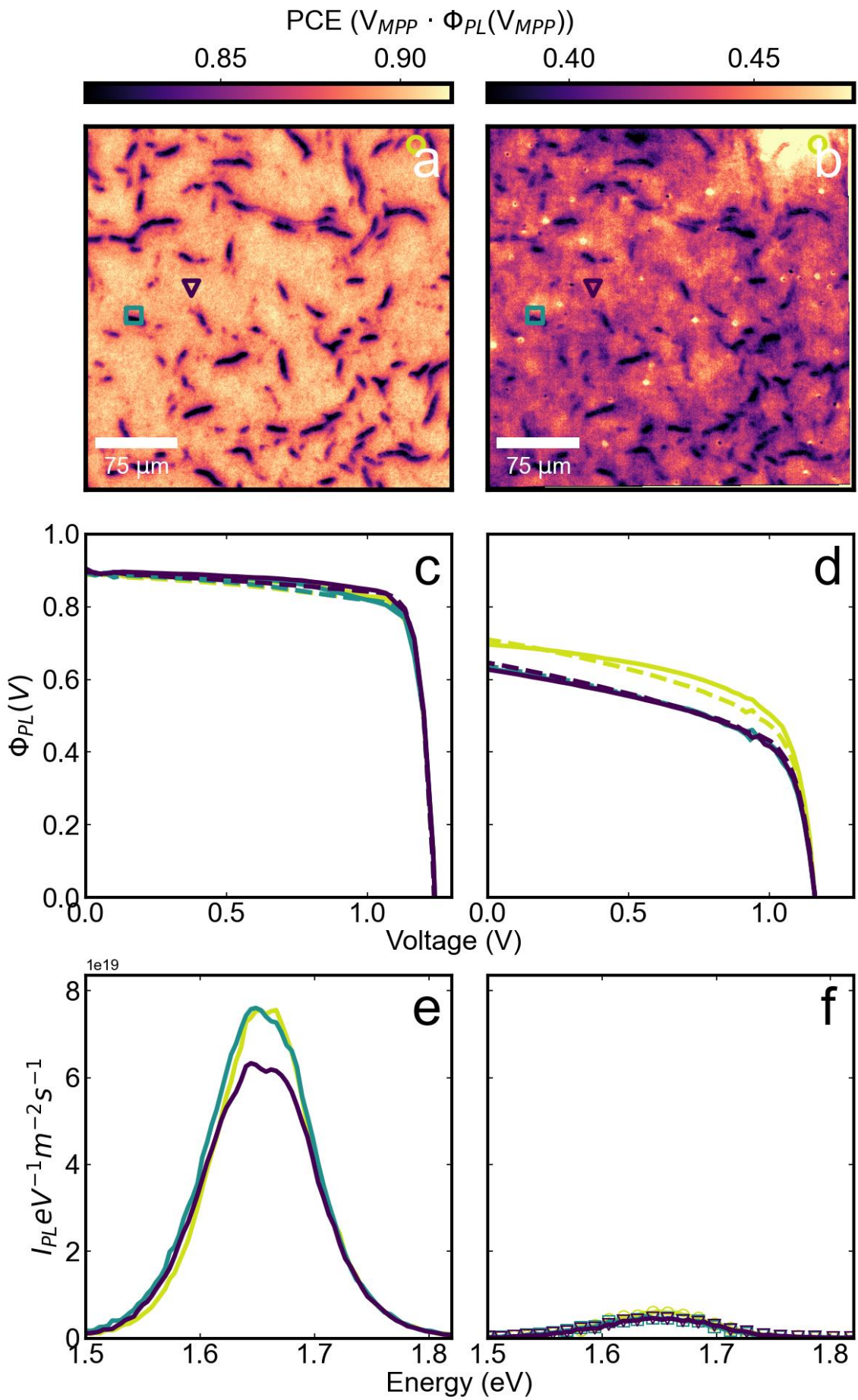

**Supplementary Figure 64. Optical and electronic spectra of TCTH solar cell 2PACz passivated with LiF before and after.** Optical PCE maps of the same area of a fresh a) and operated b) PI passivated TCTH solar cell after 100 hours at $V_{OC}$, 65 °C and 1 sun illumination. c) and d) Optical JV curves before and after ageing from the points marked in panels a and b. Solid lines are reverse scans, dashed lines are forward scans. e) and f) PL spectra from the same marked areas before and after ageing respectively.

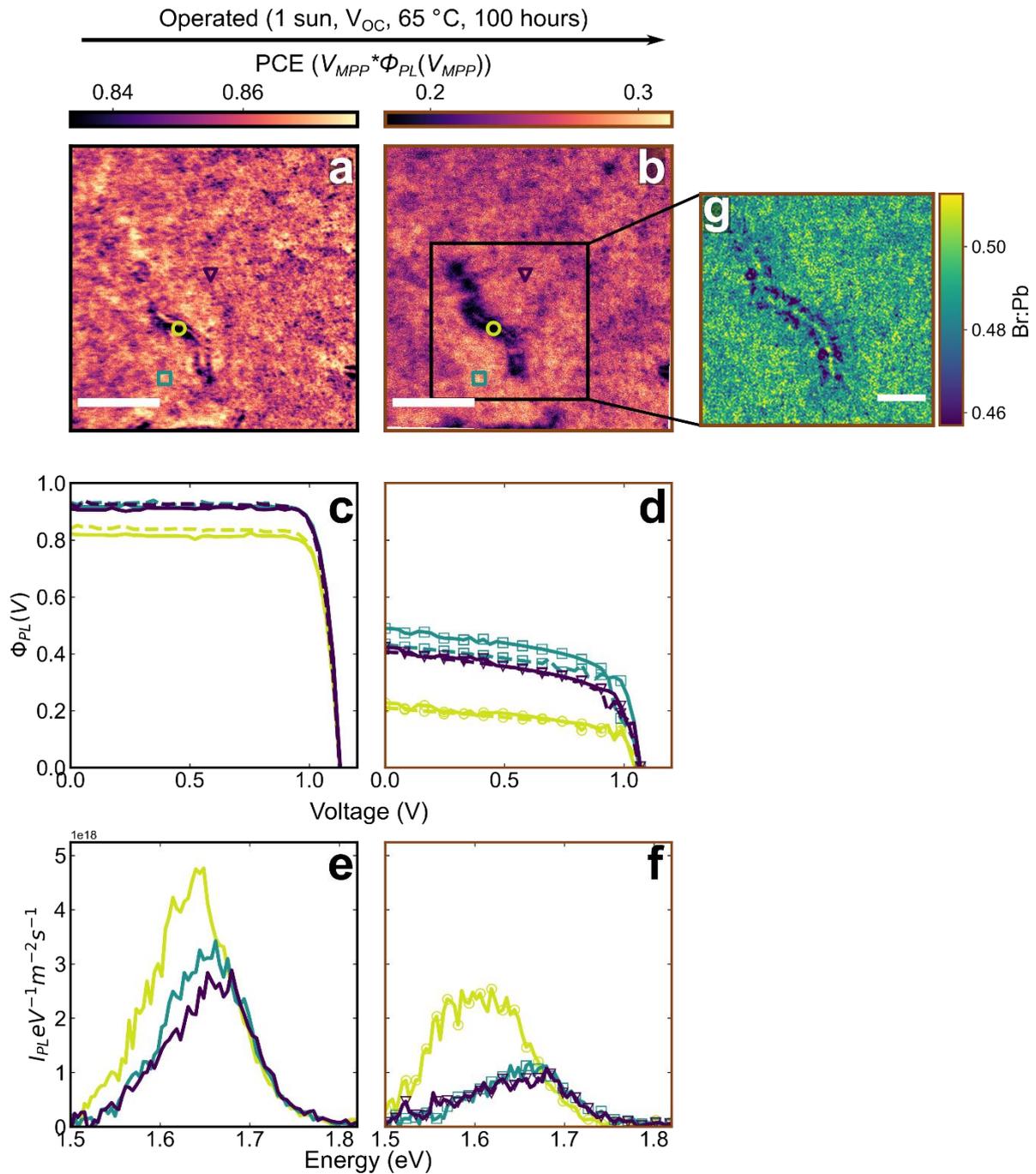

**Supplementary Figure 65. Optical and electronic spectra of TCTH solar cell on MeO-2PACz at high magnification before and after operation away from an edge.** Optical PCE maps of the same area of a fresh a) and operated b) TCTH solar cell after 100 hours at $V_{OC}$, 65 °C and 1 sun illumination. c) and d) Optical JV curves before and after ageing from the points marked in panels a and b. Solid lines are reverse scans, dashed lines are forward scans. e) and f) PL spectra from the same marked areas before and after ageing respectively. g) Br:Pb map of region marked in panel b. Scalebars in a and b are 25 µm, scalebar in g is 10 µm.

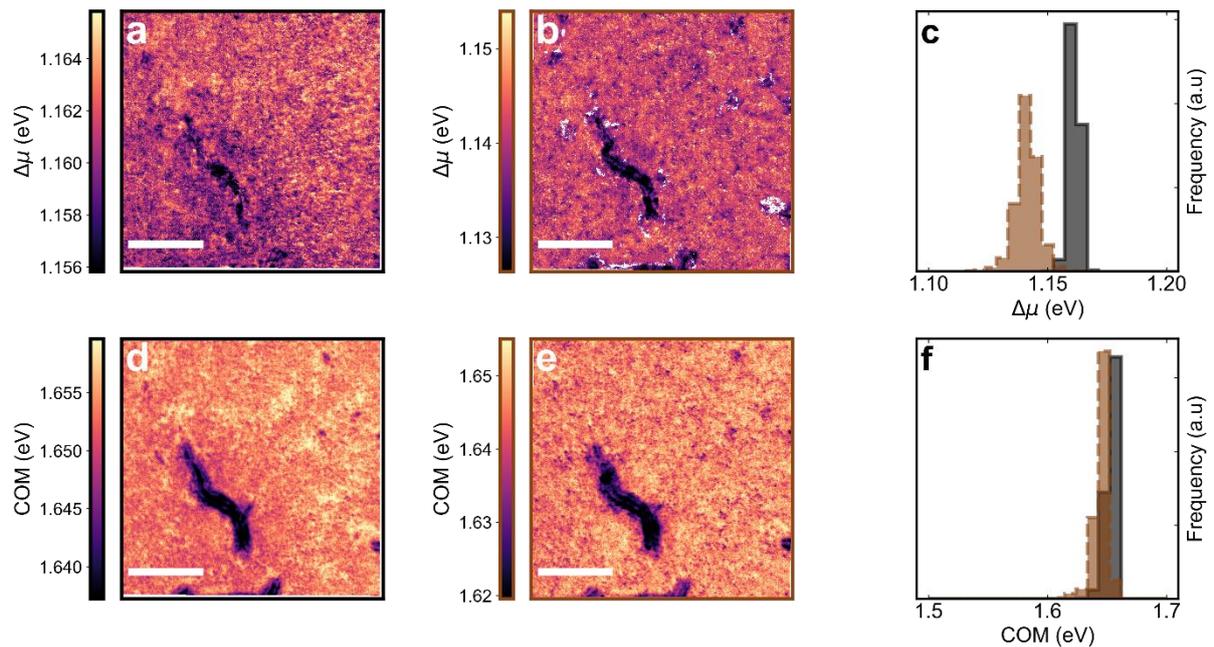

**Supplementary Figure 66. Δµ and COM of TCTH on MeO-2PACz solar cell before after accelerated ageing.** Δµ a) before and b) after accelerated ageing, distributions summarised in panel c). COM d) before and e) after accelerated ageing, distributions summarised in panel f). The scan area is the same as for the voltage dependent measurement shown in Supplementary Figure 32. Scalebars are 25 µm.

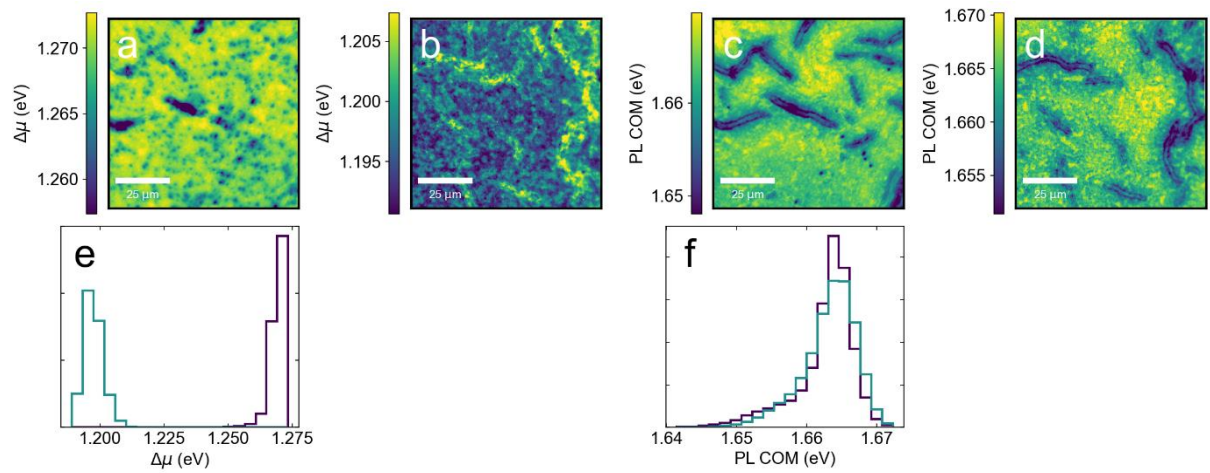

**Supplementary Figure 67: Comparison of optical properties of reference and PI passivated TCTH devices.** Δμ maps of the a) passivated and b) unpassivated 2PACz/DCTH devices. COM maps of the c) passivated and d) unpassivated devices. Histograms of the e) Δμ and f) COM distributions comparing passivated (blue) and unpassivated (purple).

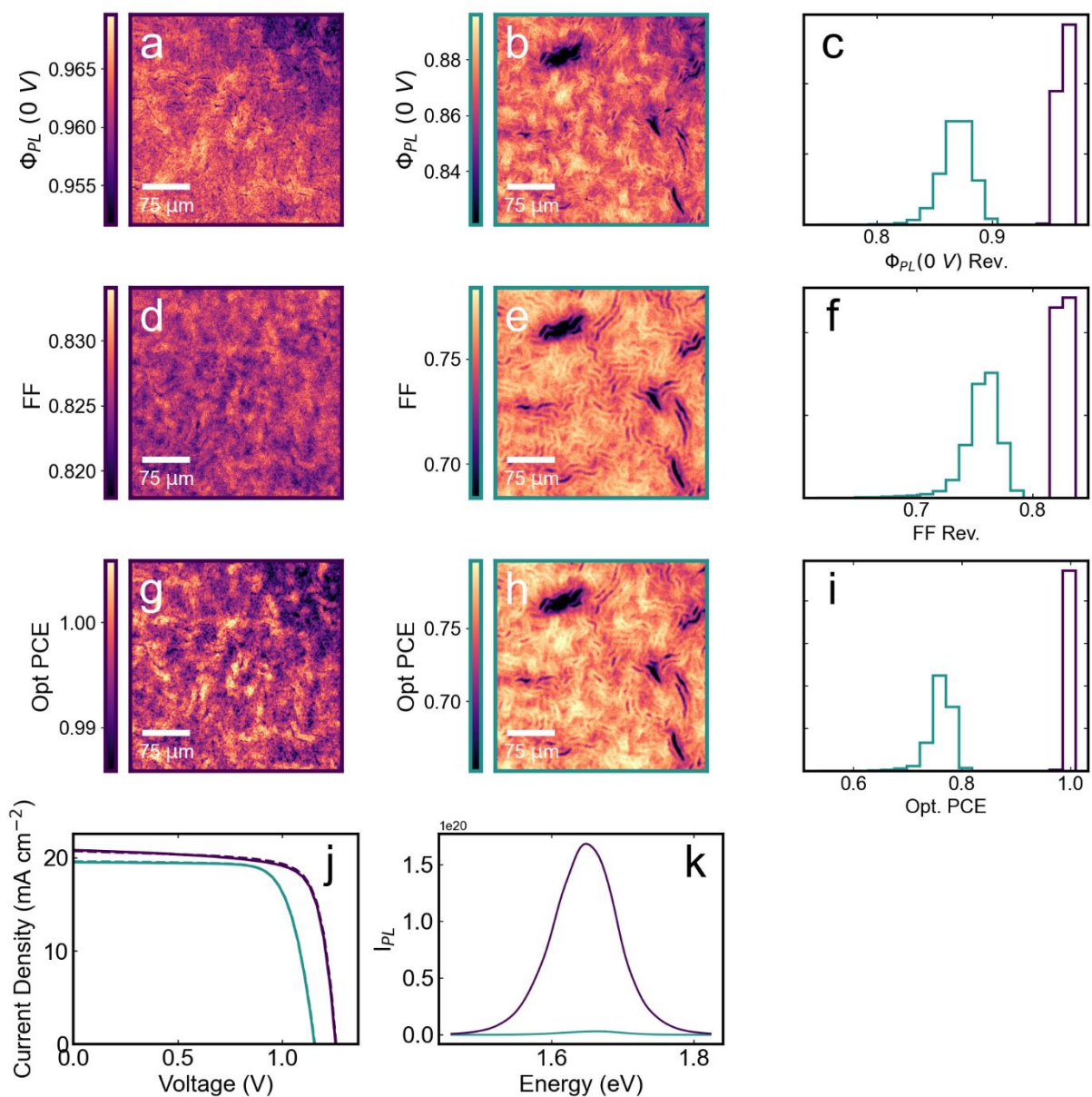

**Supplementary Figure 68**: **Microscopic comparison of reference and PI passivated TCTH devices.** Comparison of reference and PI passivated 2PACz/TCTH device. Maps of the optical extraction efficiency for a) PI passivated and b) reference devices. c) Histograms of the spatial distributions shown in a and b. Maps of the optical fill factor for d) PI passivated and e) reference devices. f) Histograms of the spatial distributions shown in d and e. Maps of the optical PCE for g) PI passivated and h) reference devices. i) Histograms of the spatial distributions shown in g and h. Comparisons of the j) electrical JV curves and k) spatially averaged PL spectra of the PI passivated (purple) and reference (blue) devices.

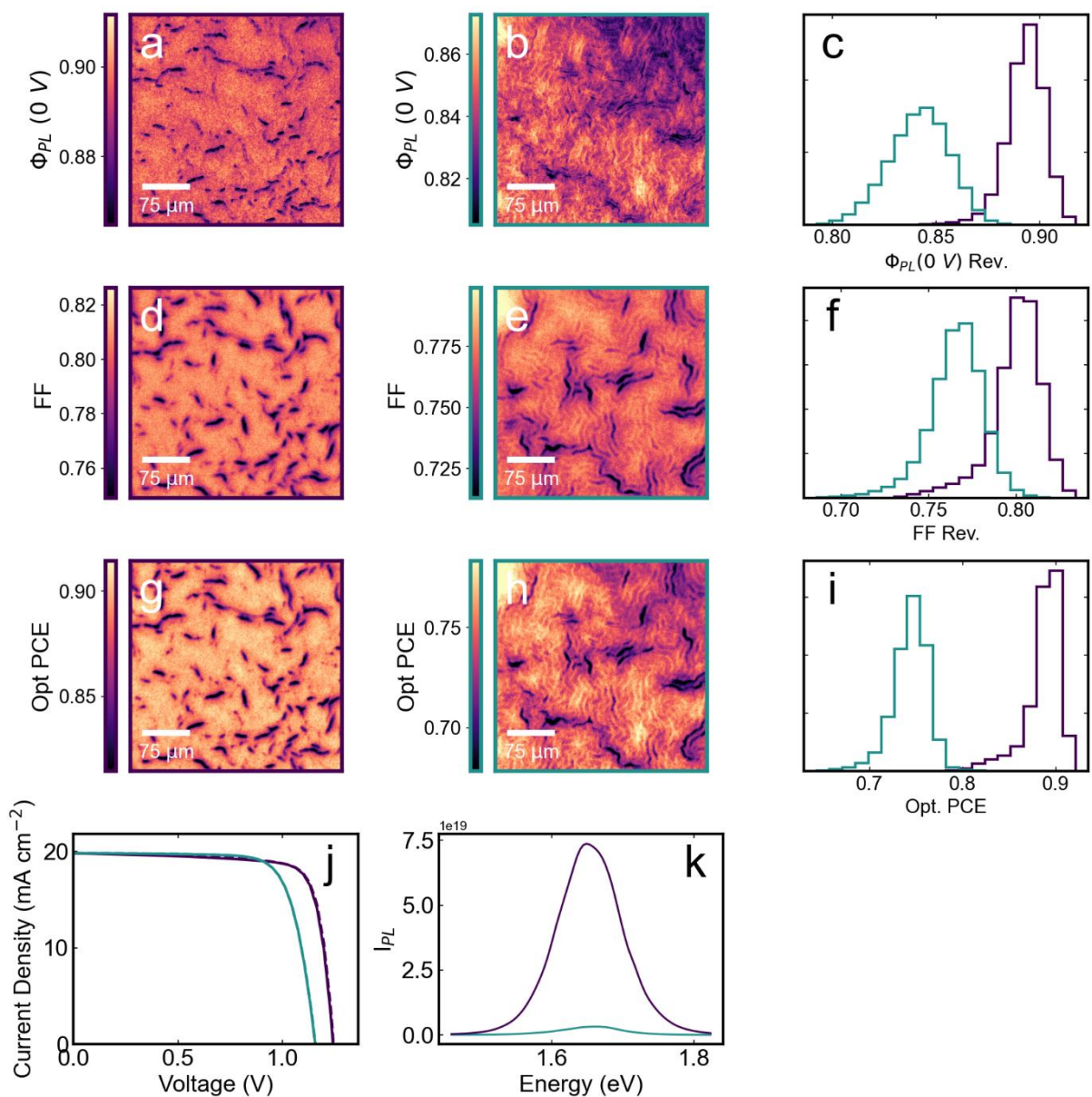

**Supplementary Figure 69**: **Microscopic comparison of reference and LiF passivated TCTH devices.** Comparison of reference and LiF passivated 2PACz/TCTH device. Maps of the optical extraction efficiency for a) LiF passivated and b) reference devices. c) Histograms of the spatial distributions shown in a and b. Maps of the optical fill factor for d) LiF passivated and e) reference devices. f) Histograms of the spatial distributions shown in d and e. Maps of the optical PCE for g) PI passivated and h) reference devices. i) Histograms of the spatial distributions shown in g and h. Comparisons of the j) electrical JV curves and k) spatially averaged PL spectra of the LiF passivated (purple) and reference (blue) devices.

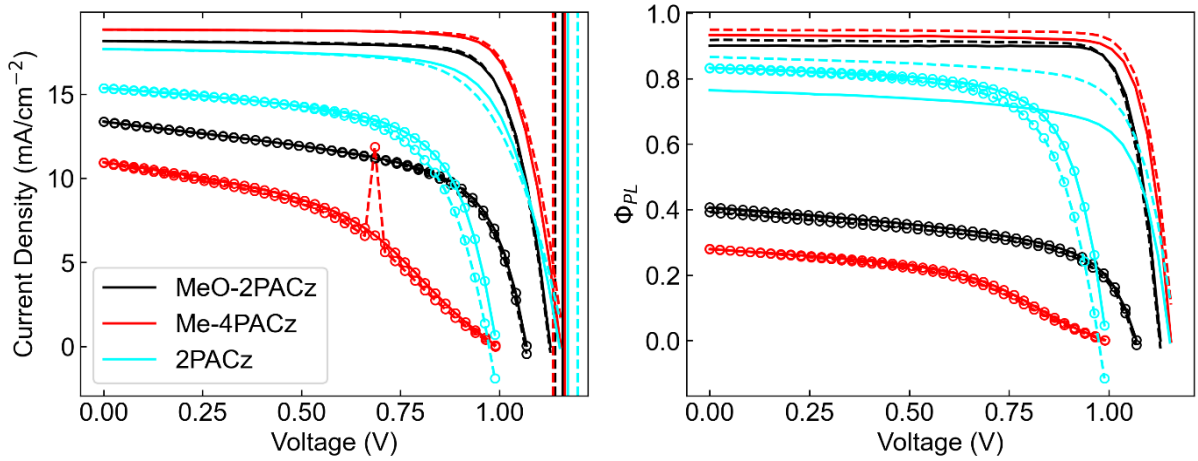

**Supplementary Figure 70. Averaged electrical and optoelectronic properties of the HTL series before and after operation.** a) Shows the average electronic JV curves for the HTL series before (no markers) and after (markers) operation. b) Shows the spatially averaged optical JV curves for the HTL series before (no markers) and after (markers) operation.

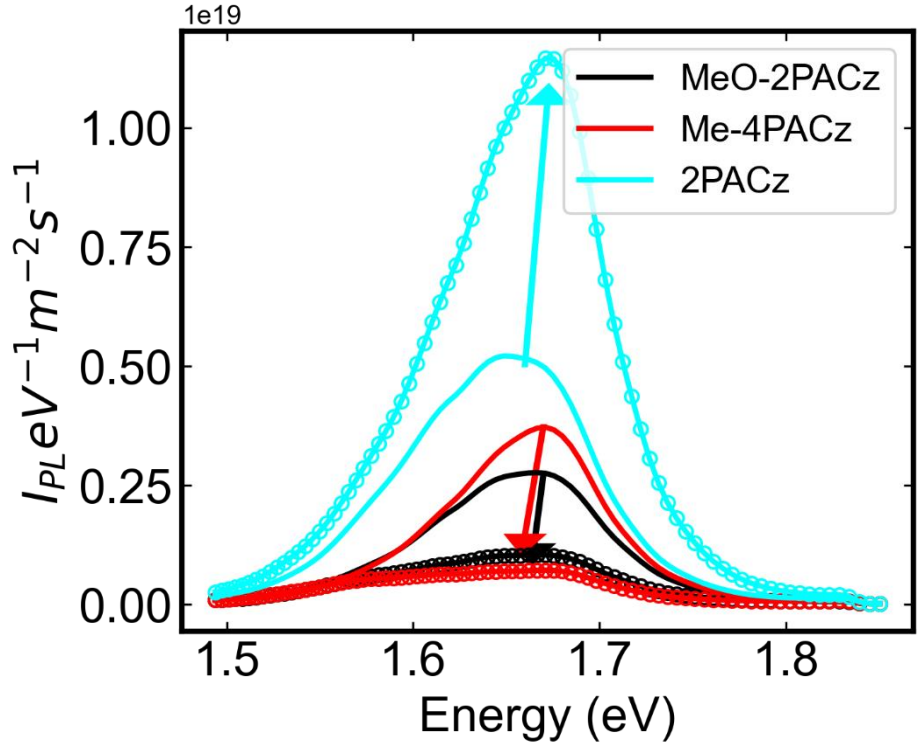

**Supplementary Figure 71.** Spatially averaged PL spectra at $V_{OC}$ before (solid lines no markers), and after (solid lines with markers) accelerated ageing. All devices are TCTH with varying HTL layers.

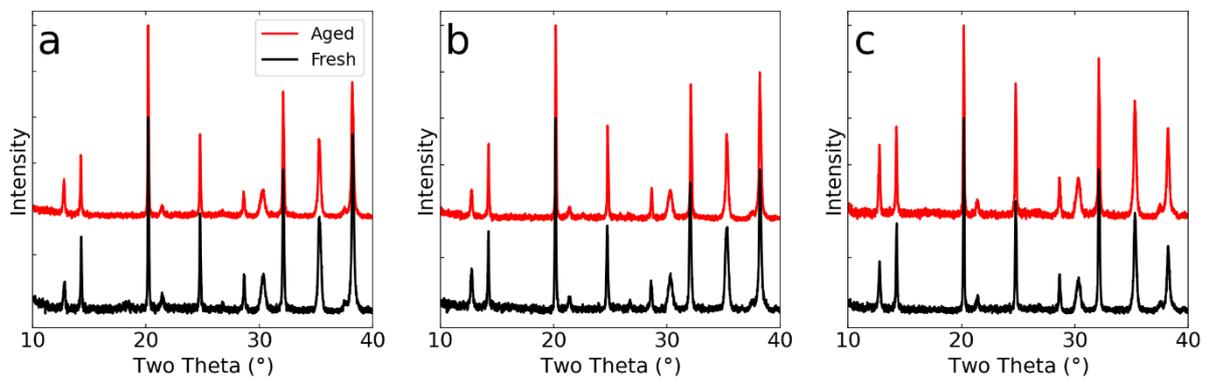

**Supplementary Figure 72:** Bulk XRD patterns of a) reference, b) PI passivated and c) LiF passivated 2PACz/TCTH devices before (black line) and after (red line) operational stress.

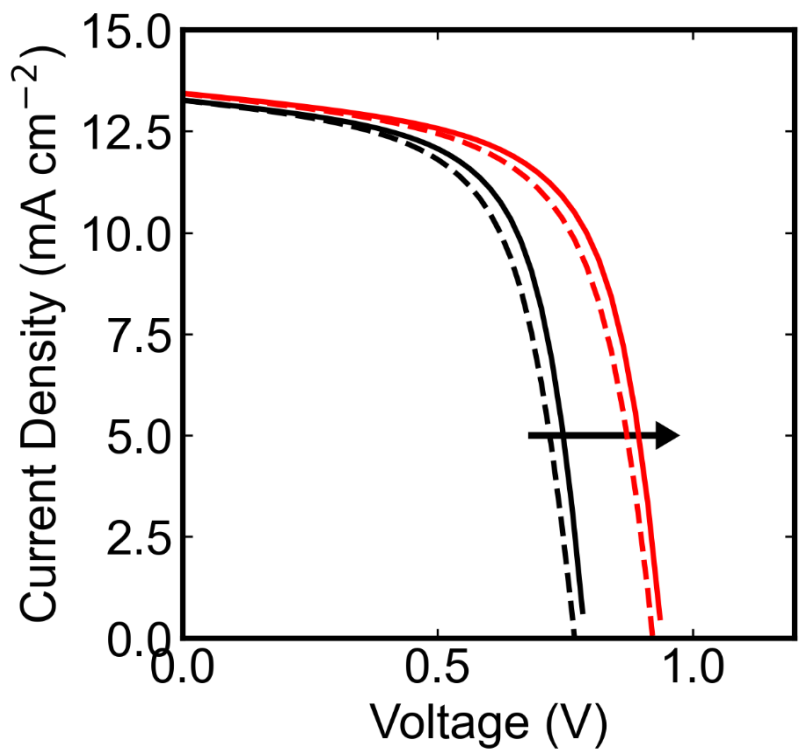

**Supplementary Figure 73. Consecutive JV curves in DCTH perovskite solar cell.** JV curves were measured ~1 hour apart with continuous light soaking showing that the $V_{OC}$ is highly transient after accelerated ageing.

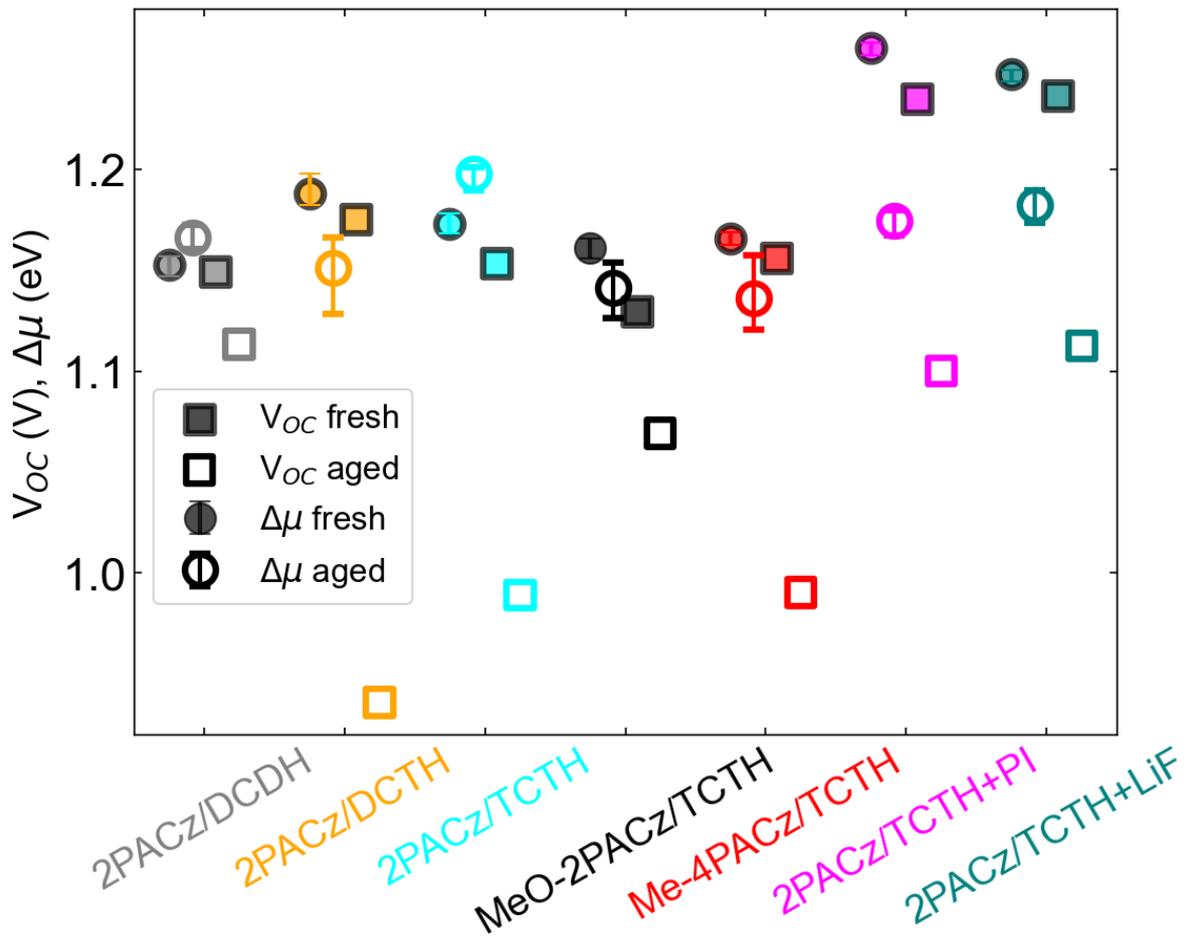

**Supplementary Figure 74:** Scatter plot of internal voltages (Δμ) and external voltages ($V_{OC}$) for representative devices of each type before and after operational stress test.

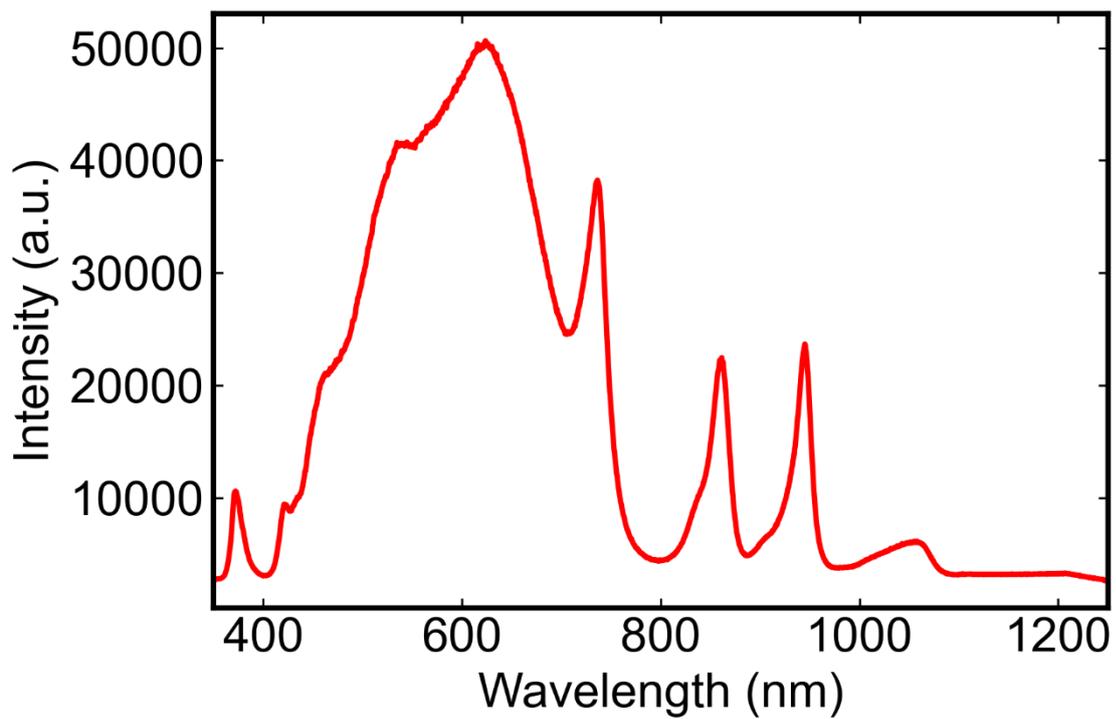

**Supplementary Figure 75. Spectrum of solar simulator used for accelerated ageing.**

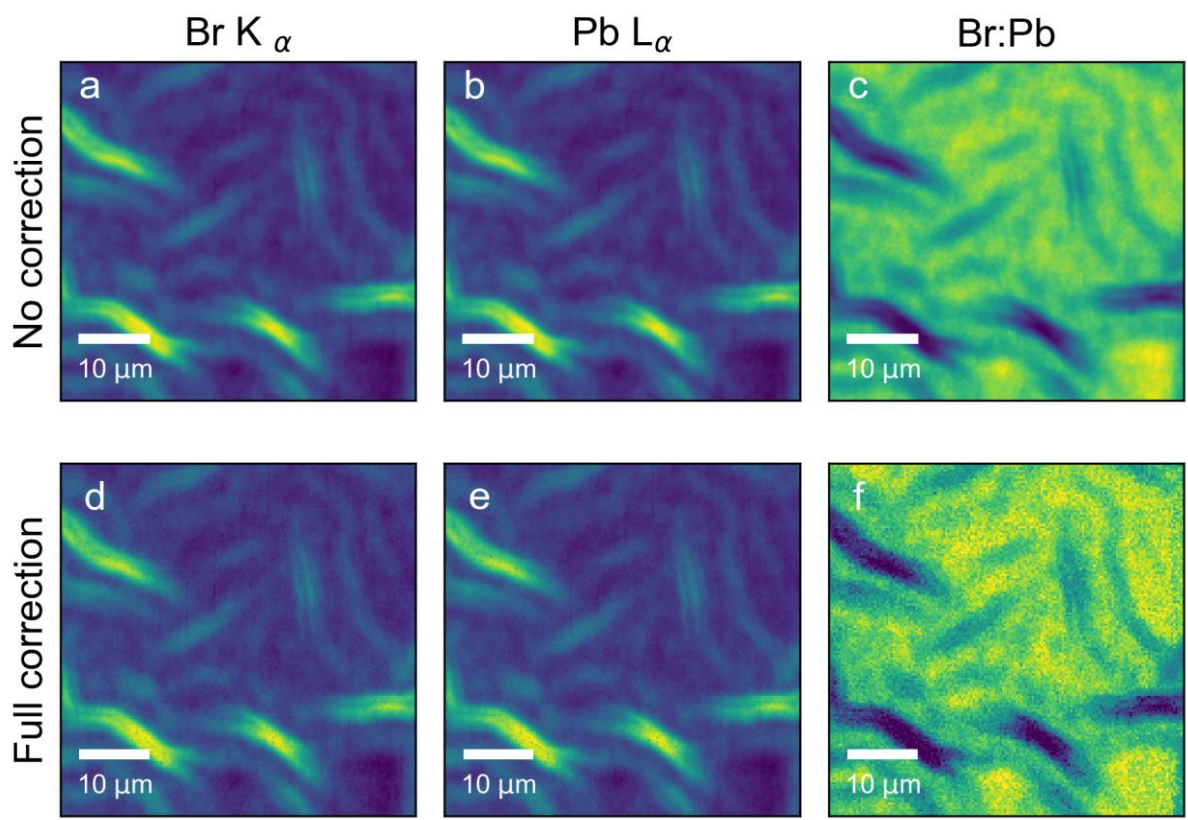

**Supplementary Figure 76**: a) Br K$_\alpha$, b) Pb L$_\alpha$ and c) Br:Pb ratio of a 2PACz/DCDH solar cell with uncorrected data. d) Br K$_\alpha$, e) Pb L$_\alpha$ and f) Br:Pb ratio of a 2PACz/DCDH solar cell with fully corrected data.